\pdfoutput=1 
\documentclass[twocolumn,showpacs,preprintnumbers,amsmath,amssymb]{revtex4}

\usepackage{graphicx}
\usepackage{dcolumn}
\usepackage{bm}

\def\bea{\begin{eqnarray}}
\def\eea{\end{eqnarray}}

\def\pp{\mbox{$p$-$p$}}
\def\ee{\mbox{$e^+$-$e^-$}}

\def\auau{\mbox{Au-Au}}
\def\cucu{\mbox{Cu-Cu}}
\def\pbpb{\mbox{Pb-Pb}}
\def\aa{\mbox{A-A}}
\def\nn{\mbox{N-N}}
\def\ss{\mbox{S-S}}
\def\pt{$p_t$}
\def\yt{$y_t$}
\def\mt{$m_t$}

\begin{document} 


\title{A critical review of RHIC experimental results
}

\author{Thomas A.\ Trainor}\affiliation{CENPA 354290, University of Washington, Seattle, WA 98195}


\date{\today}

\begin{abstract}
The relativistic heavy ion collider (RHIC) was constructed to achieve an asymptotic state of nuclear matter in heavy ion collisions, a near-ideal gas of deconfined quarks and gluons denoted quark-gluon plasma or QGP. RHIC collisions are indeed very different from the hadronic processes observed at the Bevalac and AGS, but high-energy elementary-collision mechanisms are also non-hadronic. The two-component model (TCM) combines measured properties of elementary collisions with the Glauber eikonal model to provide an alternative asymptotic limit for \aa\ collisions. RHIC data have been interpreted to indicate formation of a {\em strongly-coupled} QGP or ``perfect liquid.'' In this review I consider the experimental evidence that seems to support such conclusions and alternative evidence that may conflict with those conclusions and suggest different interpretations.
\end{abstract}


\maketitle

\section{Introduction}

The experimental program addressing high-energy heavy ion (\aa) collisions initiated at the Berkeley Bevalac and later extended to the Brookhaven alternating gradient synchrotron (AGS) reported collective motion (flow) of nucleon clusters, nucleons and hadronic resonances for center-of-mass collision energies below 10 GeV~\cite{bevalac}. Arguing by analogy with the thermodynamics of ordinary matter and QCD asymptotic freedom it was conjectured that increased beam energy might lead to energy densities and temperatures sufficiently high to ``melt'' colorless hadrons to produce a thermalized medium of deconfined weakly-coupled colored quarks and gluons referred to as a quark-gluon plasma (QGP)~\cite{edwards}. Detection of collective flow in the produced medium might support the inference of thermalization required to conclude that a new form of QCD matter had been achieved. The Brookhaven relativistic heavy ion collider (RHIC) was proposed to achieve such energy densities and to produce a QGP.

A different context (HEP) emerges from observations of high-energy elementary hadronic collisions and their QCD description. Two mechanisms appear to dominate high-energy \pp\ collisions:  (a) projectile dissociation to low-transverse-momentum $p_t$ hadrons extending over a broad longitudinal rapidity $y_z$ interval according to a parton distribution function (PDF) and (b) infrequent large-angle scattering of some partons according to perturbative QCD (pQCD) followed by nonperturbative fragmentation to collimated jets of hadrons with a broad distribution of momenta, possibly extending to large as well as small values. That scenario is the two-component (soft+hard) model (TCM) of hadronic collisions.

Prior to acquisition of RHIC data the possibility existed that \aa\ collisions might be just linear superpositions of \nn\ collisions following the TCM scenario. The null hypothesis for hadron production would then be fragmentation, either longitudinal (soft) or transverse (hard). The HEP limiting case balances the QGP alternative in which most hadrons emerge from a thermalized quark-gluon medium by a little-understood process. The reality of high-energy \aa\ collisions might lie somewhere between, depending on imposed conditions, and analysis methods should anticipate the full range of possibilities.

The RHIC theoretical and experimental programs have evolved to emphasize the context extrapolated from lower energies. A QGP is formed by early thermalization of a gluon condensate (color glass condensate or CGC) also called a glasma~\cite{cgc1}. Hadron production is dominated by freezeout from the resulting flowing bulk medium composed of quarks and gluons. Jet manifestations are confined to a small fraction of all hadrons, those emerging as fragments from high-energy scattered partons.

The hadron transverse momentum spectrum is accordingly divided into three intervals. Low $p_t$ 0-2 GeV/c should include manifestations of soft processes within the bulk medium including radial and elliptic flows.  Statistical models of hadron species abundances are applicable to the thermalized system. Fluctuations of conserved quantities may reveal proximity to a phase boundary separating hadronic and partonic phases. Hydrodynamic descriptions of flows may reveal a QCD equation of state.

Intermediate \pt\ 2-5 GeV/c is assumed to represent a transition interval between bulk medium freezeout and pQCD jet production. The theoretical description is uncertain and may be driven by data phenomenology. Quark recombination or coalescence models applied to jet-medium interactions are proposed.

High $p_t$ above 5 GeV/c should be dominated by hadrons from jets. A pQCD description of parton scattering and fragmentation to jets is applicable. Evidence for modified jet production (jet quenching) resulting from parton energy loss and other interactions  in the dense colored medium is sought. Isolation of jets from a large combinatoric background presents technical challenges.

After three years of data acquisition and analysis the theoretical consensus was developed that not a weakly-coupled but a {\em strongly-coupled} QGP or sQGP is formed in central 200 GeV \auau\ collisions. The material apparently exhibits a very small viscosity and little dissipation, suggesting the term ``perfect liquid.''

The experimental and theoretical output from RHIC to date is quite large. The scope of this review is restricted as follows. Topics dealing directly with claims for formation of a QGP or ``perfect liquid'' are preferred. Results from three of the four experiments, those emphasizing the central rapidity region (collision center of momentum) are preferred. A small sample of figures and papers for each topic is presented, those seeming to illustrate major results for the topic. Conventional analysis according to assumptions outlined above is compare with alternative analysis that may be more compatible with the HEP context.

This paper is organized as follows:
Sec.~\ref{meth} reviews conventional and alternative analysis methods,
Sec.~\ref{first} reviews results from the first three years of RHIC operations,
Sec.~\ref{white} summarizes white papers from the four RHIC collaborations,
Sec.~\ref{perfect} reviews theoretical arguments leading to claims for formation of a strongly-coupled QGP or ``perfect liquid.''
Sec.~\ref{spectra} reviews subsequent progress on spectrum analysis,
Sec.~\ref{flow} summarizes results for elliptic flow analysis,
Sec.~\ref{jets} reviews jet studies and
Sec.~\ref{flucts1} reviews fluctuation analysis.
Sec.~\ref{alternate} presents results from alternative analysis methods.
Secs.~\ref{disc} and~\ref{summ} present discussion and summary.

\section{Analysis Methods} \label{meth}

Many analysis methods developed for RHIC data prior to start-up assumed an established theoretical context. The \aa\ final state should emerge from an expanding bulk medium, the result of ``freezeout'' or decoupling of that medium to free-streaming particles. The properties of the medium, whether a hadronic or quark-gluon scenario better described it, were of primary interest. Access to medium properties should be through hadron yields and spectrum analysis, with the so-called blast-wave model preferred for spectra. Bulk properties should be accessed primarily at low $p_t < 2$ GeV/c.

Some properties of the medium might also be probed by high-energy scattered partons fragmenting to jets. Would some jets be absorbed by the medium, would some jets survive with modification by the medium? How would the medium in turn be modified by jets? Jet information and theoretical definition should be best accessed at high $p_t >$ 5 GeV/c where a pQCD description of jet production and jet fragmentation (DGLAP) is valid, and jet fragments dominate the soft component of spectra.

That QGP scenario is an alternative to the high-energy physics description of \aa\ collisions in terms of ``direct reactions'' dominated by projectile-nucleon dissociation and in-vacuum parton scattering to jets with fragment distributions extending down to small hadron momenta. In the latter case a different combination of analysis methods might be more appropriate.

\subsection{Conventional methods}

\aa\ collision centrality is measured by charge multiplicity density $dn_{ch}/d\eta$ or participant-nucleon number $N_{part}$ which emphasizes the upper half of the \aa\ total cross section and tends to obscure the lower half. Spectra are presented on transverse momentum $p_t$ which tends to emphasize structure at higher \pt\ for a given \pt\ acceptance interval. Spectra from more-central \auau\ collisions are compared directly to \pp\ or d-Au collisions. There is no reference model for transparent \aa\ collisions (i.e., for no partonic or hadronic {\em secondary} interactions).
Jet modification is measured by spectrum ratios  $R_{AA}$ or $R_{CP}$ which include soft as well as hard spectrum components. Sensitivity to jet modifications (hard component) is then confined to $p_t > 4$ GeV/c. 

Structure (anisotropy) in 1D azimuth correlations is measured primarily by $v_2$, the square root of a {\em per-pair} correlation measure (defined below), interpreted to represent elliptic flow. Other correlation structure (e.g., representing jets) is isolated by subtracting a background based on numerical $v_2$ measurements. Jet-related 1D and 2D angular correlations are further conditioned by {\em trigger-associated} $p_t$ cuts based on assumptions about the relative contribution of jets to spectra and correlations. Jets may be assumed to retain the same form as in \pp\ collisions. Any deviations from the \pp\ jet form are attributed to {\em nonjet} contributions (e.g., a ``ridge''). Fluctuations are measured by {\em per-pair} statistical quantities that include a trivial but dominating $1/n_{ch}$ or $1/N_{part}$ trend.

\subsection{Alternative methods} \label{alt}

\aa\ collision centrality is measured by fractional cross section $\sigma/\sigma_0$, fractional impact parameter $b/b_0$ or mean participant-nucleon path length $\nu = 2N_{bin} / N_{part}$. Those centrality measures provide balanced comparisons of peripheral and central collision systematics over the full centrality range down to \nn\ (nucleon-nucleon) collisions. Spectra are plotted on transverse rapidity $y_t = \ln[(m_t + p_t)/m_h]$ ($m_h$ is the hadron mass, with default $m_\pi$) which provides more-balanced visual comparison of low-$p_t$ vs high-$p_t$ structure. Jet modifications are measured by differential ratio $r_{AA}$ which compares only the hard components of spectra, providing significant access to jet structure and modifications down to 0.5 GeV/c~\cite{hardspec}.

Correlations are measured with {\em per-particle} measure $\Delta \rho / \sqrt{\rho_{ref}}$ which is by definition independent of system size under combination of {\em un}\,correlated subsystems~\cite{anomalous}. Four charge combinations are studied: LS (like-sign), US (unlike-sign), CI = LS + US (charge-independent), CD = LS $-$ US (charge-dependent). Significant features in correlations are modeled by simple functional forms not motivated by a priori physical models~\cite{axialci,anomalous}.

 Initial studies focus on {\em minimum-bias} correlations (no $p_t$ cuts). Thus, 100\% of the jet structure for all centralities is considered. A Glauber linear superposition (GLS) reference (transparent \aa\ collisions) is defined based on \pp\ measurements and the Glauber model of \aa\ collisions. Significant deviations from GLS are emphasized.

A {\em nonjet quadrupole} component $\langle \cos[2(\phi_1 - \phi_2)] \rangle$ is isolated from jet structure by 2D model fits. A {\em quadrupole spectrum} inferred from its $p_t$ dependence can be compared directly with hydro boost predictions. Fluctuations are also measured with a per-particle quantity. The scale (bin-size) dependence of fluctuations so measured can be inverted to infer underlying angular correlations that are directly and simply comparable with theory.

\section{The first three years}  \label{first}

The main goal of the RHIC in 2000 remained detection and study of the QGP as a near-ideal gas of quarks and gluons by observing certain ``signals'' in heavy ion collision data. A variety of analysis methods was prepared to search for those signals. Analysis methods were tested on large volumes of Monte Carlo data prior to first RHIC data in August, 2000. In this section we review some of the more prominent early results that seemed to support claims for formation of a QGP in \auau\ collisions. 

\subsection{Hadron production}

Total charge multiplicity $n_{ch}$ for most-central \aa\ collisions, the event-number minimum-bias distribution on $n_{ch}$ and the charge multiplicity variation with centrality provide initial indications of new physics that might be accessed by a higher collision energy.

\begin{figure}[h]
  \includegraphics[width=1.65in,height=1.65in]{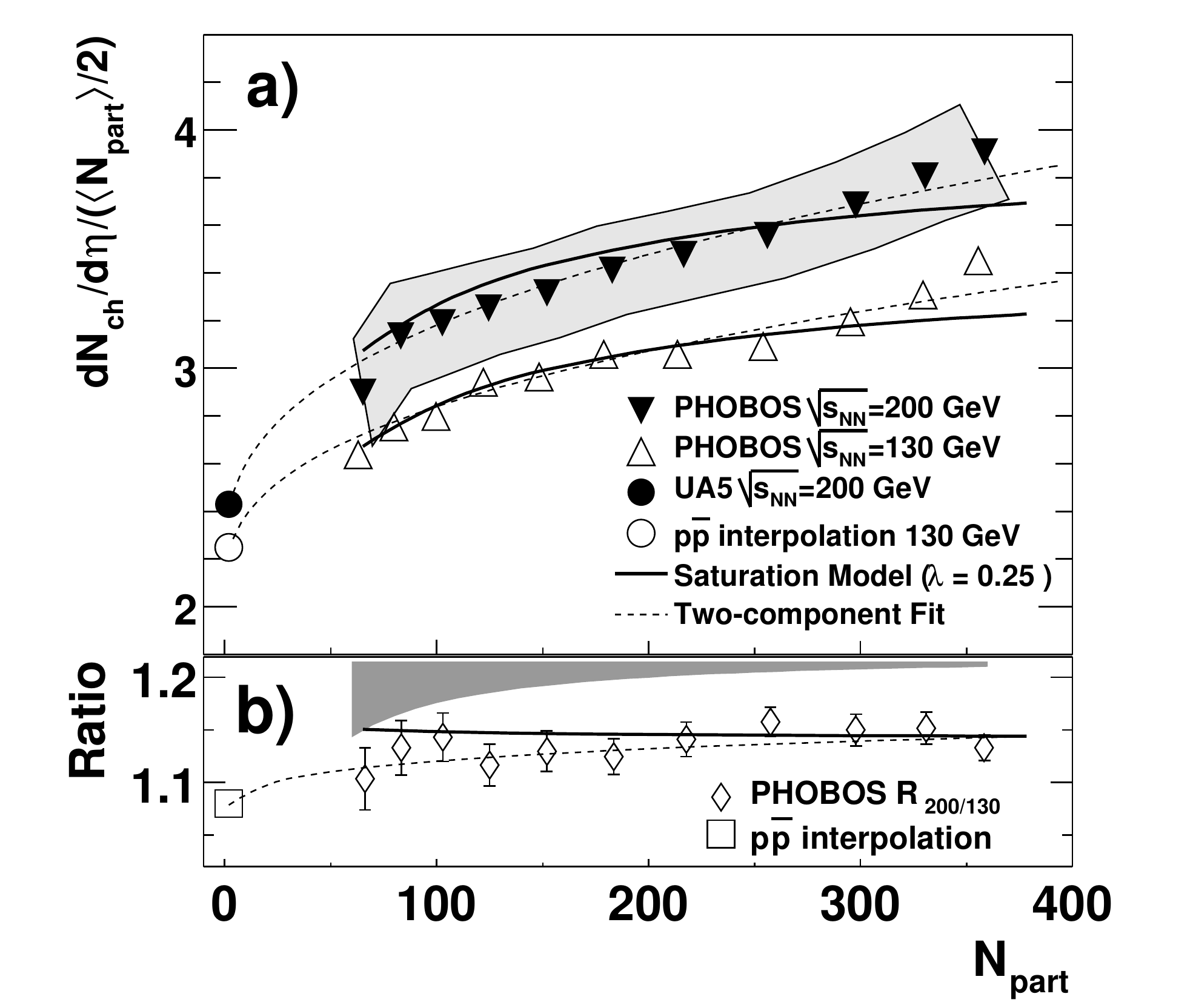}
  \includegraphics[width=1.65in,height=1.63in]{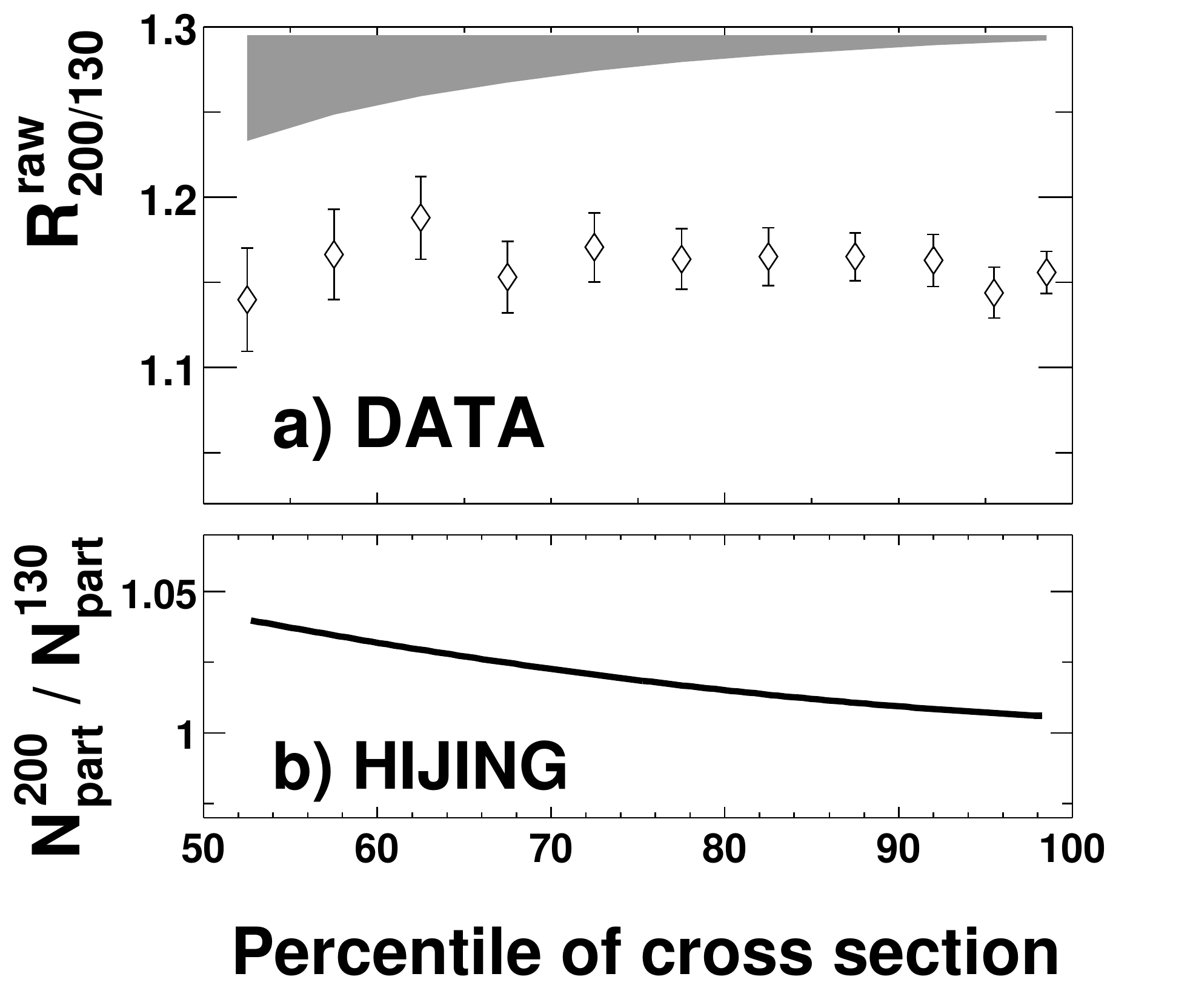}
\caption{\label{global}
Left: Charge multiplicity density per participant nucleon vs  $N_{part}$ for 130 and 200 GeV \auau\ collisions (points) compared to theoretical predictions (curves)~\cite{global3}. The lower panel indicates that the energy ratio does not vary significantly with centrality.
Right: The energy ratio for data (points) compared to that predicted from the HIJING Monte Carlo (curve) based on minijet production in a two-component model, showing apparent disagreement.
 }  
 \end{figure}

Figure~\ref{global} (left) shows charge multiplicity per participant-nucleon pair vs $N_{part}$ near $\eta = 0$ for 200 and 62 GeV \auau\ collisions compared to theoretical predictions~\cite{global3}. The data span the most-central 40\% of the \auau\ total cross section and increase relative to a \pp\ value near 2.4 to a value near 3.8 for central collisions. The theoretical curves represent a two-component (minijet) model~\cite{kn} and a saturation-scale or CGC model~\cite{cgcmult}.

Figure~\ref{global} (right) shows the ratio of 200 to 62 GeV yield data compared to HIJING~\cite{hijing} results representing the two-component model. The conventional HEP description of \aa\ collisions assuming linear superposition of \nn\ collisions includes minijet production in a TCM.  The HIJING Monte Carlo is elected to represent that model~\cite{liwang}. The different trends in Fig.~\ref{global} (right) are then interpreted to reject the TCM in favor of a  color glass condensate (CGC) model~\cite{cgc1}. Such yield trends are considered essential to establish initial conditions for hydrodynamic theory calculations and to test the CGC hypothesis. The apparent deviation of the hadron production data from HIJING is characterized as a ``too slow'' increase of the hadron yield with increasing centrality and collision energy~\cite{global1,global2}.

\subsection{Low-$\bf p_t$ spectra and radial flow}

In the low-\pt\ interval 0-2 GeV/c particle production is assumed to be dominated by freezeout (chemical and kinetic) from a flowing bulk medium, possibly a thermalized QGP. The statistical model should predict hadron species abundances corresponding to a chemical freezeout temperature $T_{chem}$ and relevant chemical potentials $\mu$. Radial flow should be manifested as a deviation of \pt\ or \mt\ spectra from a Maxwell-Boltzmann reference function. So-called ``blast-wave'' fits are used to model spectra with parameters $T_{kin}$, the kinetic freezeout temperature and $\langle \beta_t \rangle$, the mean radial speed.

 \begin{figure}[h]
\includegraphics[width=3.in,height=1.7in]{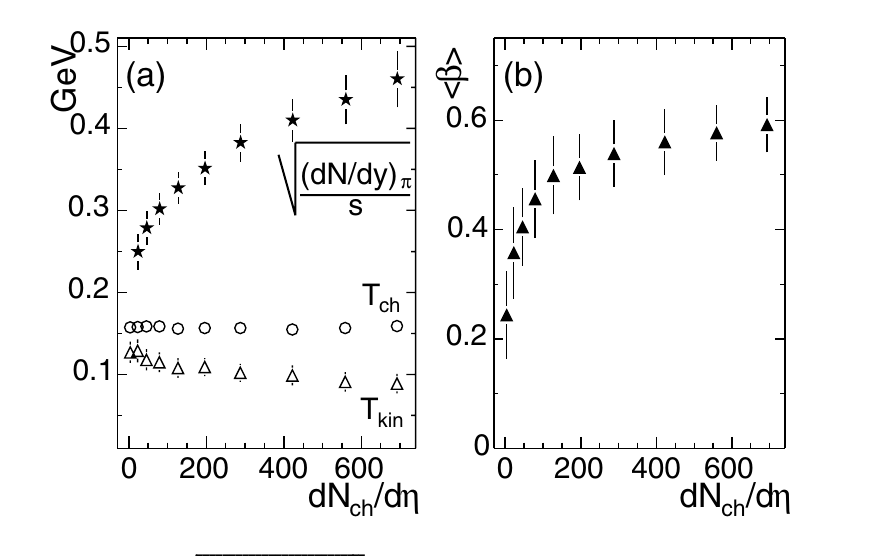} \hfill
\caption{\label{bulk}
Bulk-medium parameters inferred from statistical model and blast-wave model fits to yields and $p_t$ spectra plotted vs total charge density for identified hadrons from 200 GeV \auau\ collisions~\cite{blastwave}. 
Left: Fitted chemical $T_{ch}$ and kinetic $T_{kin}$ freezeout temperatures.
Right: Fitted mean transverse speed $\langle \beta_t \rangle$.
 }  
 \end{figure}

Figure~\ref{bulk} (left) shows chemical $T_{ch}$ and kinetic $T_{kin}$ freezeout temperatures (open points) vs \auau\ centrality (charge multiplicity) inferred from blast-wave fits to PID spectra for pions, kaons and protons below $m_t - m_h = 0.6$ GeV/$c^2$~\cite{blastwave}. The fit results are typical:   $T_{ch}$ remains near 150 MeV approximately independent of centrality. $T_{kin}$  falls from a higher value near 130 MeV for peripheral collisions to a lower value near 90 MeV for central collisions.

Figure~\ref{bulk} (right) shows mean radial speed $\langle \beta_t \rangle$ inferred from the same fits. The data increase from 0.25 for \pp\ collisions to about 0.6 for central \auau\ collisions. The combination of temperature and radial speed trends is interpreted to indicate isentropic expansion of a thermalized medium. The hadron species ratios are determined earlier, and the hadronic system then continues to do thermodynamic work until kinetic freezeout at a later time.

\subsection{Elliptic flow}

``Elliptic flow'' is the interpretation of the correlation structure represented by measure $v_2 = \langle \cos[2(\phi - \Psi_r)]\rangle$ inferred from two-particle correlations on azimuth relative to estimated reaction-plane angle $\Psi_r$.  The magnitude of elliptic flow at RHIC in comparison to  lower-energy data from the SPS has been of major interest. The $v_2$ values at the SPS fell below hydrodynamic predictions. Would (ideal) hydro apply to RHIC collisions? 

 \begin{figure}[h]
  \includegraphics[width=1.65in,height=1.75in]{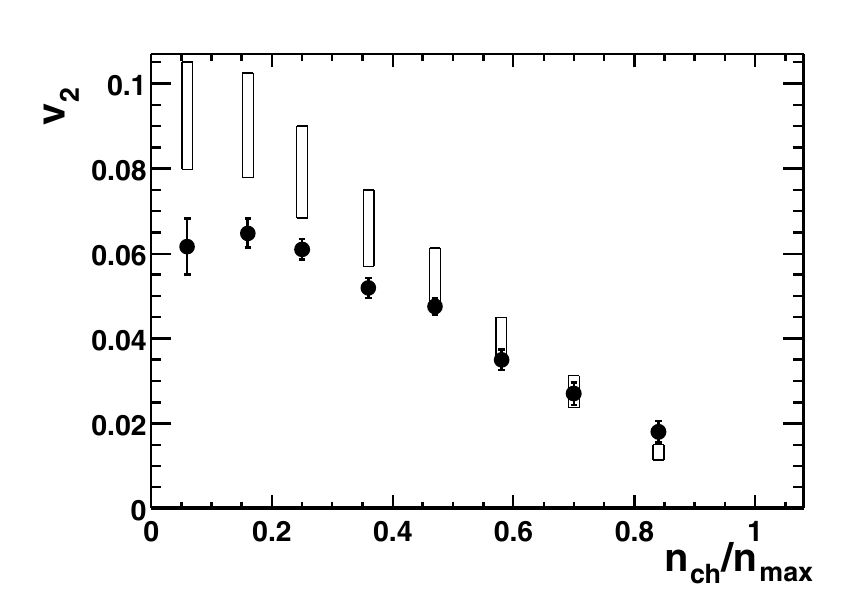}
  \includegraphics[width=1.65in,height=1.65in]{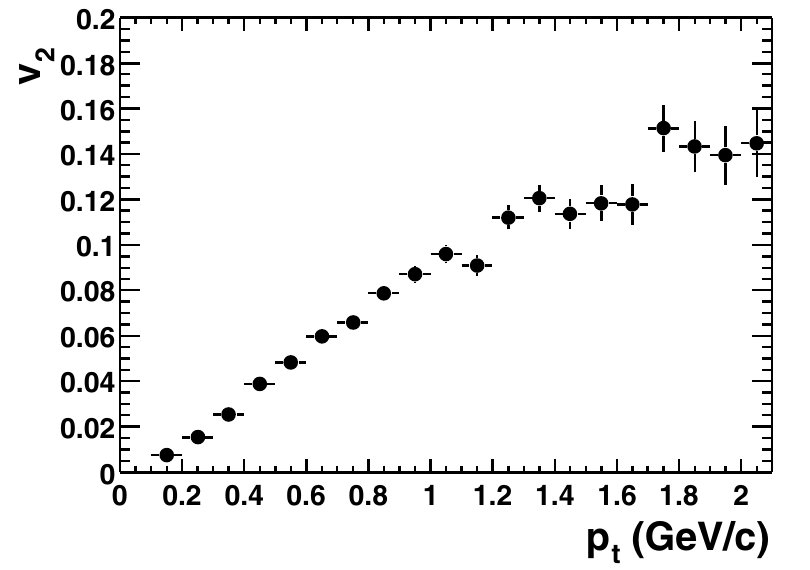}
\caption{\label{v2}
Elliptic flow $v_2$ data from 130 GeV \auau\ collisions~\cite{eflow1}.
Left: $p_t$-integral $v_2$ vs centrality measured by total charge multiplicity relative to the maximum observed value.
Right:  Minimum-bias $v_2$ vs transverse momentum $p_t$.
 }  
 \end{figure}

Figure~\ref{v2} shows the first $v_2$ measurements on centrality and \pt\ (and the first paper on any topic) from RHIC~\cite{eflow1}. The surprising result was the near equivalence of $v_2$ data (solid points) and (ideal) hydro theory (open boxes). The data also increase strongly with hadron $p_t$ (right panel). The $v_2(p_t)$ results are minimum-bias (averaged over centrality). In Ref.~\cite{eflow2} $v_2$ data are inferred from Fourier series fits to two-particle azimuth correlations and plotted vs fractional cross section for several $p_t$ bins. The data are interpreted to indicate transformation of initial-state \aa\ overlap geometry to final-state momentum asymmetry. Deviations of $v_2/\epsilon$ scaling at larger \pt\  are interpreted to indicate a possible change in the EoS. 

A similar analysis in Ref.~\cite{v2jets1} reported $v_2(p_t)$ results and apparent jet structure derived from azimuth correlations. Possible saturation of $v_2$ above 3 GeV/c would deviate from expectations for ideal hydrodynamics. The jet-like correlation structure may represent the first direct evidence for jets at RHIC.

\begin{figure}[h]
  \includegraphics[width=1.65in,height=1.55in]{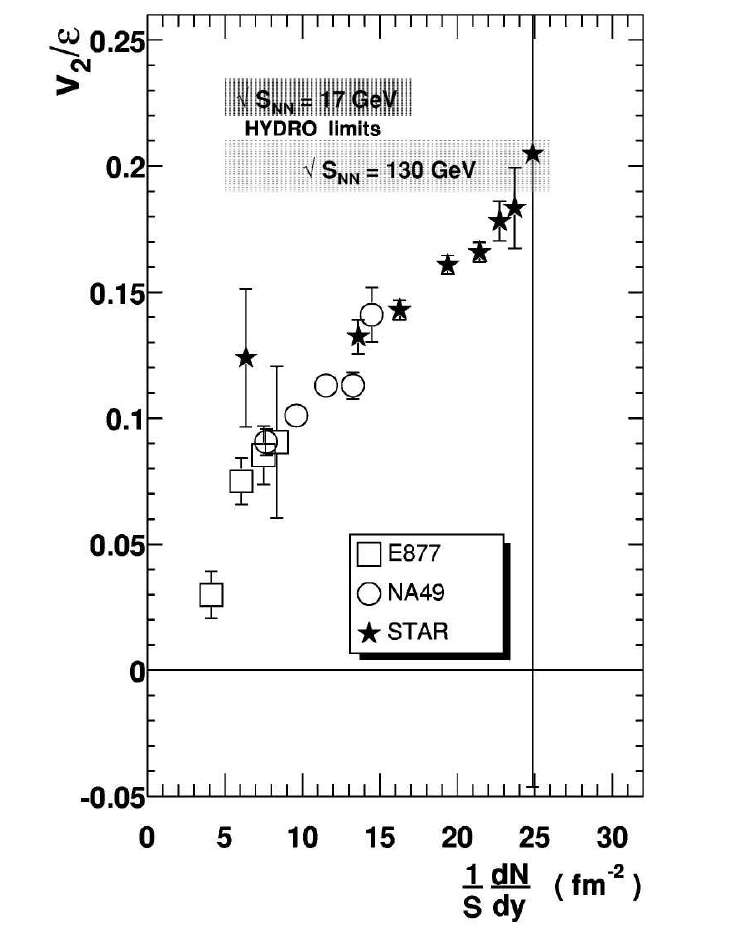}
  \includegraphics[width=1.65in,height=1.7in]{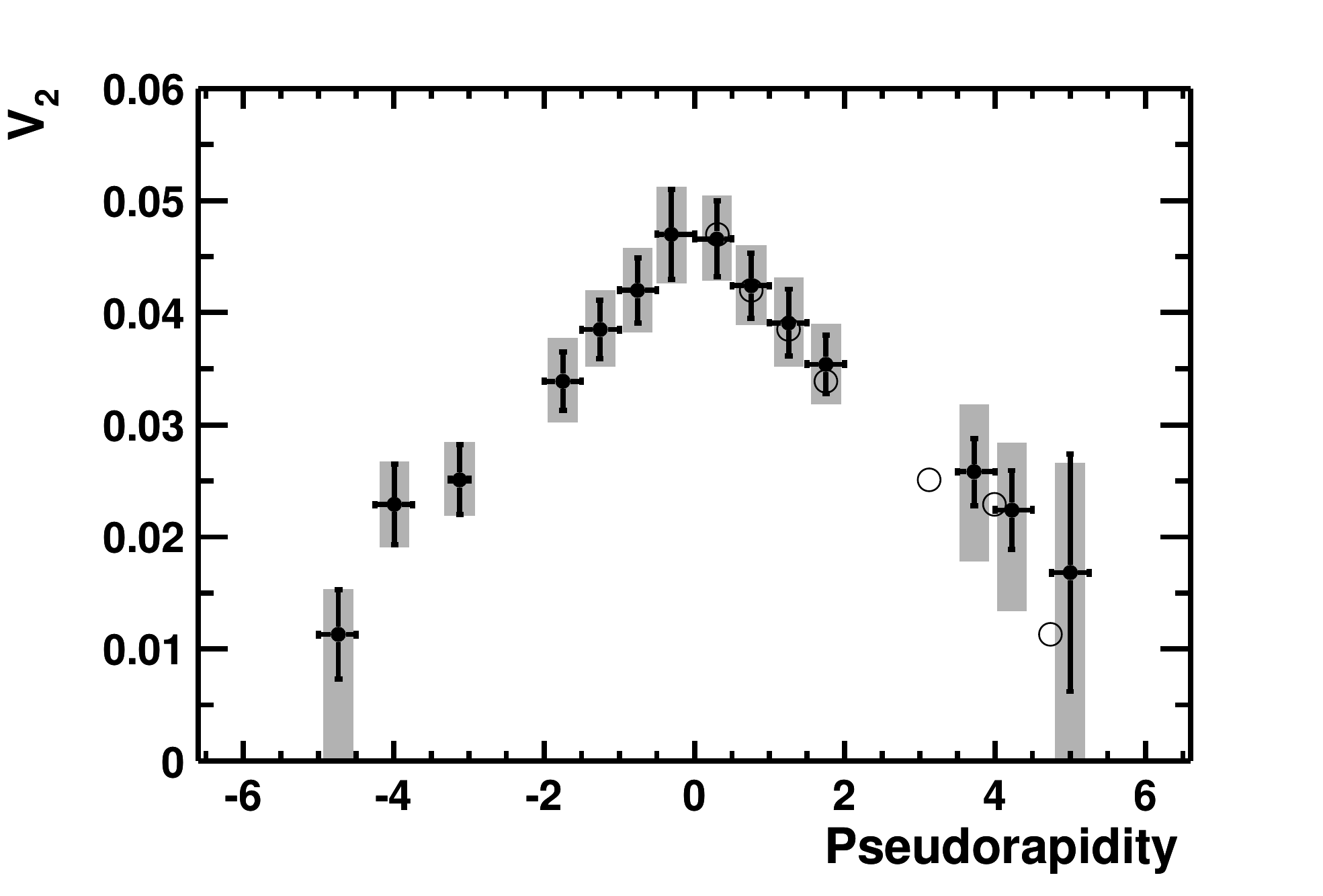}
\caption{\label{highway}
Left:  Ratio $v_2/\epsilon$ vs particle density $(1/S)dn_{ch}/d\eta$ for several collision energies, where $S$ is the \aa\ overlap area~\cite{v2prc1}. The data are compared with an ideal-hydro prediction (band).
Right: Distribution of $v_2$ on $\eta$ showing a sharply-peaked distribution~\cite{v2phob}.
 }  
 \end{figure}

Figure~\ref{highway} (left) compares ratio $v_2 / \epsilon$ from the AGS, SPS and RHIC vs hadron density that together seem to confirm a conjecture in Ref.~\cite{volposk}: The $v_2/\epsilon$ ratio (possibly measuring transport of initial-state nucleon spatial anisotropy to final-state hadron momentum anisotropy) should increase from near zero for peripheral collisions and low hadron densities (little secondary scattering) to a large value (possible saturation) for central collisions and a possible QGP~\cite{v2prc1}. $v_2/\epsilon$ values for most-central RHIC collisions appear to achieve a predicted ideal-hydrodynamic limit (band).

Figure~\ref{highway} (right) shows $v_2$ variation over a large $\eta$ interval (averaged over centrality)~\cite{v2phob}. Hydro theory has difficulty with such a trend (strong variation on $\eta$)~\cite{perfliq2}. And $v_2^2$ (what is actually inferred from two-particle correlations) would be even more sharply peaked. 

A critical test of the hydro interpretation of $v_2$ data is PID data for identified hadrons. If hadrons emerge from a radially-expanding bulk medium then ``mass ordering'' (hadron mass increase from left to right on $p_t$) should be observed below 2 GeV/c. The same mass trend is expected for $m_t$ spectra as a manifestation of radial flow.

 \begin{figure}[h]
  \includegraphics[width=1.65in,height=1.65in]{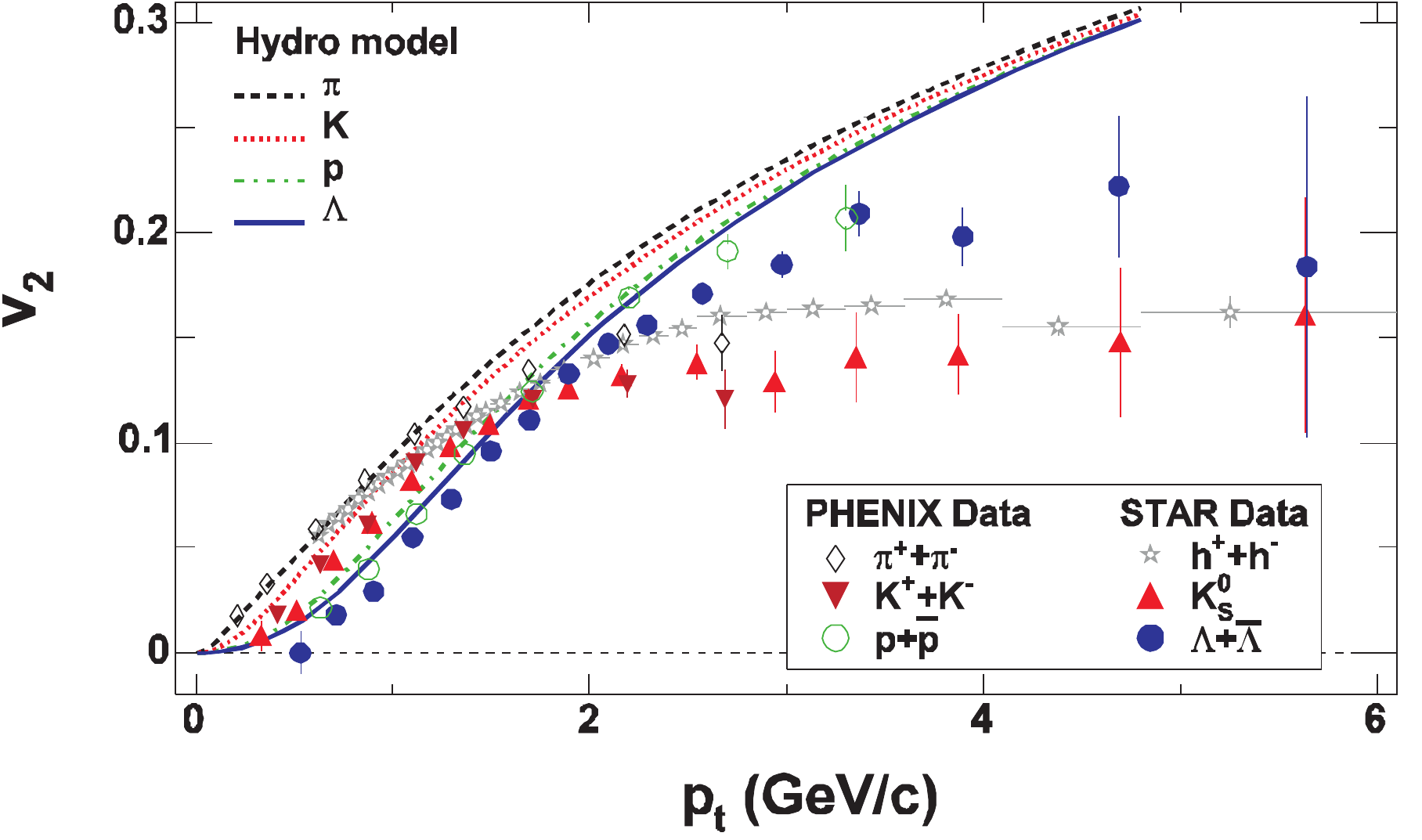}
  \includegraphics[width=1.65in,height=1.65in]{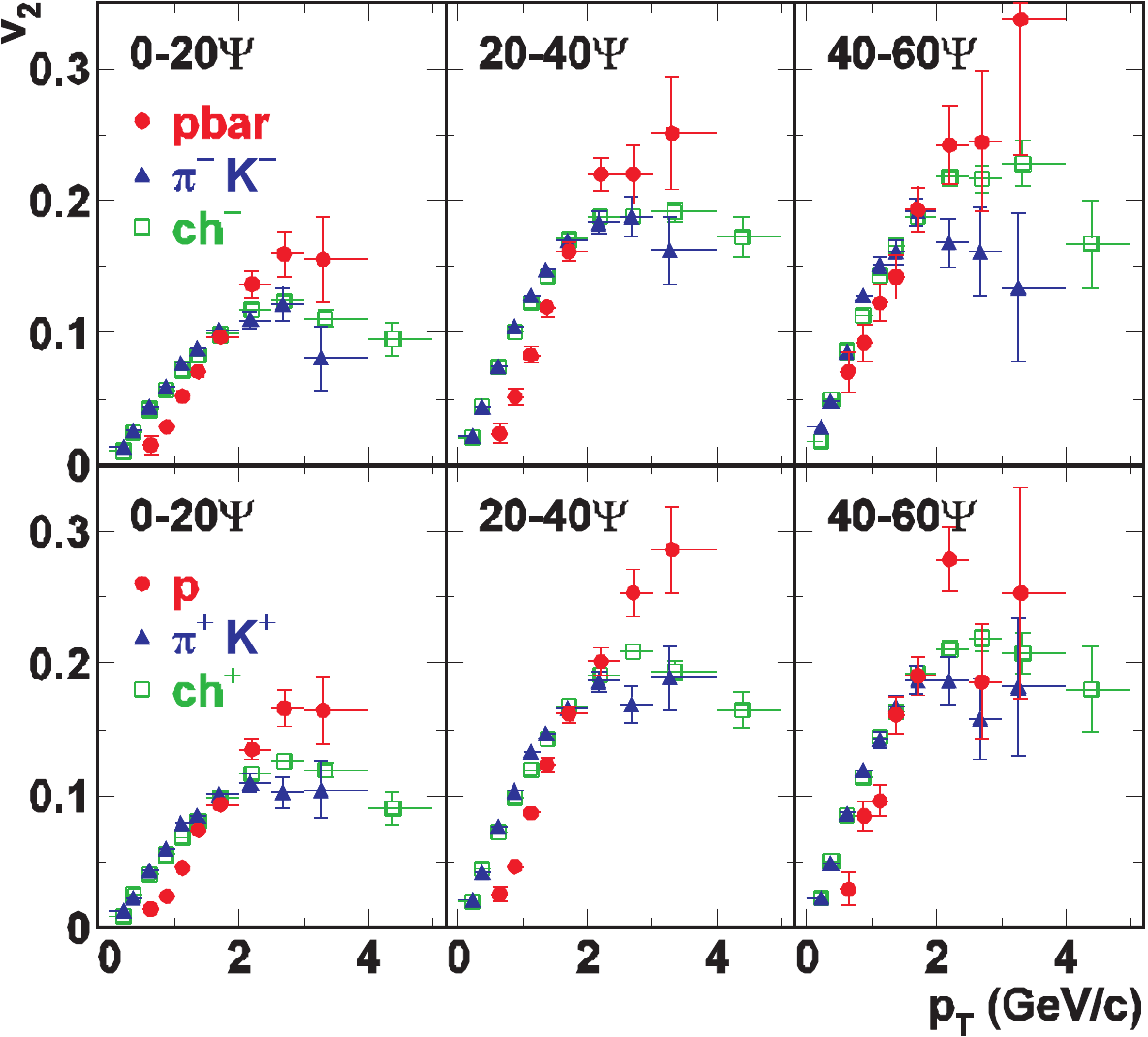}
\caption{\label{v2pid}
Left: $v_2$ vs $p_t$ for identified hadrons from minimum-bias 200 GeV \auau\ collisions (points) compared to ideal hydro predictions (curves)~\cite{2004}. Possible saturation of data above 2 GeV/c is indicated.
Right: PID $v_2(p_t)$ data for several \auau\ centrality intervals showing similar structure~\cite{v2ph1}.
 }  
 \end{figure}

Figure~\ref{v2pid} (left) shows minimum-bias PID $v_2$ data from Ref.~\cite{2004}. The requisite mass ordering below 2 GeV/c is observed, seeming to confirm a hydro interpretation. Above 2 GeV/c the data (points) deviate strongly from ideal hydro theory (curves), possibly indicating saturation. Fig.~\ref{v2pid} (right) shows similar spectra for several  \auau\ centrality bins and for (anti)particles at (top)bottom. Although the amplitude trend increases with decreasing centrality (as for $p_t$-integral data) the trends on \pt\ are very similar over a large centrality and \pt\ interval.

\subsection{High-$\bf p_t$ spectra and jet correlations}

If a color-deconfined medium is formed in \aa\ collisions scattered high-energy partons, as QCD colored objects, should interact with the medium and lose energy. The result is termed jet quenching: most jets are modified and some jets may be entirely absorbed by the medium. Jets should be manifested in single-particle \pt\ spectra (by assumption only at higher \pt) and in two-particle angular correlations. The reference for spectrum analysis is \pt\ spectra from \pp\ collisions. The spectrum ratio $R_{AA} = (1/N_{bin}) \rho_{AA}(p_t) / \rho_{pp}(p_t)$ compares jet manifestations in \aa\ collisions to {\em in vacuum} jets in \pp\ collisions. In case of linear superposition of \nn\ collisions within \aa\ collisions $R_{AA}$ should be 1 over a \pt\ interval where dijet production dominates spectra. Values substantially less than 1 would indicate jet quenching.

\begin{figure}[h]
  \includegraphics[width=1.6in,height=1.65in]{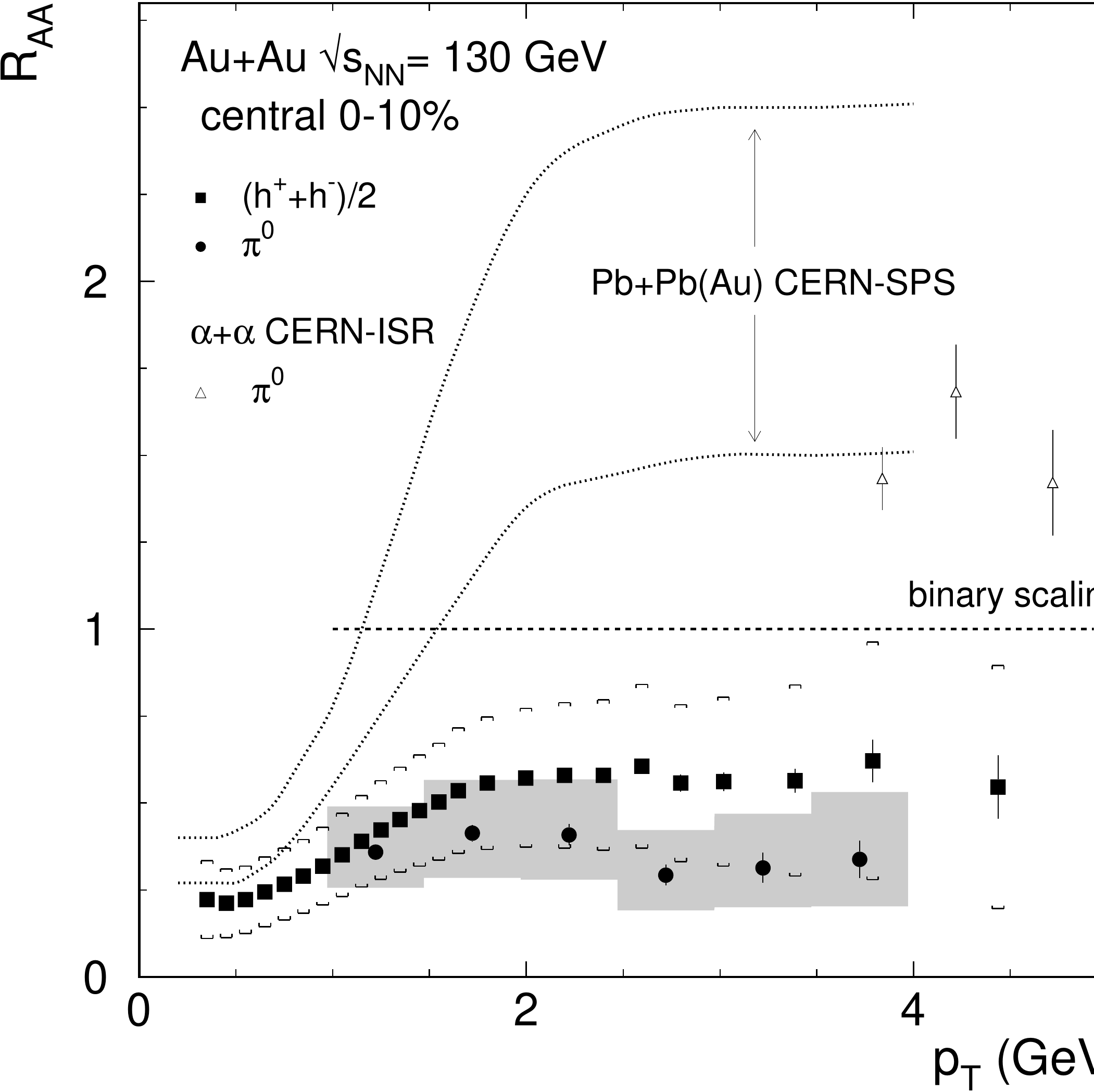} 
  \includegraphics[width=1.65in,height=1.7in]{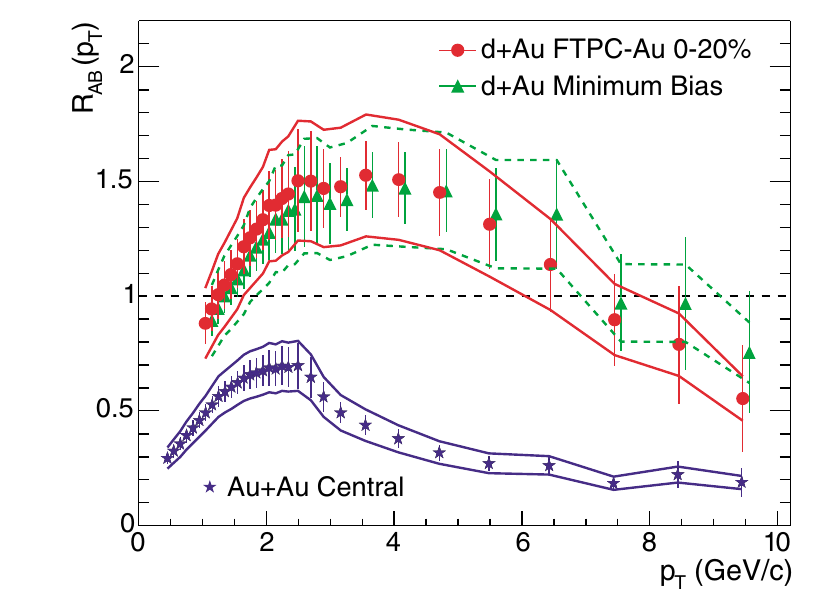}
\caption{\label{jetquench}
Left:   Spectrum ratio $R_{AA}$ vs \pt\ for 130 GeV \auau\ collisions (points) showing suppression (jet quenching) above 2 GeV/c in marked contrast to substantial enhancement (Cronin effect) for \pbpb\ collisions at 17 GeV (solid curves).~\cite{raaph1}
Right: Comparison between strong suppression in central 200 GeV \auau\ collisions (lower points) and no suppression in 200 GeV d-Au collisions (upper points), confirming no jet suppression in cold nuclear matter~\cite{raav21}
 }  
\end{figure}

Figure~\ref{jetquench} (left) shows data from Ref.~\cite{raaph1} (early results for $R_{AA}$)  that reveal values of $R_{AA}$ substantially less than 1 above 2 GeV/c, seeming to confirm substantial jet quenching in more-central \auau\ collisions. The RHIC results are compared with \pbpb\ results from the SPS at 17 GeV (solid curves) that {\em exceed} 1 as expected for initial-state effects (Cronin effect) and no jet quenching.

Figure~\ref{jetquench} (right) shows a comparison between \auau\ (lower points, central) and d-Au (upper points) showing that partons passing through cold nuclear matter (the latter case) do not manifest jet  quenching~\cite{raav21}. It is concluded that jet quenching is only observed for the hot and dense medium created in more-central \auau\ collisions.
 
\begin{figure}[h]
  \includegraphics[width=1.65in,height=1.65in]{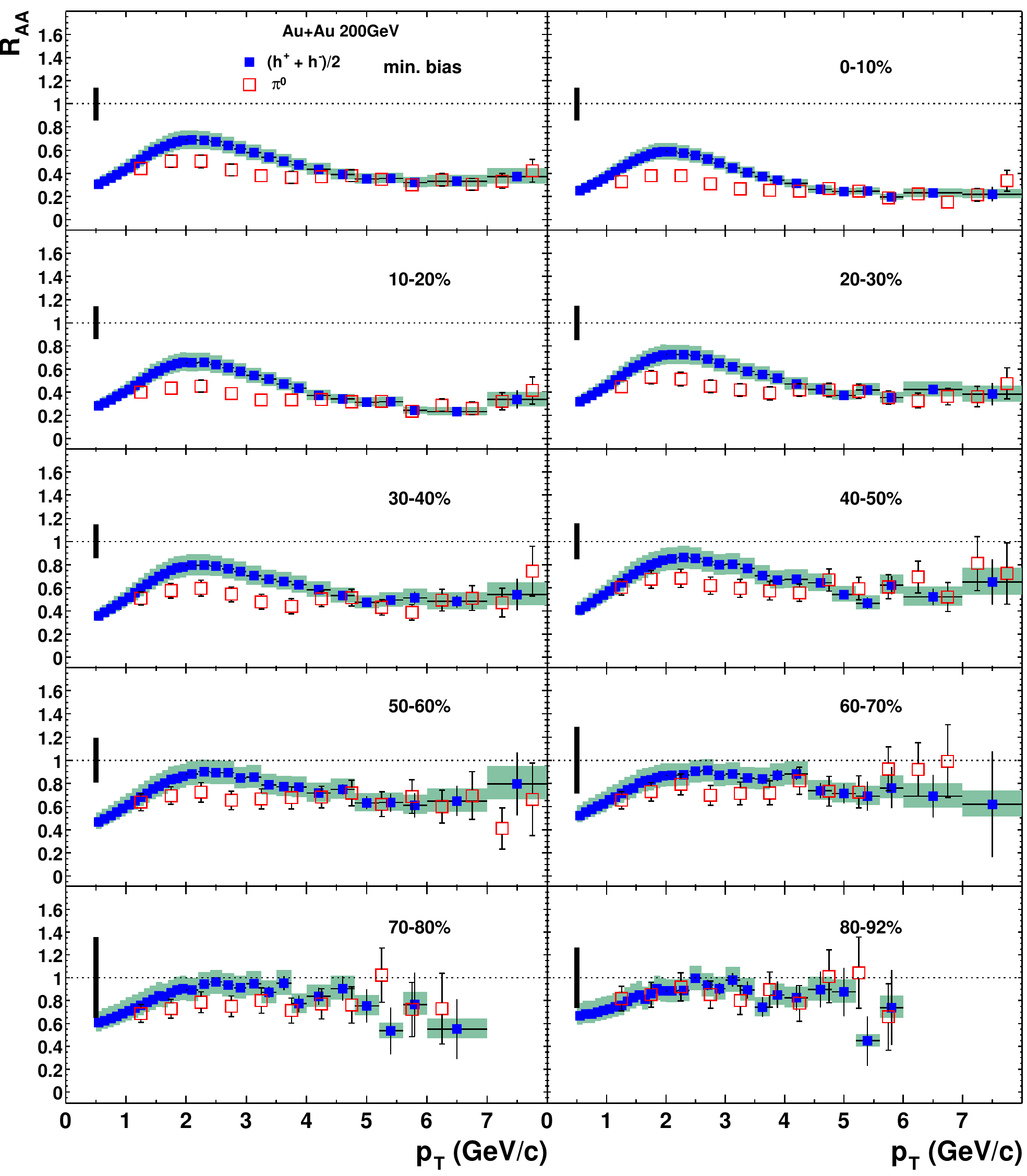}
  \includegraphics[width=1.6in,height=1.68in]{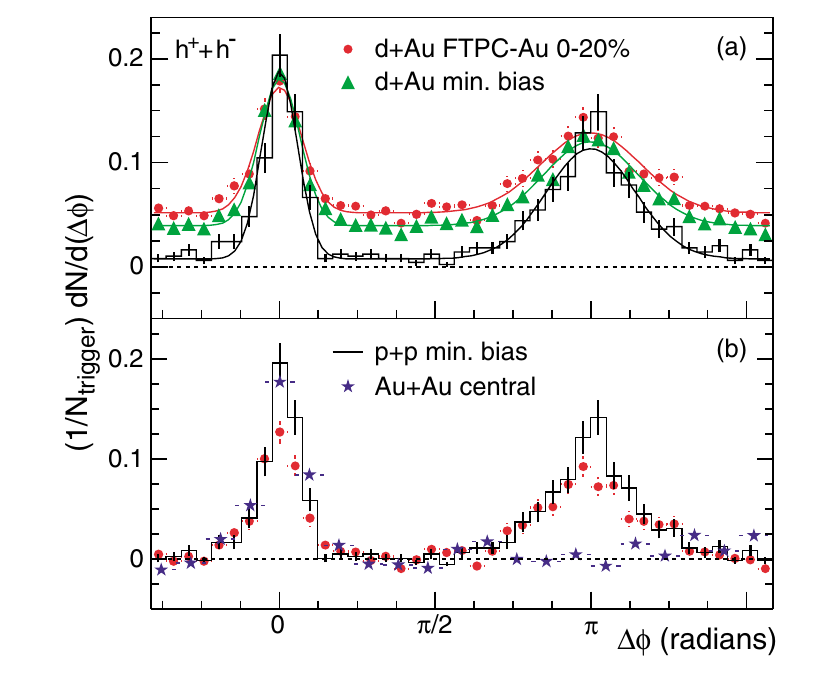} \hfill
\caption{\label{jets2}
Left:  Centrality dependence of $R_{AA}(p_t)$ from 200 GeV \auau\ collisions~\cite{raaph}. Strong suppression for central collisions reduces to negligible suppression for peripheral collisions.
Right: Jet-related dihadron azimuth correlations demonstrating disappearance of the away-side jet ($\Delta \phi = \pi$) in central 200 GeV \auau\ collisions (blue points) compared its presence for \pp\ collisions (histogram)~\cite{raav21}. The away-side peak is also present in d-Au collisions (red and green points).
 }  
\end{figure}

Figure~\ref{jets2} (left) from Ref.~\cite{raaph} shows $R_{AA}$ spectrum ratios vs \auau\ centrality from central (upper left) to peripheral (lower right). Jet quenching proceeds from strong ($R_{AA} \approx 0.2$) to negligible ($R_{AA} \approx 1$) over that centrality range, apparently illustrating the centrality evolution of QCD medium formation. Similar results were reported in Ref.~\cite{raast2}.

Figure~\ref{jets2} (right) from Ref.~\cite{raav21} shows dihadron correlations from \pp, d-Au and \auau\ collisions. A narrow same-side peak representing {\em intra}\,jet correlations from ``triggered'' jets is expected and observed at $\Delta \phi = 0$ for all cases. The critical question is whether an {\em inter}\,jet away-side peak representing recoil-partner jets appears at $\Delta \phi = \pi$. For \pp, d-Au and {\em peripheral} \auau\ collisions an away-side peak is observed representing the recoil partner of a dijet pair. For central \auau\ collisions the peak representing the away-side jet has disappeared.

Such high-\pt\ spectrum and correlation results are interpreted to demonstrate formation of a dense colored medium in which some fraction of jets is fully absorbed (opaque medium). Triggered jets are believed to be biased to emerge only near the medium surface (surface bias), and the recoil partners are then biased to pass through most of the dense medium and lose much or all of their energy in more-central \auau\ collisions. The evidence from $R_{AA}$ and dihadron correlation data in several collision systems is seen as consistent with that scenario. There is apparently 80\% reduction of all jet fragments in central \auau\ collisions, and triggered jets have no detectable recoil partner {\em given the imposed \pt\ conditions}.

\subsection{Minimum-bias 2D angular correlations}

Although 1D dihadron analysis on azimuth imposes specific trigger-associated \pt\ cuts considered appropriate for high-\pt\ jet analysis it is possible to study  angular correlations on both pseudorapidity and azimuth with no \pt\ conditions, resulting in {\em minimum-bias} or $p_t$-integral 2D angular correlations. 
There are no a priori expectations for correlation structure. For a thermalized system there might be no significant jet correlations when averaged over all combinatoric pairs.
The per-particle measure $N (r - 1)$ in the following plots is related to the later per-particle measure $\Delta \rho / \sqrt{\rho_{ref}}$~\cite{anomalous} by
\bea
\frac{\Delta \rho}{\sqrt{\rho_{ref}}} &=& \frac{N}{2 \pi \Delta \eta} (r - 1)
\eea
where $N / 2\pi \Delta \eta \approx \rho_0(b)$ is the average single-particle 2D angular density near midrapidity, as noted in~\cite{anomalous}. The updated measure is a density approximately independent of detector acceptance near mid rapidity.

\begin{figure}[h]
  \includegraphics[width=1.65in,height=1.65in]{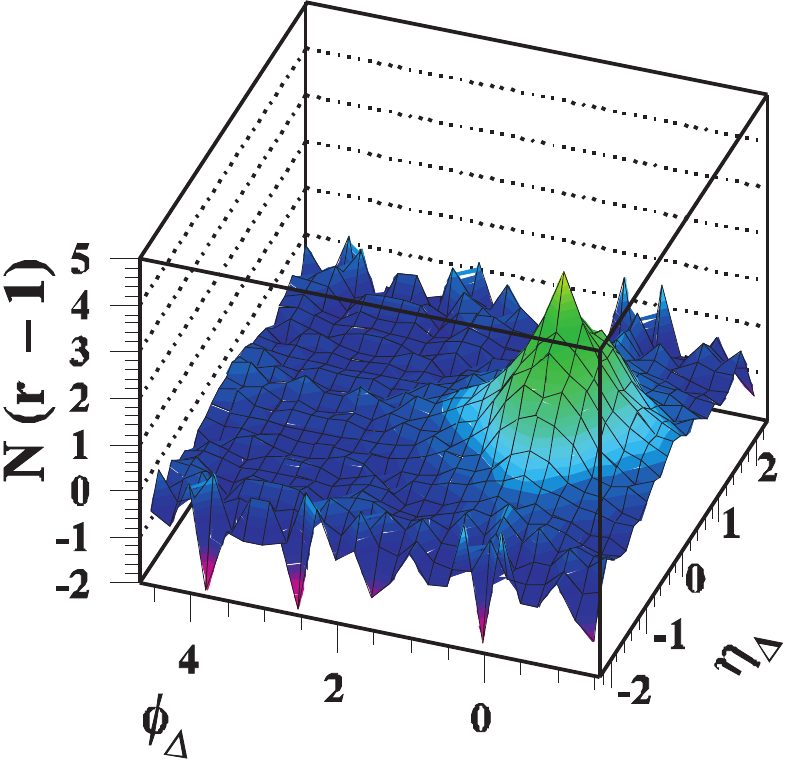}
  \includegraphics[width=1.65in,height=1.65in]{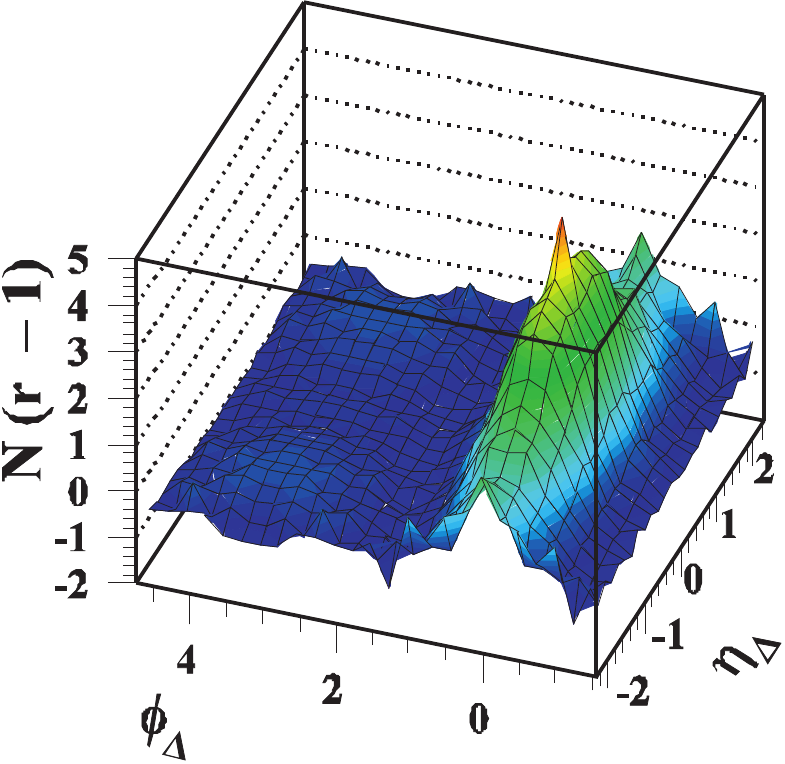}
\caption{\label{axialci1}
Minimum-bias ($p_t$-integral) CI (charge-independent) 2D angular correlations on angle differences $(\eta_\Delta,\phi_\Delta)$ from 130 GeV \auau\ collisions~\cite{axialci}. A fitted broad away-side 1D peak at $\phi_\Delta = \pi$ and nonjet quadrupole $\cos(2\phi_\Delta)$ have been subtracted. The remaining same-side 2D peak is strongly elongated proceeding from peripheral (left, consistent with \pp\ data~\cite{porter2}) to central (right) \auau\ collisions.
 }  
\end{figure}

Figure~\ref{axialci1} shows CI correlations for peripheral (left) and central (right) 130 GeV \auau\ collisions~\cite{axialci}. Angular {\em difference variables} (e.g., $\eta_\Delta = \eta_1 - \eta_2$ and similar for azimuth) are appropriate for such untriggered combinatoric correlations where no trigger-particle reference angles apply. Those minimum-bias 2D angular correlations include three prominent structures: (a) a SS 2D peak modeled by a 2D Gaussian, (b) a broad AS 1D peak on azimuth modeled by dipole $\cos(\phi_\Delta - \pi)$ and (c) a {\em nonjet azimuth quadrupole} modeled by $\cos(2\phi_\Delta)$. In Fig.~\ref{axialci1} fitted features (b) and (c) have been subtracted to emphasize evolution of the SS 2D peak with centrality. 

Correlation features (a) and (b) are expected for dijet pairs: intrajet correlations (a) + interjet correlations (b). But the correlations in Fig.~\ref{axialci1} were formed from particles with $p_t < 2$ GeV/c where hydro (elliptic and radial flows) is expected to describe spectra and correlations. The jet-like result was quite surprising. Feature (c) is nominally consistent with elliptic flow expectations, and its amplitude serves as an alternative measure of $v_2$ as inferred from model fits to 2D angular correlations (Sec.~\ref{njquad})~\cite{davidhq}. The narrow peaks at the origin represent conversion electron pairs and Bose-Einstein (quantum) correlations.

Along with the surprise of large-amplitude jet-like angular correlations below 2 GeV/c (Sec.~\ref{starwht}) came evolution of the SS 2D peak shape from strongly elongated on $\phi$ (\pp\ and peripheral \auau)~\cite{porter2,porter3} to strongly elongated on $\eta$ (more-central \auau)~\cite{axialci,anomalous}. The $\eta$ elongation (so-called same-side ``ridge'') has become a central problem for theory and experiment in later RHIC and LHC analysis. 

The plots in Fig.~\ref{axialci1} result from subtracting a fitted nonjet quadrupole and a large-amplitude AS 1D peak fully consistent with back-to-back jet correlations relative to the SS 2D peak. Given that interpretation the ``away-side'' peak representing jet-jet correlations does not {\em disappear} in central \auau\ collisions but is significantly {\em modified} (mean \pt\ is reduced). A similar AS dipole is expected for {\em combinatoric hadronic} (not dijet) momentum conservation, but the amplitude and centrality dependence for that feature are very different. If the combination (a) + (b) is indeed (mini)jets then arguments against a two-component minijet model based on HIJING simulations~\cite{liwang} can be questioned (Sec.~\ref{nomini}).

\begin{figure}[h]
 \hfill \includegraphics[width=1.65in,height=1.65in]{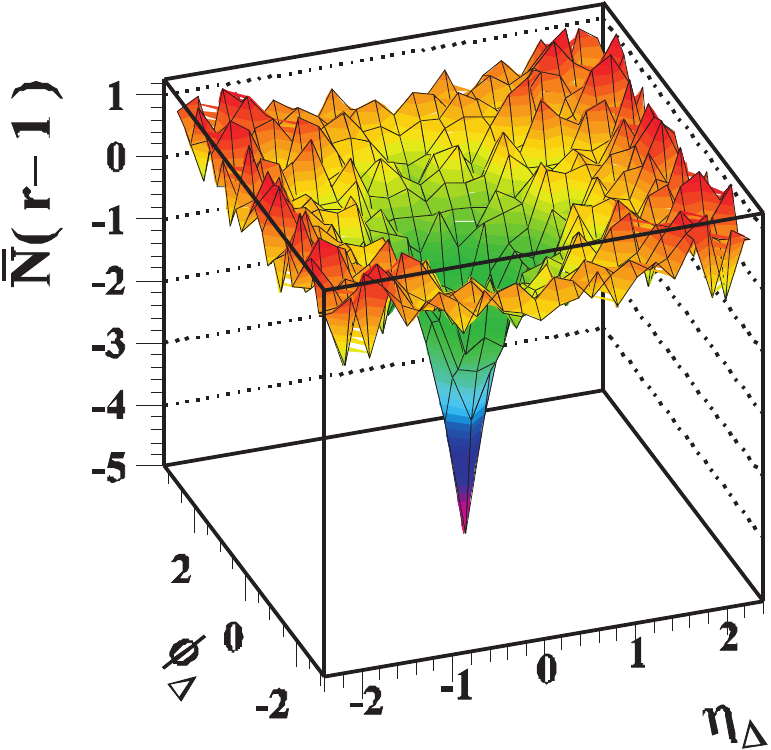}
  \includegraphics[width=1.65in,height=1.65in]{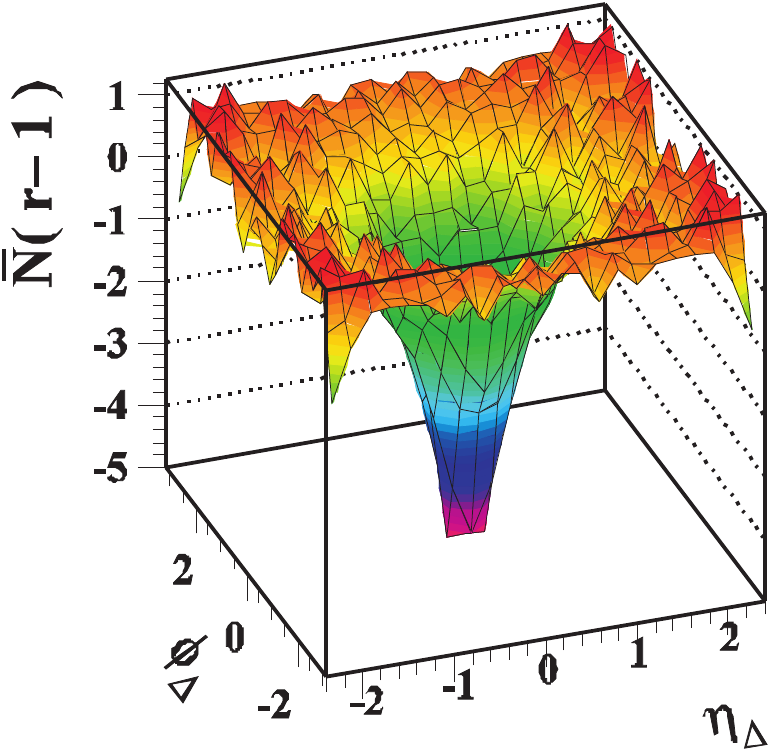} \hfill
\caption{\label{axialcd1}
Left: Minimum-bias ($p_t$-integral) CD (charge-dependent) 2D angular correlations from 130 GeV \auau\ collisions~\cite{axialcd}. A narrow 2D exponential peak at the origin is common to all centralities. A broader 1D peak on $\eta_\Delta$ is observed in more-peripheral collisions (left). An extension narrow on $\phi_\Delta$ is observed in more-central collisions (right) that may be associated with the $\eta$ elongation in Fig.~\ref{axialci1}.
 }  
\end{figure}

Figure~\ref{axialcd1} shows complementary CD angular correlations from the same 130 GeV \auau\ collision data for peripheral (left) and central (right) collisions. Negative values reflect ``canonical suppression'' of net charge within small angle differences. A large-amplitude 2D peak at the angular origin may be interpreted as representing local charge conservation during parton fragmentation to (mini)jets (local charge conservation within a fragmentation cascade)~\cite{axialcd,opalcd}. For peripheral collisions (left) an additional smaller-amplitude broad 1D peak on $\eta_\Delta$ can be interpreted in terms of projectile-nucleon fragmentation to charge-neutral hadron (mainly pion) pairs. For central collisions a small-amplitude structure narrow on $\phi_\Delta$ and extended on $\eta_\Delta$ corresponds to $\eta$ elongation of the SS 2D peak (``ridge'').

The charge {\em balance function} (BF) was proposed to detect ``delayed hadronization'' from a partonic medium by broadening of CD angular correlations on $\eta$. The prediction is based on arguments relating to longitudinal expansion of the QCD medium~\cite{prattbf}. It was shown that the BF is a 1D projection of CD 2D correlations in Fig.~\ref{axialcd1} onto 1D $\eta_\Delta$~\cite{tombf}. The narrow structure on azimuth is lost in such projections. The possible dominance of jet contributions to CD structure is then overlooked. These CD 2D angular correlations are consistent with the TCM for hadron production: longitudinal projectile fragmentation and transverse parton fragmentation, both with {\em local} charge conservation. There is no significant CD structure corresponding to the nonjet azimuth quadrupole prominent in CI correlations.

\subsection{$\bf p_t$ and number fluctuations}

If RHIC \aa\ collisions achieved sufficient energy densities and temperatures to cross the QCD phase boundary and form a QGP event-wise critical fluctuations in some event properties might be detectable based on analogies with {\em macroscopic} critical phenomena. A candidate event property is event-wise mean $p_t$ denoted $\langle p_t \rangle$ interpreted as a proxy for kinetic temperature $T_{kin}$. The question is posed: Do RHIC collisions exhibit critical temperature fluctuations for some collision conditions?

\begin{figure}[h]
  \includegraphics[width=1.65in,height=1.5in]{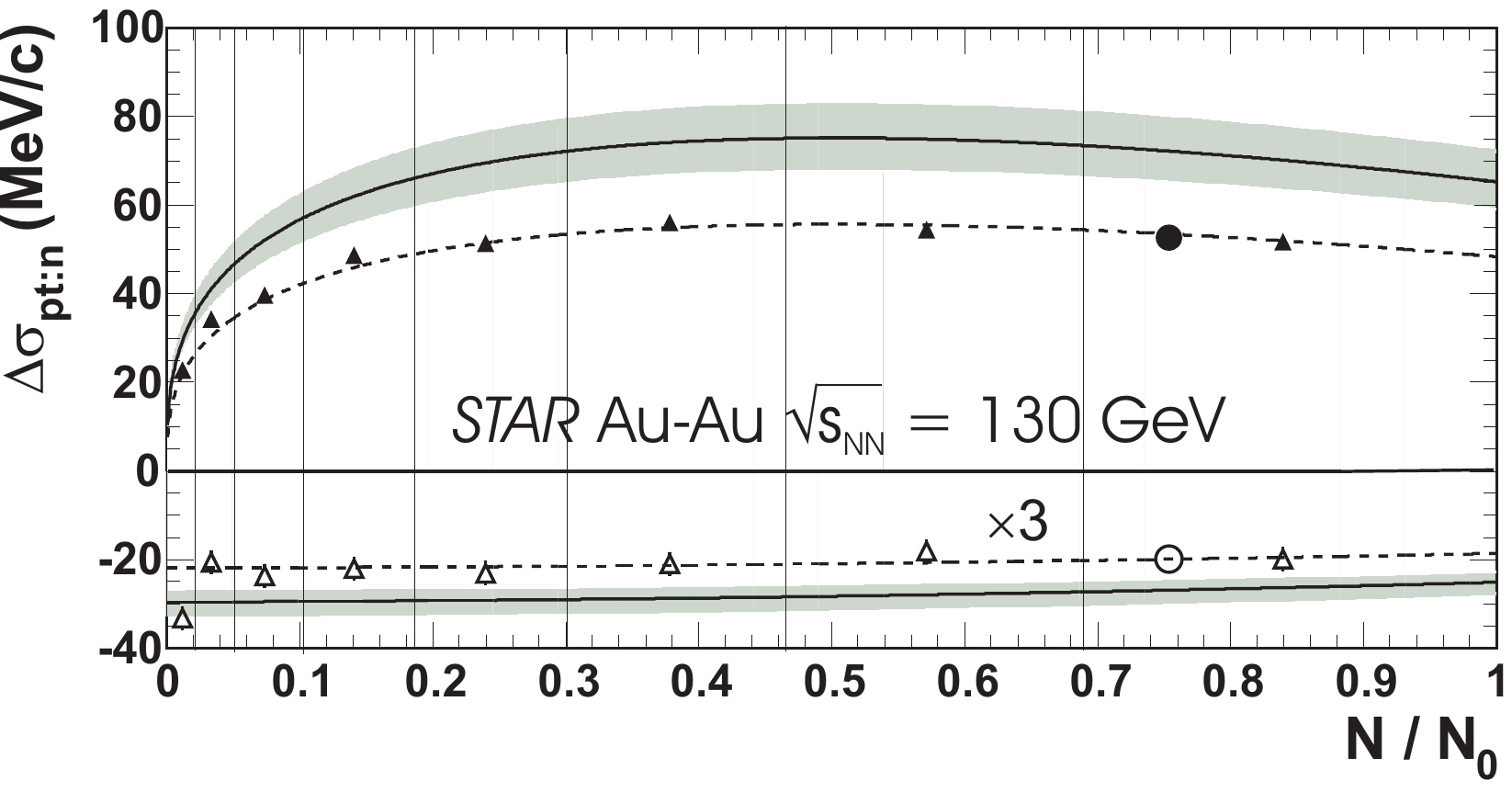}
  \includegraphics[width=1.65in,height=1.65in]{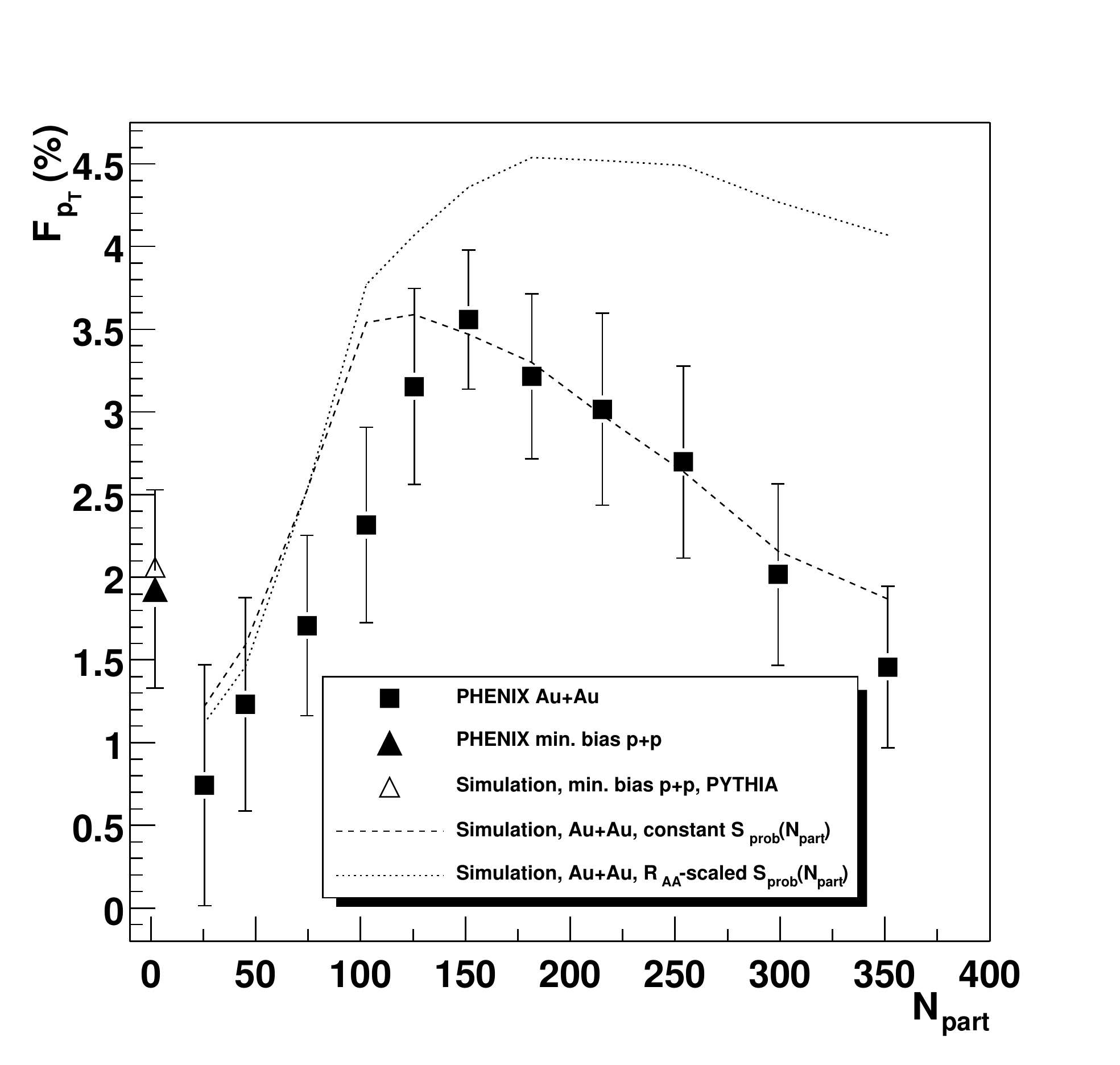}
\caption{\label{mpt}
Left: Mean-$p_t$ fluctuations measured by r.m.s.\ quantity $\Delta \sigma_{p_t:n}$ vs centrality measured by $n_{ch}$ from 130 GeV \auau\ collisions~\cite{ebyept1}. Upper points represent uncorrected CI $\langle p_t \rangle$ fluctuations, lower points represent CD (related to net-charge) fluctuations. The solid curves are corrected for detector inefficiencies.
Right: Mean-$p_t$ fluctuations measured by alternative r.m.s.\ measure $F_{p_t}$ vs centrality measured by $N_{part}$ showing a similar non-monotonic trend~\cite{ebyeptph}.
 } 
 \end{figure}

Figure~\ref{mpt} (left) shows fluctuation measure $\Delta \sigma_{p_t:n}$, a {per-particle} r.m.s.~measure of mean-\pt\ fluctuations, vs centrality measured by relative charge multiplicity $N/N_0$, where $N_0$ estimates the mean multiplicity corresponding to $b =0$~\cite{powerlaw}, for 130 GeV \auau\ collisions~\cite{ebyept1}. Fluctuations are measured at the ``acceptance scale,'' the angular acceptance of the STAR TPC. The positive points correspond to CI fluctuations, the negative to CD (net-charge) fluctuations. The centrality variation is smooth from very peripheral to very central collisions. There is no discontinuity suggesting a phase boundary. The fluctuation amplitude is large, prompting the question---what mechanism can produce such large $\langle p_t \rangle$ fluctuations over all \auau\ centralities?

Figure~\ref{mpt} (right) shows mean-$p_t$ fluctuations measured by an alternative statistic $F_{p_t}$~\cite{ebyeptph}. A relative-variance measure $\omega_{p_t}^2 = \sigma^2_{\langle p_t \rangle} / \bar p_t^2$ is evaluated for single events and for a mixed-event reference. Ratio $r = \omega_{p_t,single} / \omega_{p_t,mixed}$ is defined and $F_{p_t} \equiv r-1$. The same general trend of increase then decrease with centrality is followed, but the numerical values are quite different. The RHIC and SPS communities were confronted with a variety of fluctuation measures with very different properties and interpretations~\cite{tfluctmeth}.  The lack of conventions for fluctuation measures, the dependence on specific detector geometries and imposed kinematic conditions, remained an unresolved problem. More-recent fluctuation measures and interpretations are discussed in Sec.~\ref{newflucts}. In general, no sharp variations in fluctuation trends that might correspond (by analogy) to a macroscopic phase transition were observed in early RHIC fluctuation data.

\section{The RHIC white papers} \label{white}

During 2004 the four RHIC experiments reviewed the analysis performed and data obtained during the first three years of RHIC operation. The review results were presented as white papers~\cite{whitebrahms,whitephob,whitestar,whitephen}. The central question: had convincing evidence emerged to support discovery of a QGP? If not, what evidence was missing and what might be required to obtain it. If so, what further studies were required for its complete description. In this section I present brief highlights from each of the RHIC white papers.

\subsection{STAR} \label{starwht}

The QGP is first defined:  ``...a (locally) thermally equilibrated
state of matter in which quarks and gluons are deconfined from
hadrons, so that color degrees of freedom become manifest over
nuclear, rather than merely nucleonic, volumes.'' The expression ``strongly-interacting'' QGP is said not to differentiate from that expected in the temperature regime just above $T_c$ accessed by RHIC collisions, a highly-correlated quark-gluon system.

Standards of scientific proof are identified: necessity -- do data require a QGP explanation with no other alternatives; consistency -- are specialized theories applied to different data aspects mutually compatible; endurance -- are incompatibilities with some data indicative of inevitable small adjustments or a failure of the basic paradigm?

Experimental results and relevant theories are reviewed for hadron production, hadron species abundances, spectrum structure (flows and jets), elliptic flow $v_2$ systematics, fluctuations and jet-related correlations. 

Resulting argument in favor of QGP includes ``...the unprecedented success of hydrodynamics in providing a reasonable quantitative account for collective flow at RHIC, and of the statistical model in reproducing hadron yields through the strange sector, together argue for an early approach toward thermalization spanning the u, d and s [quark] sectors.''

Also, ``The fitted parameters of the statistical model analyses, combined with inferences from the produced transverse energy per unit rapidity, suggest attainment of temperatures and energy densities at least comparable to the critical values [e.g., $T_c$] for QGP formation in LQCD calculations of bulk, static strongly interacting matter,'' implying that colored quarks and gluons are the relevant dynamical degrees of freedom.

However, the paper cautions: ``On the other hand, measurements of angle difference distributions for soft hadron pairs [e.g., Fig.~\ref{axialci1} of  this paper] reveal that some (admittedly heavily modified) remnants of jetlike dynamical correlations survive the thermalization process, and indicate its incompleteness.''

Among several prominent goals for future work are: (a) establish that jet quenching is an indicator of parton, and not hadron, energy loss and (b) extend RHIC \auau\ measurements down toward SPS measurements
in energy, to test quantitative predictions of the energy dependence.  At present (a) has been accomplished and (b) is in progress as the RHIC beam energy scan (BES).

\subsection{PHENIX}

Based on a summary of experimental results it is concluded that ``...hydrodynamical calculations which reproduce the magnitude of elliptic flow observed at RHIC require local thermalization to occur very quickly, typically by 1 fm/c or earlier.... If the system does reach local equilibrium on this time scale then the energy density of the first thermalized state would be in excess of 5 GeV/fm$^3$, well above the amount required to create the QGP.''

Concerning minijets and the TCM or HEP context: ``Pre-RHIC expectations that $E_t$ and charged particle production would be dominated by factorized pQCD processes [jets] were contradicted by data, which showed only very modest increases with centrality and beam energy. A new class of models featuring initial-state gluon saturation compares well with RHIC multiplicity and $E_t$ data, and are also consistent with our Bjorken style arguments for estimating energy densities at early times.'' The observation and its implications are discussed in Sec.~\ref{nomini}.

``The observed suppression of high-$p_t$ particle production at RHIC... provides direct evidence that Au+Au collisions at RHIC have produced matter at extreme densities, greater than ten times the energy density of normal nuclear matter and the highest energy densities ever achieved in the laboratory.''

The paper concludes that ``...there is compelling experimental evidence that heavy-ion collisions at RHIC produce a state of matter characterized by very high energy densities..., strong collective flow, and early thermalization. ...this matter modifies jet fragmentation and has opacity that is too large to be explained by any known hadronic processes. [...] The most economical description is in terms of the underlying quark and gluon degrees of freedom.''

\subsection{PHOBOS}

Emphasis is placed on particle production and the large $\eta$ acceptance of the PHOBOS detector.  Inferred scaling behaviors, factorizations (e.g., centrality and energy) and manifestations of the initial-state \aa\ geometry in the data are featured. Descriptions of the data in terms of hadronic degrees of freedom are ruled out.

``...the data clearly demonstrate that proportionality to the number of participating nucleons, $N_{part}$, is a key concept which describes much of the phenomenology. [...] Further, the total particle yields per participant from different systems are close to identical when compared at the same available energy.... [...] ...and many characteristics of the produced particles factorize to a surprising degree into separate dependences on centrality and beam energy.
[...] All of these observations point to the importance of the geometry of the initial state and the very early evolution of the colliding system in determining many of the properties of the final observables.''

The paper concludes ``...these simple scaling features will constitute an integral component or essential test of models which attempt to describe the heavy ion collision data at ultrarelativistic energies. These unifying features may, in fact, provide some of the most significant inputs to aid the understanding of QCD matter in the region of the phase diagram where a very high energy density medium is created.''

\subsection{BRAHMS}

Concerning evidence for the QGP in data it is pointed out that  ``...experimental signatures can be roughly grouped into two classes: 1) evidence for bulk properties consistent with QGP formation, e.g.\ large energy density, entropy growth, plateau behavior of the thermodynamic variables, unusual expansion and lifetime properties of the system, presence of thermodynamic equilibration, fluctuations of particle number or charge balance etc, and 2) evidence for modifications of specific properties of particles thought to arise from their interactions with a QGP, e.g. the modification of widths and masses of resonances, modification of particle production probabilities due to color screening (e.g. $J/\Psi$ suppression) and modification of parton properties due to interaction with other partons in a dense medium (e.g. jet quenching), etc.''

The paper asks ``...if there are any specific features that may falsify the conclusion [of QGP formation]? To our knowledge no tests have been proposed that may allow falsification of either a partonic scenario or a hadronic scenario, but it would be important if any such exclusive tests were to be formulated.''

Based on a summary of available data the paper concludes ``The overall scenario is therefore consistent with particle production from a color field, formation of a QGP and subsequent hadronization. Correlation and flow studies suggest that the lifetime of the system is short ($< 10$ fm/c) and, for the first time, there is evidence suggesting thermodynamic equilibrium already at the partonic level.''

\section{Announcement of Perfect liquid} \label{perfect}

BNL and other RHIC-related institutions announced in 2005 formation of  a ``perfect liquid'' in high-energy \auau\ collisions~\footnote{http://www.bnl.gov/newsroom/news.php?a=1303}. The term refers generally to a fluid with very low shear viscosity $\eta$, perhaps approaching a theoretical quantum lower limit denoted by $\eta/s = 1/4\pi$ (natural units), where $s$ represents entropy density. The QCD fluid reportedly formed in \auau\ collisions is described as a {\em strongly-coupled} QGP or sQGP, in contrast to the weakly-coupled QGP (ideal gas of quarks and gluons) anticipated at large energy densities and temperatures corresponding to QCD asymptotic freedom. Theoretical arguments favoring perfect liquid formation focus on inference from data of strong hydrodynamic flows and the systematic properties of those flows, especially elliptic flow represented by quantity $v_2$. Low viscosity implies hydrodynamic expansion with little dissipation. Additional evidence for sQGP is sought in modifications to QCD jet formation and bulk properties of the medium. 

This section reviews specific theoretical arguments presented in Refs.~\cite{perfliq1,perfliq2} as representative.
In Ref.~\cite{perfliq2} it is acknowledge that interpretation of particle data can be ambiguous: ``...a healthy bit of `luck' was essential in order to find the `needles in the haystack' that are least distorted by uncertain non-equilibrium hadronic final-state dynamics.'' 
The conjectured 200 GeV collision scenario is (i) interpenetration of two thin sheets of ``color glass'' (gluon condensate within projectiles Lorentz contracted to 0.1 fm) (ii) initial 1D (longitudinal) Hubble expansion of collision products, (iii) produced quarks and gluons thermalize locally, (iv) the resulting QGP develops collective flows described by hydrodynamics, (v) quarks and gluons finally hadronize, (vi) hadrons may continue to scatter until kinetic decoupling and free streaming. 

The initial conditions (e.g., gluon and energy densities) are inferred from the final-state hadron multiplicity and transverse energy $E_t$. Extrapolation backward from the final state assumes the system density falls as $1/\tau$ (proper time). Expansion without work (no secondary scattering of particles) would maintain a constant energy per particle. In a thermalized system entropy is conserved (isentropic expansion), entropy per particle remains constant and the medium cools. Onset of 3D expansion occurs shortly before freezeout.  That argument combined with lattice QCD (LQCD) results provides a basis to claim sufficiently high initial energy densities for QGP formation. The principal empirical evidence for conjectured sQGP follows.

{\bf Collective flow} -- A claim of thermalized bulk matter requires observation of collective flow, the primary observable. Flow tests two conditions: (a) thermalization and (b) validity of an equation of state or EoS (a relation between energy density and pressure). The observed flow is said to be consistent with a QCD EoS. Elliptic flow $v_2$ is emphasized because it is thought to develop early in the collision and to be more sensitive to QGP formation. Measured values for central \auau\ collisions are compatible with nonviscous ideal hydrodynamics below $p_t \approx 2$ GeV/c. A QCD EoS is said to be confirmed by the extent of mass ordering of $v_2(p_t)$ for identified hadrons below 2 GeV/c. The validity of ideal hydro for observed $v_2$ implies very strong interactions of quarks and gluons early in the collision and therefore little dissipation. ``The smallness of dissipative corrections [required for hydro descriptions of $v_2$ data]...is in itself a remarkable and unexpected discovery. [...] ...the QGP at RHIC is almost a perfect liquid. [...] Elliptic flow measurements confirm...local thermal equilibrium...'' early in the collisions~\cite{perfliq1}.

{\bf Jet quenching or suppression} -- A hydrodynamic description of $v_2$ data is said to break down at higher \pt\ values. For instance, the $v_2(p_t)$ mass ordering reverses above 2 GeV/c. Local equilibrium is not achieved because of weaker coupling (and therefore more dissipation) at higher momentum scales (QCD asymptotic freedom). In the interval 2-5 GeV/c the applicable theory is uncertain. Above that interval perturbative QCD (pQCD) is applied to describe jet production and possible jet modifications (quenching) within the sQGP.

The second major discovery at RHIC is the strong suppression of hadron spectra above 4 GeV/c in more-central \auau\ collisions. Since jet fragments should dominate that $p_t$ interval the suppression effect is referred to as ``jet quenching.''  The principal mechanism is thought to be parton energy loss via gluon bremsstrahlung within the dense colored medium. Jet quenching in turn can be used to study the gluon density profile $dN_g/dy$ of bulk QCD matter. The value inferred from jet quenching is said to be compatible with that inferred from (a) the final-state hadron density, (b) the initial conditions required for hydro descriptions of elliptic flow and (c) estimated CGC (gluon saturation) initial conditions based on a {\em saturation} scale $Q_s$ value 1-1.5 GeV. The CGC inferred gluon density is much greater than that predicted by the minijet-based HIJING Monte Carlo with a default parton spectrum lower bound $p_0 = 2$ GeV (parton energy) that limits minijet production to less than 1000. One should also note the experimental lower bound $Q_0 \approx 6$ GeV ({\em dijet} energy) for dijet production in \pp\ collisions~\cite{ppprd,fragevo}.

Jet-related angular correlations may provide {\em tomographic} information about the dense medium. In \pp\ and peripheral \auau\ collisions in-vacuum dijet 1D azimuth correlations include two peaks at 0 and $\pi$. In central \auau\ collisions the so-called ``away-side'' peak at $\pi$ is greatly reduced or disappears (for certain \pt\ cuts). The result is termed a ``monojet'' and is interpreted to demonstrate that one jet of a triggered dijet is essentially absorbed by the dense medium. The away-side peak is restored in d-Au collisions, demonstrating that cold nuclear matter in a gold nucleus does not reduce the initial-state gluon density by ``shadowing.'' ``The observed jet quenching in \auau\ [collisions] is due to parton energy loss.'' ``Theoretical analysis of jet quenching...strengthens the case for multiple strong interactions of the quark and gluon constituents of the matter made at RHIC.''

{\bf CGC initial conditions} -- The collision initial conditions (IC) must be specified for any hydro description of flow data. The validity of hydro descriptions and inference of QGP formation is said to depend on the accuracy of IC estimates as well as a theoretical EoS. Minijet production (as modeled by HIJING) and the CGC prediction represent alternative IC descriptions. Initial comparisons of predictions for hadron production by the competing models with data led to preference for the CGC description. ``...the surprising very weak centrality and beam energy dependence [slow growth] observed [in the data] is most satisfactorily explained and predicted by the CGC...'' The reference is to HIJING predictions of more-rapid increases. The comparison, seen to rule out minijets and the TCM in favor of the CGC, ``...is one of the strongest lines of empirical evidence...'' for the CGC IC (see comments in Sec.~\ref{nomini}).

{\bf Conclusions} -- Criteria for QGP discovery: (a) Nuclear matter is created at large enough energy densities that it must consist of quarks and gluons when compared with LQCD. (b) The matter must be thermalized. (c) Properties of the bulk matter must be consistent with QCD predictions based on hydro, LQCD and pQCD.
The published RHIC data are said to satisfy those requirements. Ideal hydro agrees to a surprising degree with some $v_2$ data, implying that the matter must be a strongly-coupled QGP. Other evidence from data seems to indicate a prominent role for ``constituent'' or ``valence'' quarks in flow manifestations [constituent-quark scaling of $v_2(p_t)$ data] and spectrum structure (evidence for quark recombination in PID spectrum systematics).

Theoretical arguments for a strongly-coupled QGP or ``perfect liquid'' formed in RHIC \auau\ collisions were based on data from the initial RHIC running periods. Much work has been carried out since to test theoretical conjectures and provide a more-detailed view of collision dynamics and the nuclear matter created at RHIC. We review more-recent results for a  subset of topics that seem most closely related to sQGP claims.

\section{Hadron yields and Spectra} \label{spectra}

Particle production mechanisms, bulk matter thermodynamics at low \pt\ including radial and elliptic flow, unexpected hadron ratios at intermediate \pt\ and jet quenching at high \pt\ continue to be explored. With increasing beam luminosity and upgraded detector components identified-particle (PID) spectra for more-massive hadrons and extension to higher \pt\ greatly increase the available spectrum information in the three \pt\ intervals. 

\subsection{Hadron yields near mid-rapidity}

As noted, the systematics of hadron yields within some fixed angular acceptance ($\Delta \eta,\Delta \phi$) and the hadron density distribution on $\eta$ are essential to establish the IC for hydro calculations and to test conjectures about hadron production mechanisms (e.g., soft vs hard particle production in a two-component model).

\begin{figure}[h]
  \includegraphics[width=1.65in,height=1.65in]{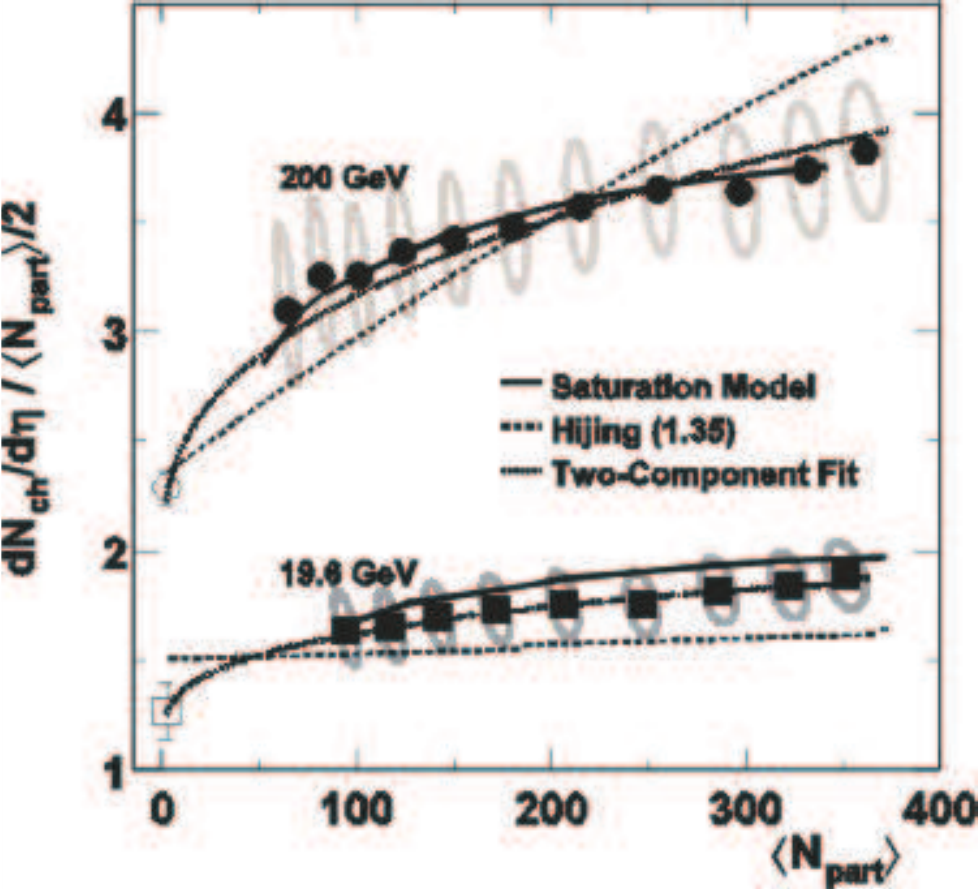}
  \includegraphics[width=1.65in,height=1.65in]{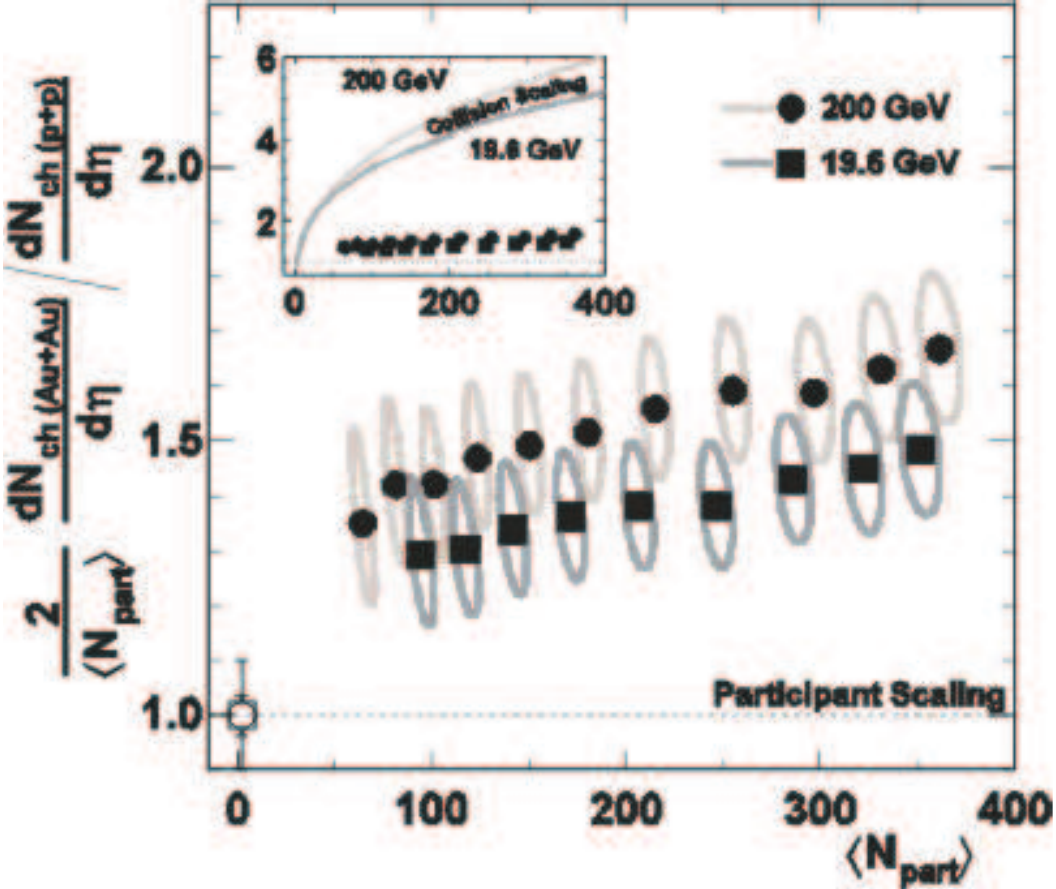}
\caption{\label{phob1}
Left: Charge multiplicity density per participant pair vs $N_{part}$ for 20 and 200 GeV \auau\ collisions (points) compared to three theoretical predictions (curves)~\cite{2compphob}.
Right: Data in the left panel scaled by \pp\ charge density showing similar trends with centrality for the two energies.
 }  
 \end{figure}

Figure~\ref{phob1} (left) shows a comparison of participant-scaled hadron yields within $|\eta|<1$ vs \auau\ centrality (points) for 20 and 200 GeV collisions~\cite{2compphob}. The data are compared with three theoretical models: (a) HIJING minijet two-component model (dashed curve), (b) fitted two-component model (dotted curve) and (c) saturation-scale (CGC) model (solid curve). It is important to note that the data span only the most-central 40\% of the \auau\ cross section. It is concluded that the CGC model best describes the data:

``We find that the Hijing calculation gives the expected increase of pQCD minijet production with centrality over this energy range, but the predicted increase is now in strong contradiction to the data. The flat centrality dependence of the [energy] ratio is relatively well described by the parton saturation [CGC] model calculation''~\cite{2compphob}.

Figure~\ref{phob1} (right) shows the same data normalized by the \pp\ values. The trends for two energies are very similar. The inset compares the HIJING prediction to the same ratios. There appears to be a dramatic difference that falsifies HIJING. If the data in the left panel are taken in ratio 200/20 the result is consistent with a constant value independent of centrality:
``The ratio of the measured yields at 200 and 19.6 GeV shows a clear geometry scaling over the central 40\% inelastic cross section and averages to R200/19.6 = 2.03±0.02(stat)±0.05(syst). A large increase in yield from hard processes, which contribute to multiplicity, is not apparent in the data, even over an order of magnitude range of collision energy.''
In Ref.~\cite{allparticlephob} Figs.~13 and 14 extend the above study to \cucu, d-Au and several energies. Relative and absolute variations of soft and hard  components with collision energy are discussed in Sec.~\ref{energy} of the present paper.

\subsection{Low-$\bf p_t$ spectra and bulk medium properties}

Hadrons from the lowest \pt\ interval 0-2 GeV/c are expected to emerge from a flowing bulk medium. The corresponding spectrum shape systematics should reveal medium properties and flow parameters. Identified-hadron abundances should test statistical-model predictions and medium properties such as chemical and kinetic freezeout temperatures $T_{chem}$ and $T_{kin}$. Certain ``scaling relations'' are sought as an indicator of underlying simplicity in the sQGP.

\begin{figure}[h]
  \includegraphics[width=3.3in,height=1.4in]{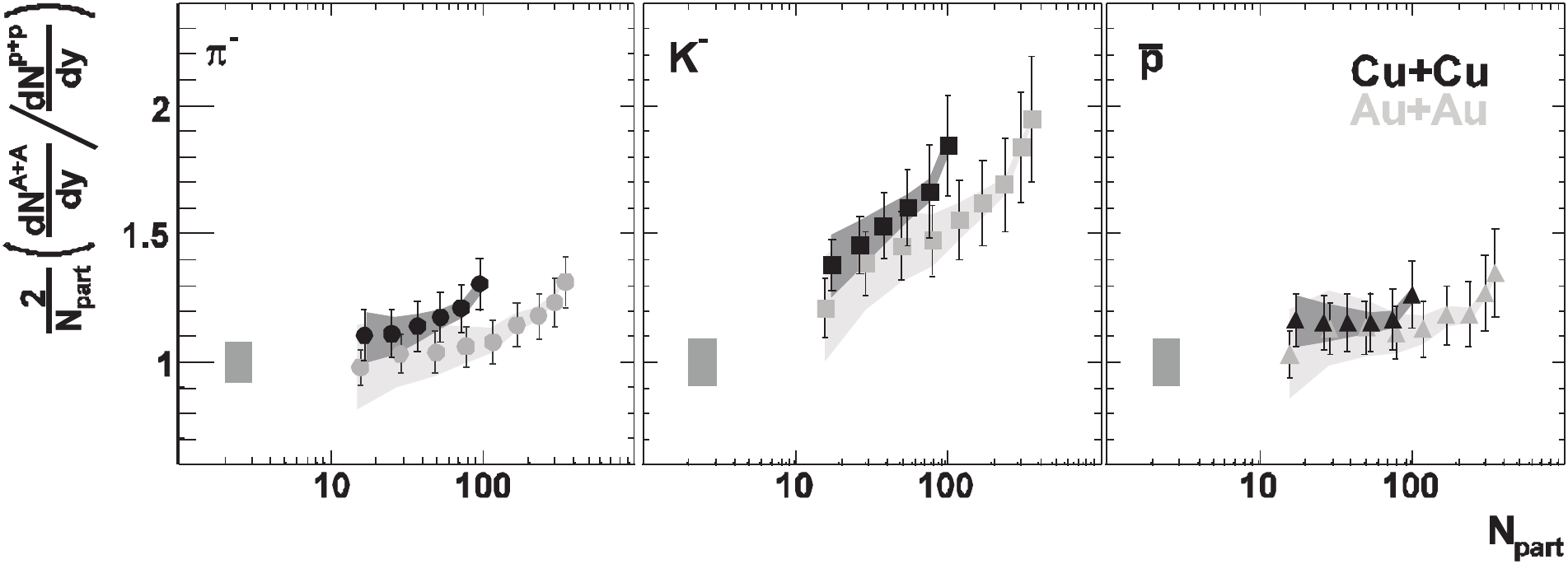} 
 \hfill \includegraphics[width=3.3in,height=1.4in]{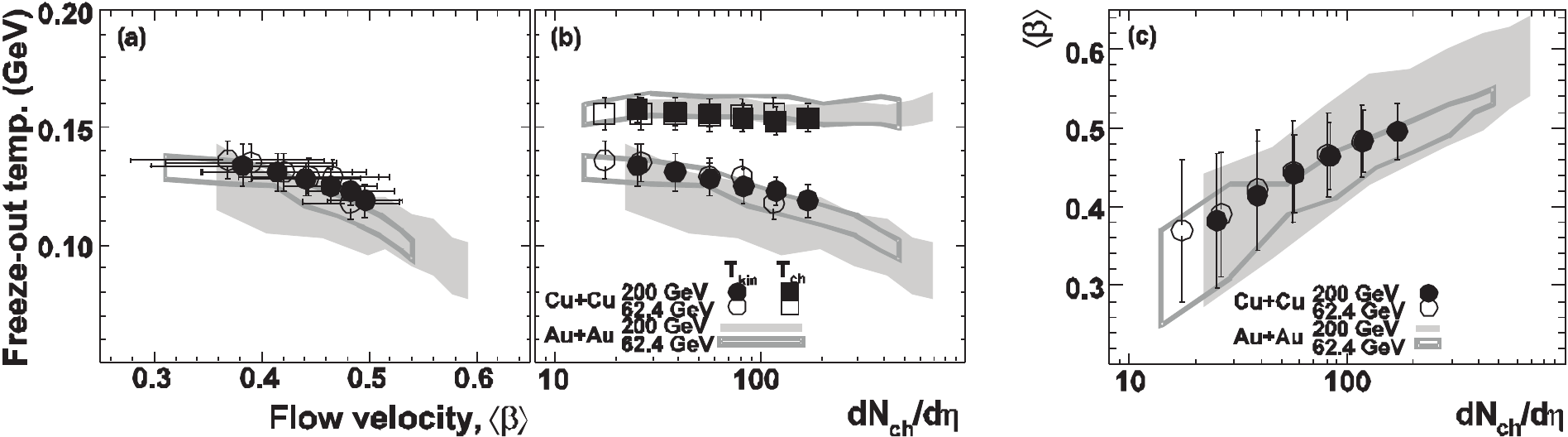} 
\caption{\label{lowpt}
Upper: Per-participant identified-hadron production relative to \pp\ vs \aa\ centrality measured by $N_{part}$ for \auau\ (light points) and \cucu\ (dark points) collisions \cite{starbulkprc}.
Lower: Bulk-medium properties $T_{kin}$, $T_{chem}$ and $\langle \beta_t\rangle$ vs centrality measured by $dn_{ch}/d\eta$ for 200 GeV \auau\ and \cucu\ collisions.
 }  
 \end{figure}

Figure~\ref{lowpt} (upper) shows per-participant yield ratios for three hadron species, comparing \auau\ and \cucu\ collisions on participant number $N_{part}$~\cite{starbulkprc}. The parameters are obtained from blast-wave (BW) fits to PID spectra for pions, kaons and protons over a limited low-$p_t$ interval $< 1$ GeV/c. The yield data for the two collision systems do not coincide when plotted on participant number $N_{part}$, do not scale.

Figure~\ref{lowpt} (lower) shows apparent scaling when BW parameters are plotted on total hadron multiplicity $dn_{ch}/d\eta$. The first panel is $T_{kin}$ vs $\langle \beta_t \rangle$, the second shows $T_{chem}$ (upper) and $T_{kin}$ (lower) vs multiplicity, the third shows $\langle \beta_t \rangle$ vs multiplicity. The trends for \cucu\ and \auau\ appear to coincide within overlapping multiplicity intervals. Those results are interpreted to indicate that a thermalized bulk-medium description is valid, and that ``...the kinetic freeze-out
properties are determined by the initial state.'' 

Reference~\cite{starbulkprc} concludes: ``These multidimensional systematic studies reveal remarkable similarities between the different colliding systems. [...] The bulk properties studied have a strong correspondence with the total particle yield. [...] Within thermal models this reflects a relation between the energy per particle at freeze-out and the entropy derived from particle yields, which reflects the initial state properties for adiabatic expansion. [...] The lack of scaling for $N_{part}$ ``...suggests that Npart does not reflect the initial state of the system accurately.''

\subsection{Intermediate $\bf p_t$ and the baryon/meson anomaly}

Hadrons from the intermediate-$p_t$ region 2-5 GeV/c may arise from thermal hadron production, parton fragmentation to jets or a hybrid combination. The possibility that hadrons are formed by ``constituent quark'' recombination or coalescence from a thermalized QGP has received much theoretical attention~\cite{rudy,duke,tamu}. The so-called ``baryon-meson'' puzzle is intriguing. In more-central \auau\ collisions the baryon-to-meson ratio departs substantially from expectations for fragmentation to jets as in elementary collisions, increasing from $\approx 0.3$ near 10 GeV/c to exceed unity at its maximum value near 2.5 GeV/c.

 \begin{figure}[h]
  \includegraphics[width=3.6in,height=2.2in]{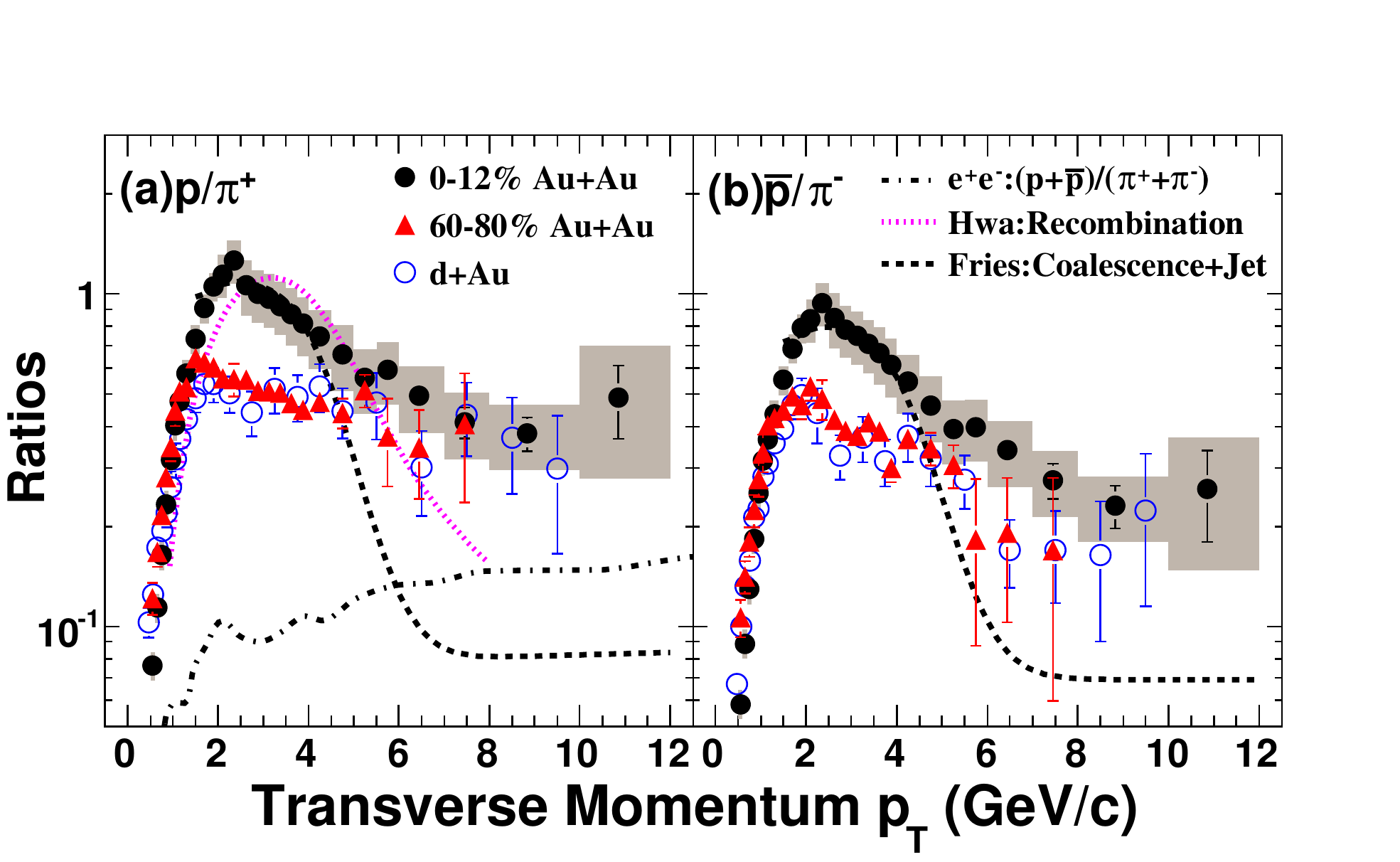} 
\caption{\label{baryonmeson}
Proton/pion spectrum rations for 0-12\% central and 60-80\% central 200 GeV \auau\ collisions for particles (left) and antiparticles (right)~\cite{barmeson}. The data (points) are compared to a fragmentation function parametrization derived from \ee\ collisions (dash-dotted curve) and to recombination theories (dashed and dotted curves).
 }  
 \end{figure}

Figure~\ref{baryonmeson} shows the proton-pion ratio plotted vs hadron $p_t$ for particles (left) and antiparticles (right) and for central (upper, black points) peripheral (lower, red points) and d-Au collisions (open points)~\cite{barmeson}. The dash-dotted curve in the left panel represents a prediction from fragmentation functions~\cite{bmfrag}. The dashed and dotted curves represent predictions from two recombination theories~\cite{rudy,duke}. Within the intermediate \pt\ region the theories seem to describe the central \auau\ data well, implying that the ratio anomaly in central collisions may signal hadron formation from a QGP in which constituent quarks play a significant role.

\subsection{High-$\bf p_t$ spectra and jet suppression}

Hadrons from the high-$p_t$ region above 5 GeV/c should be dominated by parton fragmentation to jets. The principal issue is possible jet quenching in more-central \aa\ collisions by parton energy loss in the dense colored medium or sQGP. How does jet quenching change with system size, both \aa\ centrality and A? How does jet quenching depend on the leading parton (quark or gluon)?

\begin{figure}[h]
  \includegraphics[width=1.65in,height=1.65in]{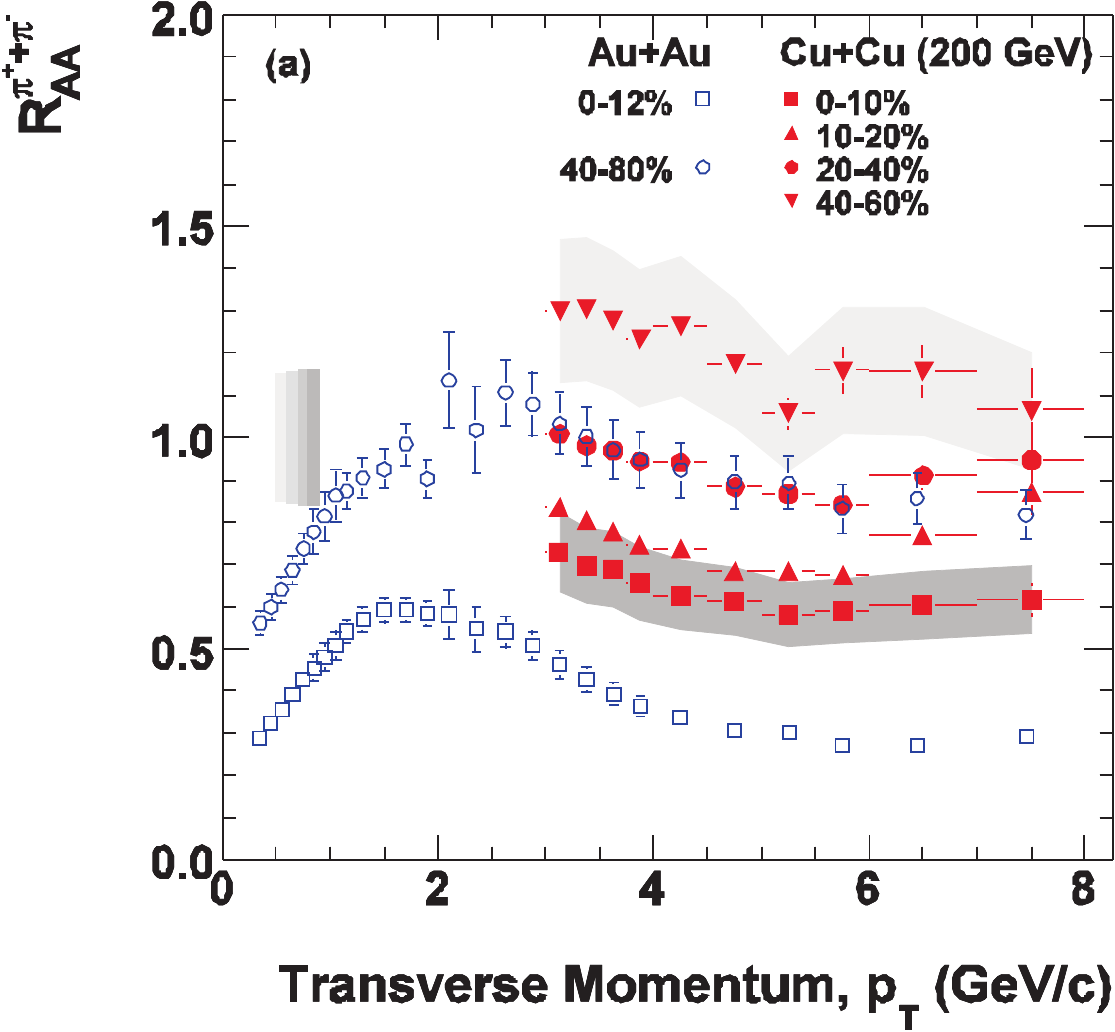} 
  \includegraphics[width=1.65in,height=1.65in]{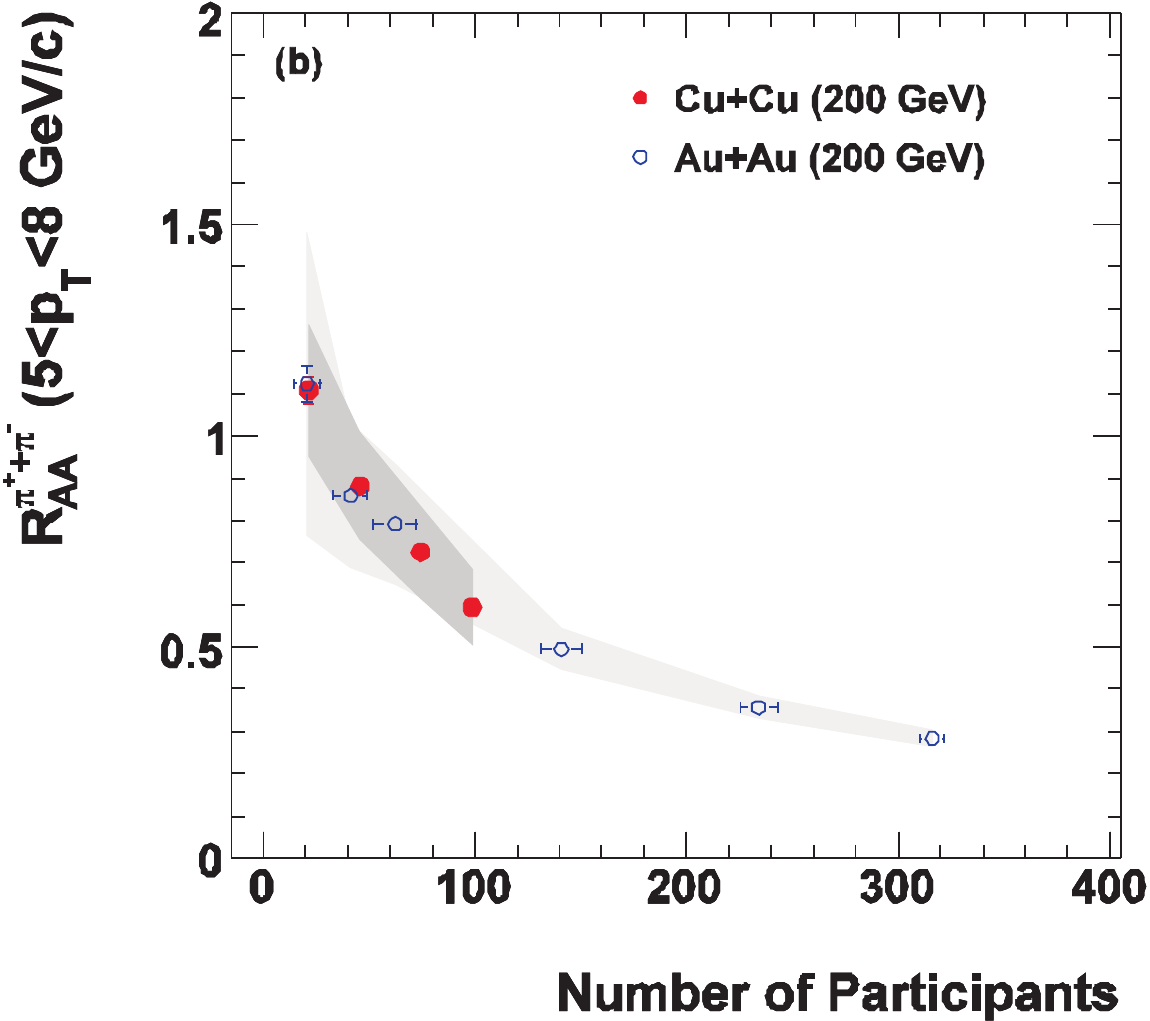} 
\caption{\label{highptraa}
Left: Spectrum ratio $R_{AA}(p_t)$ for several centralities from 200 GeV \auau\ (blue) and \cucu\ (red) collisions~\cite{starraacu}.
Right: $R_{AA}$ averaged over $p_t$ interval 5-8 GeV/c vs $N_{part}$ demonstrating apparent participant scaling of jet suppression.
 }  
 \end{figure}

Figure~\ref{highptraa} (left) shows $R_{AA}$ measurements for central and peripheral \auau\ and several centralities of \cucu\ collisions extending to 8 GeV/c~\cite{starraacu}. The suppression in most-central \cucu\ is substantially less than that for \auau, which is not surprising for the smaller collision system.
Figure~\ref{highptraa} (right) shows $R_{AA}$ mean values within 5-8 GeV/c plotted on $N_{part}$ to reveal a common centrality trend for the two collision systems. A similar pair of plots (not shown here) demonstrates that the mean $p/\pi$ ratio within the interval 3-4 GeV/c shows a similar universality when plotted on $N_{part}$.
 
Reference~\cite{starraacu} concludes: ``The [\cucu] data are found to exhibit similar systematic trends over a wide range of transverse momenta as Au-Au collisions at the same energy with a similar number of participants. [...] The participant coverage in these Cu-Cu collisions is in a region where the [parton] suppression effects are turning on. [...] A detailed study of the proton
to pion ratio reveals similar systematic dependencies to that found in Au-Au data. [...] ...these results indicate similar partonic energy loss for both gluons and quarks. [...] ...the [parton] suppression for different collision species is found to be invariant for the same number of participants.''

Reference~\cite{raaphob} introduces an alternative suppression measure $R_{AA}^{N_{part}}$ in which the factor $1/N_{bin}$ is replaced in  the spectrum ratio by $1/N_{part}$. Whereas the conventional ratio $R_{AA}$ shows strongly increasing suppression with  increasing centrality and energy above 50\% fractional cross section the $R_{AA}^{N_{part}}$ trend is nearly independent of \auau\ centrality. Reference~\cite{raaphob} remarks: ``The apparent dominance of the initial-state geometry [$N_{part}$], even for observables closely related to the dynamical evolution of heavy-ion collisions, is one of the key features of these interactions that remains to be understood.''

Reference \cite{starpidhighpt} compares PID particle spectra and ratios in \pp\ and \auau\ collisions, extending measurements up to 15 GeV/c. Variation of particle-to-antiparticle ratios confirms the expected changing mixture of quark vs gluon jets with increasing $p_t$ (as the $x$ of the hard-scattered parton  increases toward valence quarks). The available theoretical parametrizations of PID fragmentation functions describe the pion data but are not consistent with the kaon and proton data.  Several hadron species share a common $R_{AA}$ suppression above 6 GeV/c.

Some studies have employed theoretical models of parton energy-loss in the sQGP to estimate the color-charge density. Reference~\cite{phenpathlength} uses $R_{AA}$ angular variations relative to the reaction plane to infer the pathlength dependence of jet suppression. Greater suppression is observed out of plane than in plane. Suppression is approximately independent of $p_t$ above 4 GeV/c and varies with centrality approximately as $N_{part}^{2/3}$ consistent with theoretical models. The inferred path-length dependence appears to conflict with pQCD energy loss theory. Reference~\cite{phenopacity} describes an ambitious program to establish a self-consistent description of $R_{AA}$ data with global fits by energy-loss models. By relating the fit parameters to pQCD energy-loss theory the goal is to infer the systematics of the color-charge density in the sQGP.

These high-\pt\ studies emphasize jet suppression as inferred from spectrum ratio $R_{AA}$ over a $p_t$ interval above 4 GeV/c. That interval includes only a small minority of the hadron fragments within any jet~\cite{hardspec,fragevo}. From such analysis we do not learn what happens to the energy apparently ``lost'' from the leading parton, or whether it is lost at all from an {\em intact but modified} jet. Methods that access jet structure at lower $p_t$ are required to pursue such questions.

\section{$\bf v_2$ and Elliptic flow} \label{flow}

Because of its importance in providing the primary evidence for a {\em strongly-coupled} QGP with apparently very low viscosity elliptic flow $v_2$ has continued to receive a great deal of attention, with a growing number of subtopics.

\subsection{$\bf v_2$ centrality, $\bf \eta$ and energy dependence}

The initial 130 GeV \auau\ \pt-integral $v_2$ data leading to inference of ``perfect liquid'' were extended to 20, 62 and 200 GeV, to the smaller \cucu\ collision system and to a variety of hadron species. In Ref.~\cite{v2allphob} the strongly-peaked $\eta$ dependence of $v_2$ over a large $\eta$ acceptance was confirmed. The $\eta$ dependence does not change significantly with centrality within the upper 50\% of the \auau\ total cross section. The strong $\eta$ dependence is a significant problem for hydro models of elliptic flow. Collision energy dependence is also of interest. Does $v_2$ continue its rise from Bevalac-AGS to 200 GeV? Does it saturate at higher energies?

\subsection{Nonflow and flow fluctuations}

$v_2$ data may include a systematic error or bias from ``nonflow'' defined as contributions to $v_2$ not related to the \aa\ reaction plane~\cite{2004}. A number of strategies or ``methods'' denoted by $v_2\{method\}$ have been developed in attempts to reduce or eliminate nonflow. $v_2$ data may also include contributions from non-Poisson fluctuations possibly corresponding to initial-state \aa\ geometry fluctuations. The two issues are intimately related.

The origin of nonflow remains ambiguous because by hypothesis jets should not contribute to hadron production below 2 GeV/c where flows are thought to dominate, are therefore largely discounted as a nonflow mechanism. More-recent $v_2$ methods attempt to reduce nonflow bias. 

\begin{figure}[h]
  \includegraphics[width=1.55in,height=1.78in]{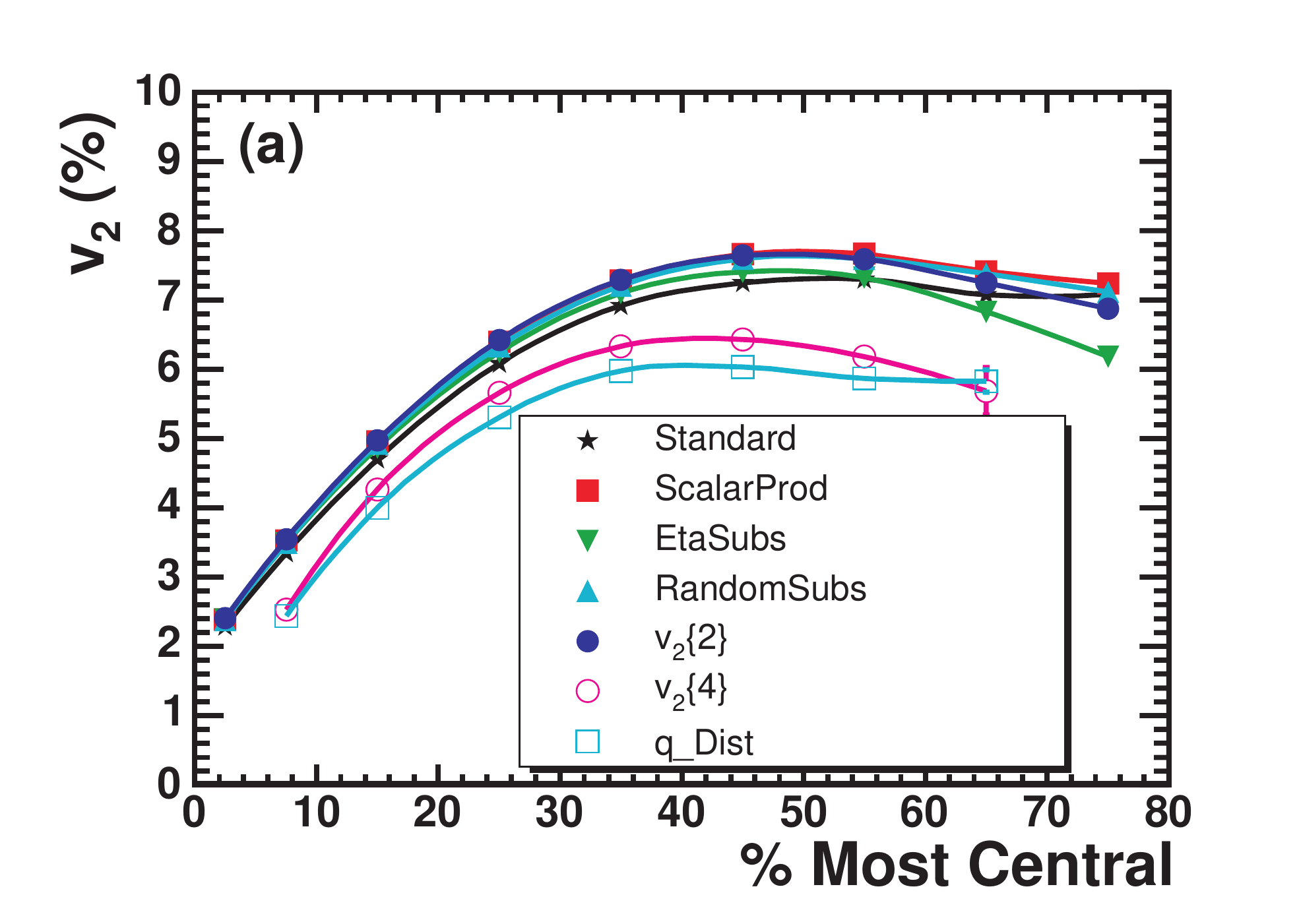} 
  \includegraphics[width=1.75in,height=1.7in]{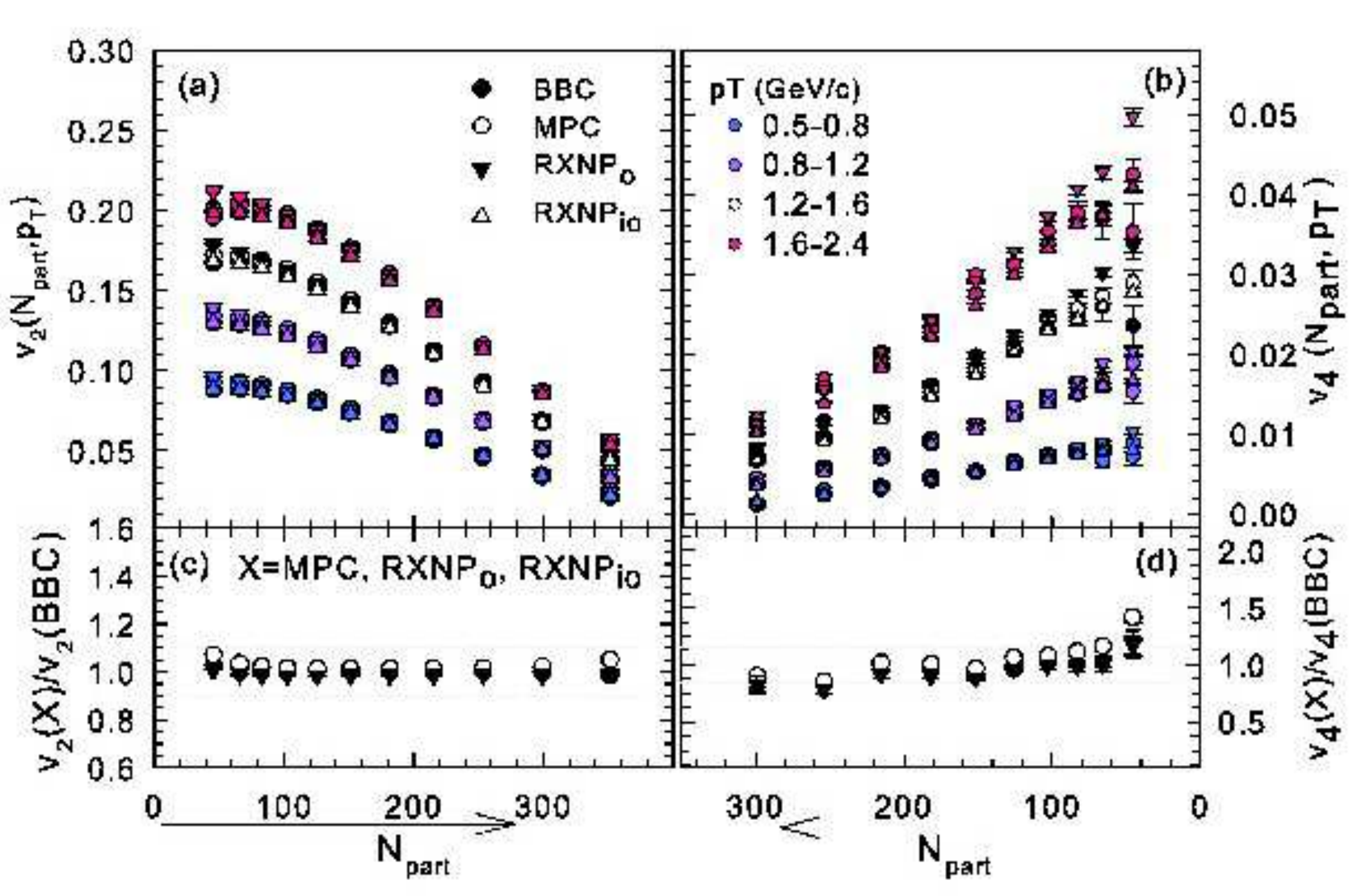} 
\caption{\label{v2nf}
Left: $p_t$-integral $v_2$ vs fractional cross section from 200 GeV \auau\ collisions for several $v_2$ methods~\cite{2004}. The differences among methods are conventionally invoked to estimate systematic uncertainties due to nonflow.
Right: Comparison of $v_2$ and $v_4$ for 200 GeV \auau\ collisions~\cite{phenv2vsv4}. The close agreement among the employed methods suggests minimal nonflow bias for those methods.
 }  
 \end{figure}

Figure~\ref{v2nf} (left) shows $v_2(b)$ data for several analysis methods~\cite{2004}. ``Standard'' denotes the original event-plane (EP) method~\cite{poskvol}. $v_2\{2\}$ and $v_2\{4\}$ respectively denote two- and four-particle cumulant methods. Other methods represent specific pair acceptances on $(\eta_1,\eta_2)$ and other numerical algorithms (e.g., scalar product, $Q$-vector). The data from different $v_2$ methods show substantial differences, mainly for most-central and most-peripheral collisions, that may be due to nonflow or flow fluctuations. Such differences are used to infer systematic uncertainties due to nonflow bias, but the various methods share certain assumptions in common.

Figure~\ref{v2nf} (right) shows $v_2$ (elliptic) and $v_4$ (hexadecapole) flow data vs $N_{part}$ for several $p_t$ bins~\cite{phenv2vsv4}. The various plotting symbols represent several $v_2$ methods in which different $\eta$ acceptances (detector elements) are used to estimate the event plane. The close agreement of the different methods is interpreted to imply that nonflow bias is made negligible by EP estimates from large-$\eta$ detectors~\cite{phenbigv2prc}. It is notable that the large-$\eta$ results agree closely with $v_2\{2\}$ data.

Elliptic flow is believed to be a final-state manifestation of the initial-state \aa\ overlap region in non-central collisions. $v_2$ data are therefore compared with estimates of the initial-state geometry in the form of  eccentricity $\epsilon$. Earlier estimates of $\epsilon$ were based on an {\em optical} Glauber model.
Reference~\cite{v2phob2} introduces the concept of a Monte Carlo Glauber or {\em participant} eccentricity to model the initial-state \aa\ geometry. Whereas the optical (or standard) Glauber eccentricity falls toward zero for central and peripheral \aa\ collisions the MC eccentricity remains significantly nonzero for central collisions and rises asymptotically toward unity for peripheral collisions, a dramatically different scenario. The resulting difference in $v_2/\epsilon$ trends has major implications for hydrodynamic interpretations.

\begin{figure}[h]
  \includegraphics[width=1.65in,height=1.65in]{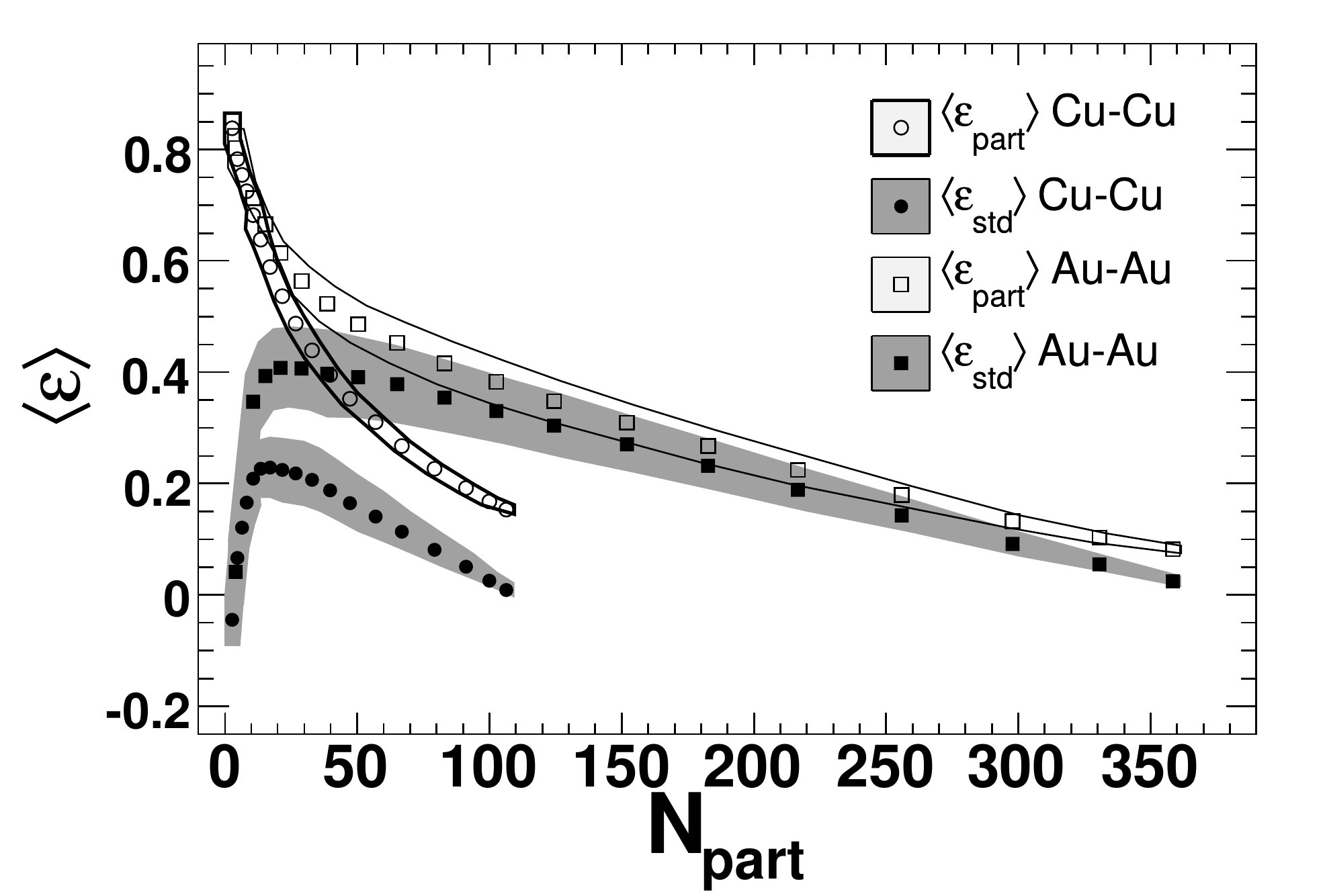} 
  \includegraphics[width=1.65in,height=1.6in]{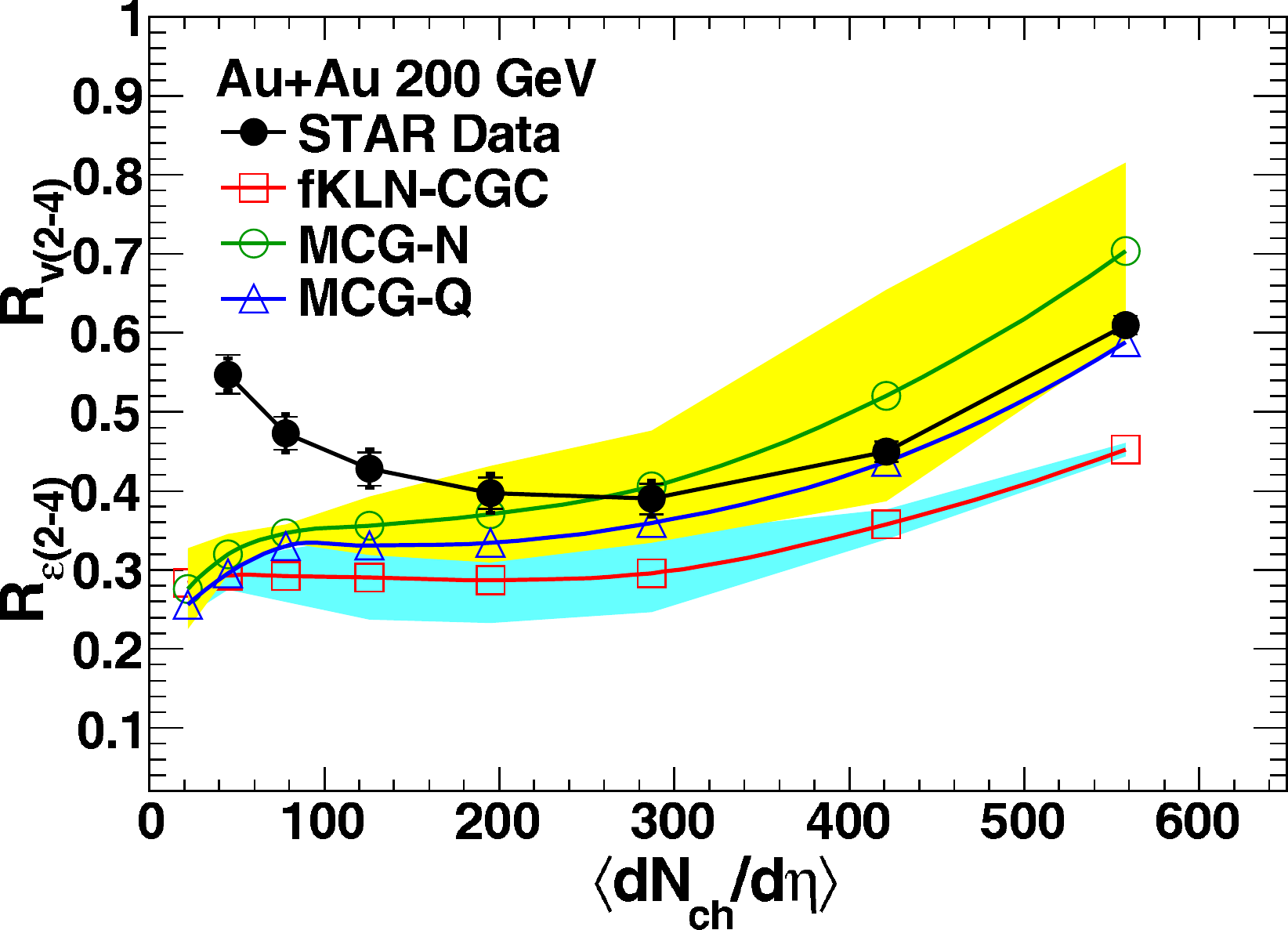} 
\caption{\label{v2flucts}
Left: \aa\ overlap eccentricity $\epsilon$ vs $N_{part}$ for optical (standard, solid points) and Monte Carlo (participant, open points) Glauber model simulations and for \auau\ and \cucu\ collisions~\cite{v2phob2}. Note the offset zero.
Right: Estimated final-state $v_2$ fluctuations measured by $R_v$ (solid points) compared to theory estimates of initial-state eccentricity fluctuations measured by $R_\epsilon$ (open points) vs \auau\ centrality measured by $dn_{ch}/d\eta$~\cite{starv2flucts}.
 }  
 \end{figure}

Figure~\ref{v2flucts} (left) shows a comparison between optical (standard) $\epsilon$ and Monte Carlo (participant) $\epsilon$ vs $N_{part}$ for \auau\ and \cucu\ collisions~\cite{v2phob2}. In that paper it is observed that $v_2/\epsilon$ for both \auau\ and \cucu\ follow the same trend vs centrality ($N_{part}$) if MC $\epsilon$ is used, suggesting that elliptic flow in \cucu\ is comparable to that in \auau\ for the same number of participant nucleons even though the \cucu\ system is much smaller. The importance of participant eccentricity (and a participant event-plane estimate) in such comparisons is emphasized.

Based on method definitions and certain assumptions the difference $v_2^2\{2\} - v_2^2\{4\}$ is said to have two main contributions: (a) a nonflow contribution $\delta_2$ from several possible sources and (b) a $v_2$ fluctuation contribution denoted by $2\sigma^2_{v_2}$~\cite{starv2flucts}. Various attempts are made to distinguish the two contributions and to estimate the r.m.s.\ measure of relative $v_2$ fluctuations $R_v = \sigma_{v_2} / \bar v_2$ in comparison with theoretical predictions of IS geometry fluctuations in the form $R_\epsilon = \sigma_\epsilon / \bar \epsilon$.

Figure~\ref{v2flucts} (right) shows comparisons between inferred $v_2$ fluctuations (solid points) and theory (open points)~\cite{starv2flucts}. The experimental data represent {\em upper limits} assuming nonflow $\delta_2$ is zero in all cases. However, there is nothing to rule out the other extreme, that $\sigma^2_{v_2}$ may be negligible in all cases. Alternative information on possible nonflow contributions to $v_2$ (e.g., $\delta_2$ arising from jets) is available from other angular correlation measurements~\cite{anomalous,multipoles}. In Ref.~\cite{v2flucts2phob} the nonflow contribution was estimated with $(\eta_1,\eta_2)$ correlations. It was assumed that the (long-range) nonflow contribution is approximately the same as for \pp\ collisions. That is a strong assumption since jet contributions may scale as $N_{bin} \approx 1000$ in central \auau\ relative to \pp\ collisions.

\subsection{PID $\bf v_2(p_t)$ and constituent-quark scaling}

Analysis of $v_2(p_t,b)$ for identified hadrons (particle ID or PID) confronts two main issues: so-called mass scaling below 2 GeV/c interpreted to confirm conjectured hydrodynamic flow and so-called constituent-quark scaling within 2-5 GeV/c interpreted to indicate that the flowing medium consists of quarks and gluons. Above 2 GeV/c $v_2$ is said to be driven more by  quark content than by hadron mass, due to hadron formation by quark coalescence~\cite{duke}.  Quark-number scaling may indicate that in RHIC heavy ion collisions collective motion is established among quarks and gluons before
hadrons are formed, apparently confirming that a strongly-coupled QGP with partonic degrees of freedom is created in RHIC collisions. Experimental questions include the accuracy or significance of inferred scaling trends and the dependence on centrality, system size and collision energy.

Reference~\cite{phenconsquark} reports comprehensive scaling with collision geometry, system size and transverse kinetic energy $KE_t\equiv m_t - m_h$. It is reasoned that early pressure gradients drive  $KE_t $ which should then replace $p_t$ as a plotting variable. At smaller \pt\ or $KE_t$ mass scaling is observed, leading to inference of hydrodynamic flow. At larger $KE_t$ mesons and baryons scale separately. If quark number $n_q$ scaling (of $v_2$ and $KE_t$) is included all hadron species seem to scale together over a large \pt\ interval.
 
 \begin{figure}[h]
  \includegraphics[width=3.3in,height=1.85in]{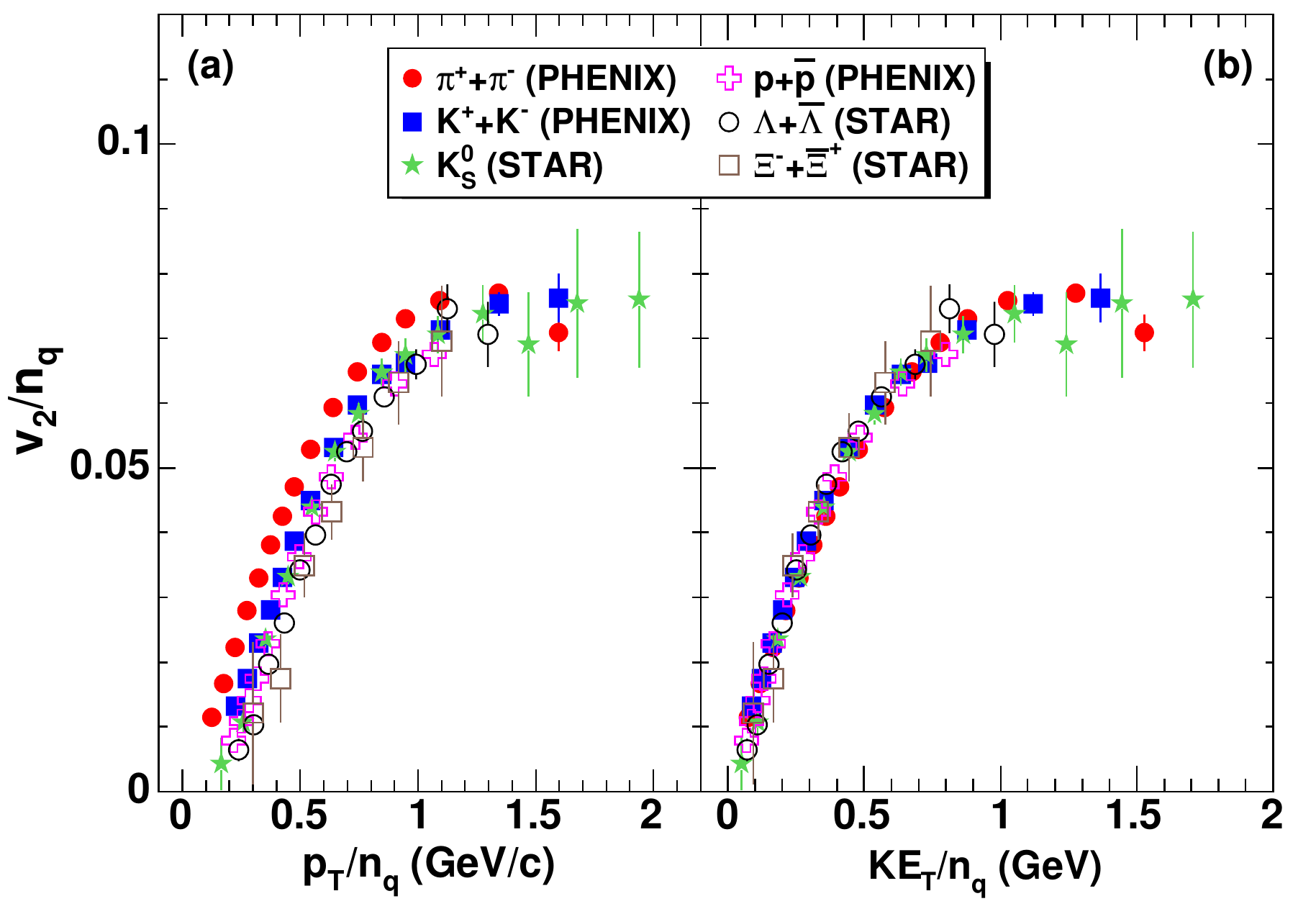} 
\caption{\label{phencq}
Left: $v_2$ for identified hadrons vs $p_t$ from minimum-bias 200 GeV \auau\ collisions, both quantities scaled by constituent-quark number $n_q$ = 2 or 3~\cite{phenconsquark}.
Right: The same data plotted vs transverse kinetic energy $KE_t = m_t - m_h$ also scaled by $n_q$ showing closer scaling.
 } 
 \end{figure}

Figure~\ref{phencq} shows constituent-quark scaling on $p_t$ (left) and $KE_t$ (right) for minimum-bias 200 GeV \auau\ data. In the right panel several hadron species appear to follow the same trend  (true for $\pi$, K, p, $\Lambda$, $\Xi$). Scaling universality is also reported between \cucu\ and \auau\ systems and for various centralities of each. Those data are observed to scale (follow a universal functional form) when plotted as $v_2(p_t,b)/v_2(b)$ vs \pt. A more recent study reports that strong deviations from such scaling in noncentral \auau\ collisions (less than 20\% central) are observed at larger $KE_t$ values ($KE_t/n_q > 1.5$ MeV/$c^2$)~\cite{phennoconsquark}.

\begin{figure}[h]
  \includegraphics[width=2.8in,height=2.2in]{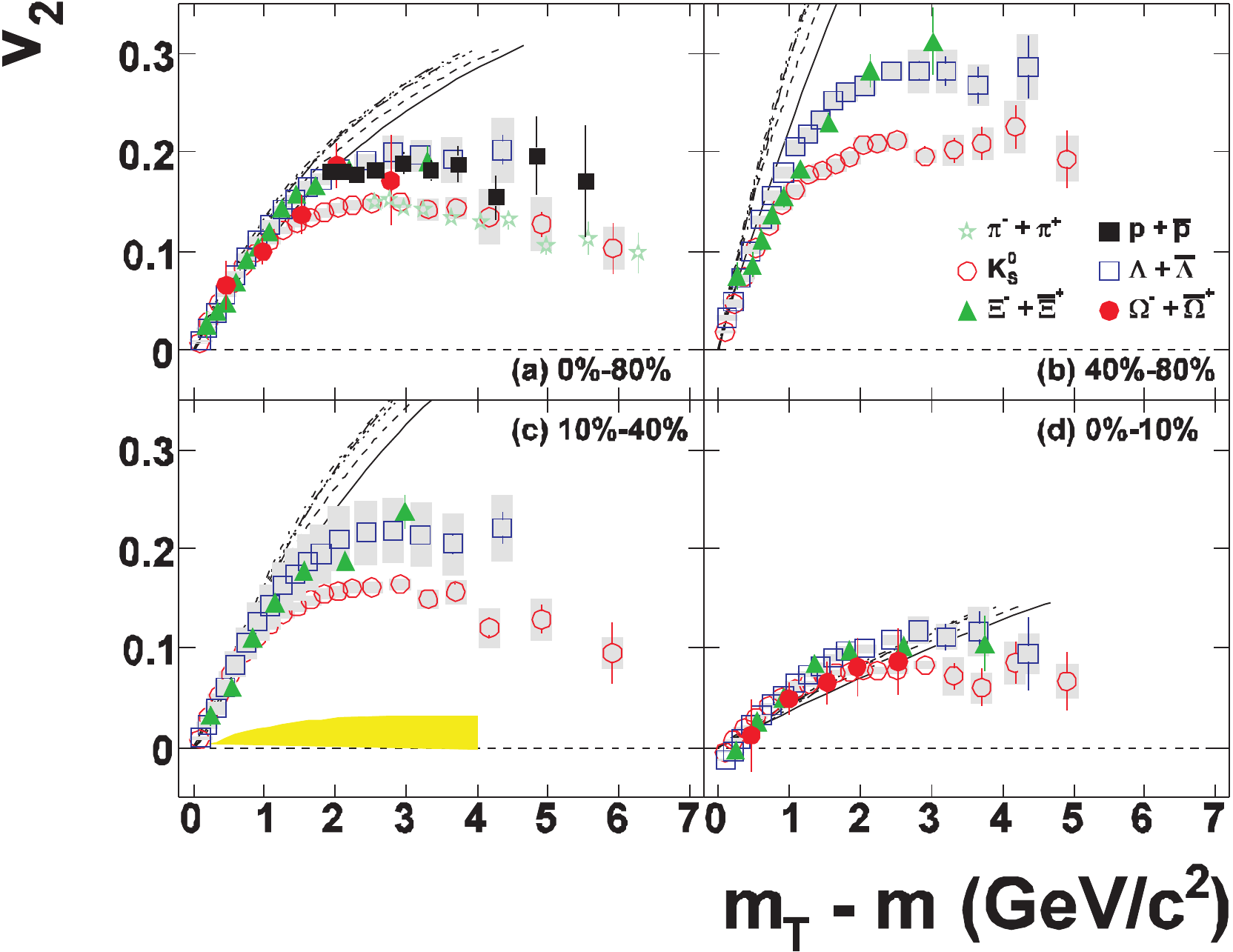} \hfill
\caption{\label{starpid1}
$v_2$ vs $m_t - m_h$ for identified hadrons and several centrality intervals from 200 GeV \auau\ collisions (points) compared with ideal hydrodynamic predictions (curves)~\cite{starpidv2}.
 }  
 \end{figure}

Reference~\cite{starpidv2} presents a comprehensive study of PID $v_2(p_t,b)$ for several hadron species for $p_t$ up to 6 GeV/c and a range of \auau\ centralities. 
Figure~\ref{starpid1} shows PID $v_2$ data for 200 GeV \auau\ collisions and several centrality intervals~\cite{starpidv2}. The figure illustrates agreement between ideal hydro (curves) and data (points) below 2 GeV/c in mass ordering and magnitude, seeming to support a hydrodynamic interpretation. It also shows separation of meson and baryon data in the intermediate $p_t$ interval 2-5 GeV/c interpreted to indicate constituent-quark scaling.

In Figs. 9-11 of Ref.~\cite{starpidv2} a differential study of deviations from a common scaling trend are shown. Substantial systematic deviations are observed, both in data and in theory predictions, leading to the conclusion that neither $v_2$ data nor hydro theory follow strict mass scaling or constituent-quark scaling as they are usually defined.

 \begin{figure}[h]
 \hfill \includegraphics[width=3in,height=4in]{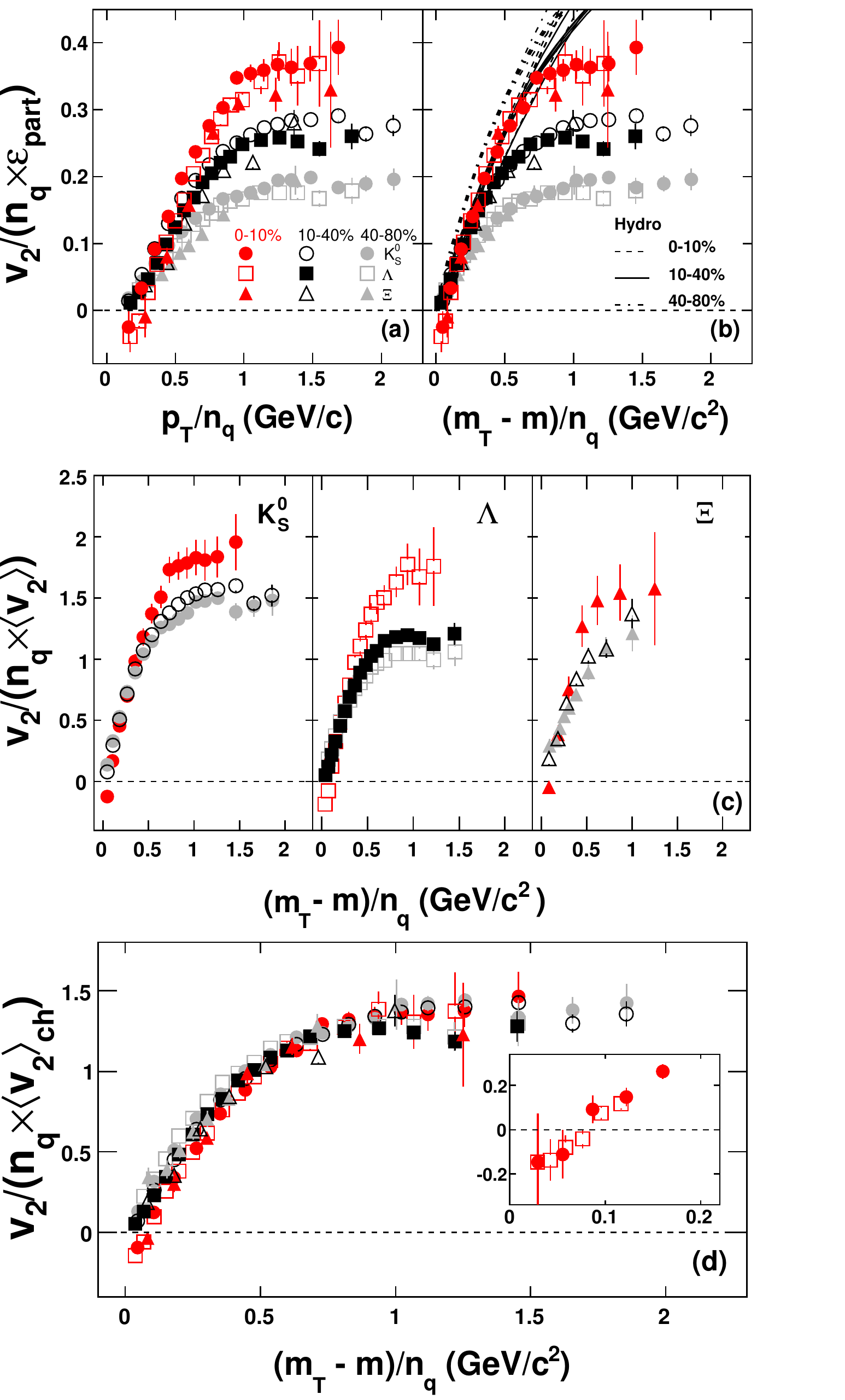} 
\caption{\label{starpid2}
$v_2(p_t)$ data plotted vs $p_t$ and $m_t - m_h$ for several hadron species and for several centrality intervals of 200 GeV \auau\ collisions scaled by constituent quark number $n_q$ and by $p_t$-integral $\langle v_2\rangle$ for individual species and all hadrons~\cite{starpidv2}.
 }  
 \end{figure}

Figure~\ref{starpid2} shows constituent-quark scaling systematics for K, $\Lambda$ and $\Xi$ hadrons~\cite{starpidv2}.  In panel (a) there is no scaling with participant eccentricity for $p_t > 2$ GeV/c. For $p_t < 2$ GeV/c there is mass scaling with $m_t - m_h$, and in the intermediate region there is constituent-quark scaling within each centrality bin but not between centrality bins. The curves at upper right are hydro predictions that approximately follow the data (with mass scaling) below 2 GeV/c but strongly overshoot above that point. Hydro theory shows no quark-number scaling. It is concluded that ``... mass ordering at low $p_t$ alone is not sufficient to claim thermalization in \auau\ collisions at RHIC.''

When the same data are also scaled by $p_t$-integral $\langle v_2 \rangle$ for each hadron species the scaling at larger $p_t$ seems to improve as in panel (c). When the scaling is done instead with $\langle v_2 \rangle_{ch}$ common to all charged hadrons (mainly pions) the scaling improves further as  in panel (d). In the same panel (inset) significantly negative $v_2$ values are shown for the first time, interpreted to support the strong bulk expansion inferred from spectrum analysis, but see Fig.~\ref{quad3} (right).

\subsection{$\bf v_2$ from the RHIC beam energy scan}

 \begin{figure}[h]
  \includegraphics[width=3in,height=1.9in]{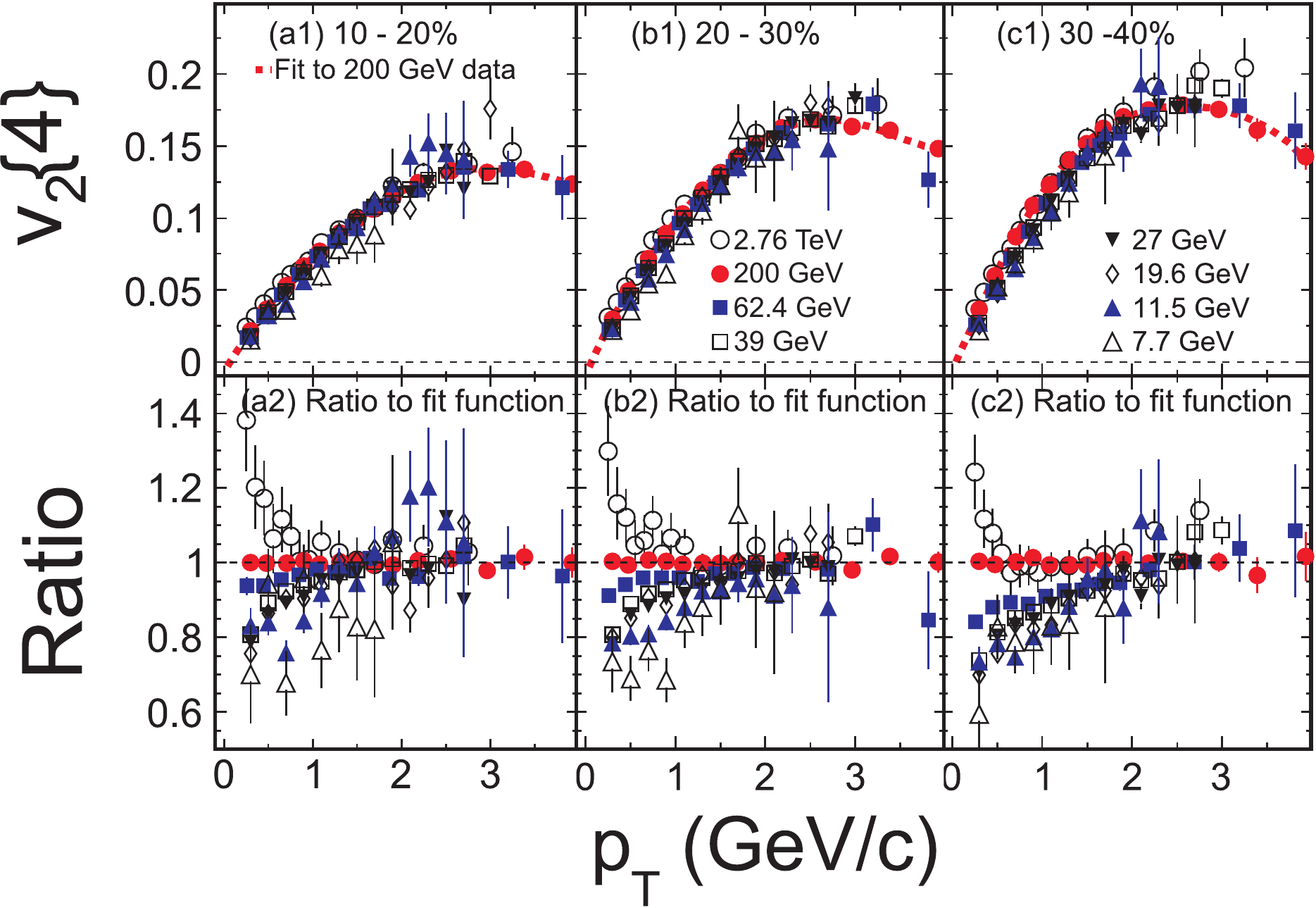} 
  \includegraphics[width=3in,height=1.9in]{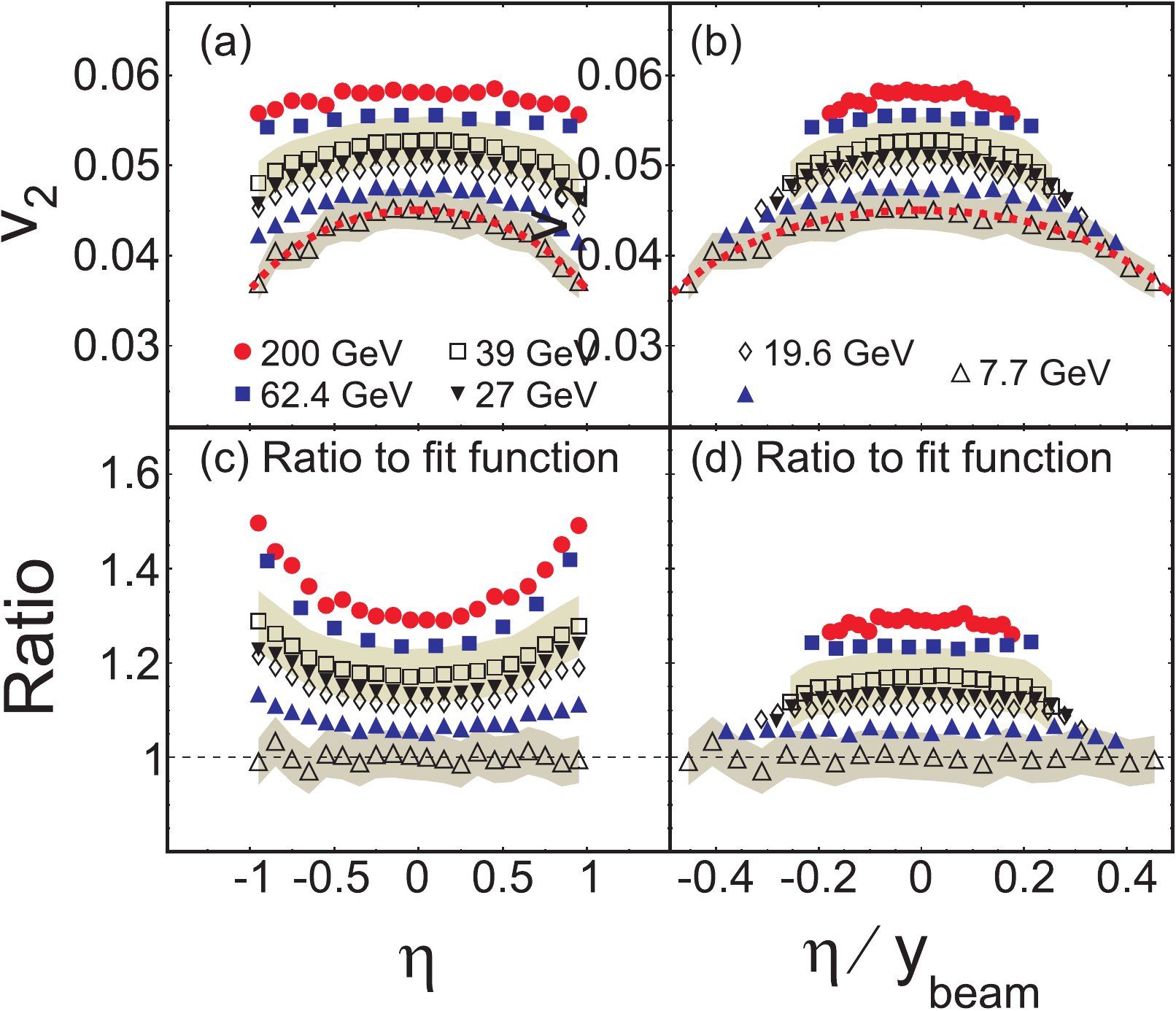} 
\caption{\label{starbes}
Upper: $v_2\{4\}(p_t)$ vs $p_t$ for several collision energies showing similar trends for different energies~\cite{starbesv2}. Significant variation apparent at lower-$p_t$ is compared to no variation at larger $p_t$.
Lower:  $v_2$ vs $\eta$ for several collision energies showing the same functional form on $\eta / y_{beam}$ for all energies.
 }  
 \end{figure}

The RHIC beam energy scan is intended to determine the energy where sQGP production ``turns off,'' and possibly locate a conjectured ``critical point'' on the QCD phase diagram. 
Figure~\ref{starbes} (upper) shows first $v_2$ results for 7.7-39 GeV from STAR~\cite{starbesv2}. The BES data are compared with other $v_2(p_t)$ data up to \pbpb\ collisions at 2.76 TeV. There is surprisingly little variation: small monotonic increase at the lowest \pt\ but no significant increase in the intermediate-\pt\ region.

Figure~\ref{starbes} (lower) shows evolution of the $\eta$ dependence of $v_2$ over the range 7.7 - 200 GeV indicating substantial variation. However, if the $v_2(\eta)$ trend is rescaled by the beam rapidity the functional form appears to be universal (right panels). It is also noted that the difference $v_2^2\{2\} - v_2^2\{4\}$ conventionally attributed to a combination of nonflow and flow fluctuations is reduced at lower energies.  Viscous hydrodynamics does not reproduce the energy dependence of $v_2(p_t)$.

\subsection{Higher harmonic flows}

Measurement of {\em higher harmonic} flows ($v_3$, $v_4$, etc.) in heavy ion collisions is an outgrowth of attempts to measure flow fluctuations as a response to initial-state geometry fluctuations~\cite{starv2flucts} and replacement of the conjectured Mach-cone interpretation of away-side 1D peak distortions (as inferred from ZYAM subtraction) with conjectured ``triangular flow'' represented by $v_3$ as one of several higher harmonic flows~\cite{gunther,luzum}.

Reference~\cite{phenhigherharm} reports measurements of $v_2$, $v_3$, $v_4$ vs \pt\ and $N_{part}$ for 200 GeV \auau\ collisions. The data are apparently well-described by hydro models incorporating a Glauber model of initial-state geometry with fluctuations below 2 GeV/c. The agreement appears to provide evidence for such geometry fluctuations. The results are interpreted to support a value for shear-viscosity measure $\eta/s \approx 1/4\pi$, the quantum lower limit consistent with ``perfect liquid.''

\begin{figure}[h]
  \includegraphics[width=1.65in,height=2.25in]{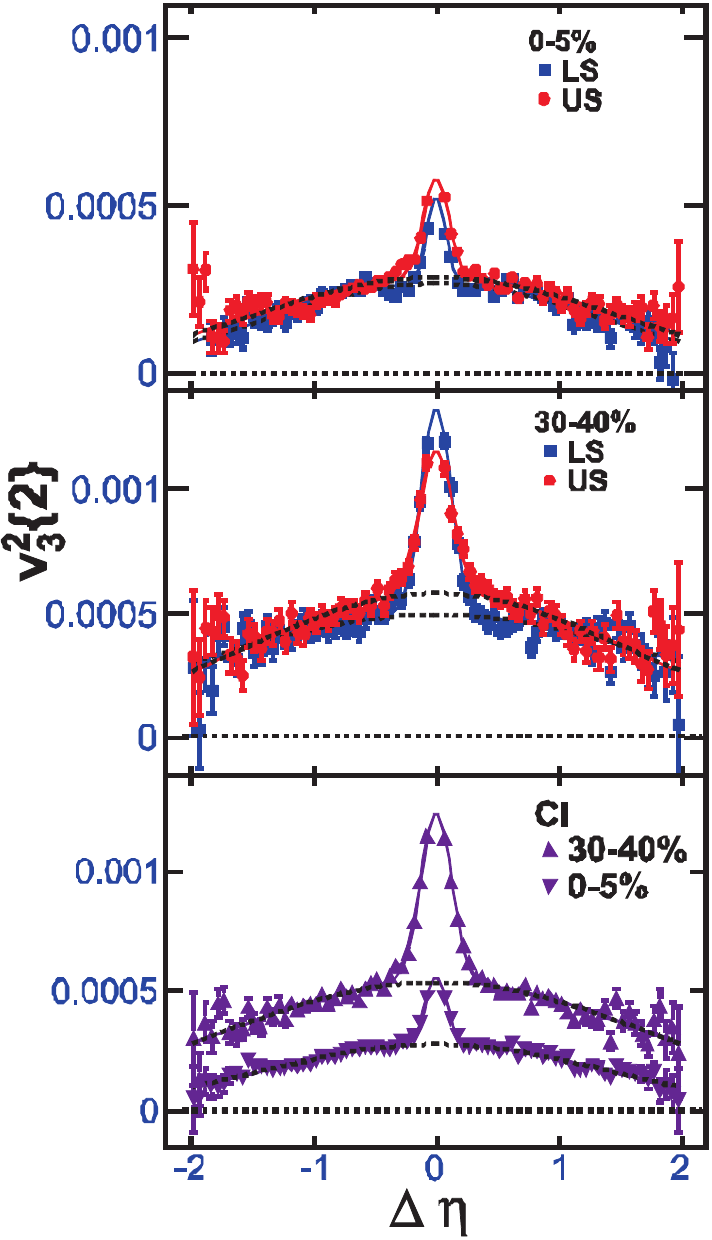} 
  \includegraphics[width=1.65in,height=2.25in]{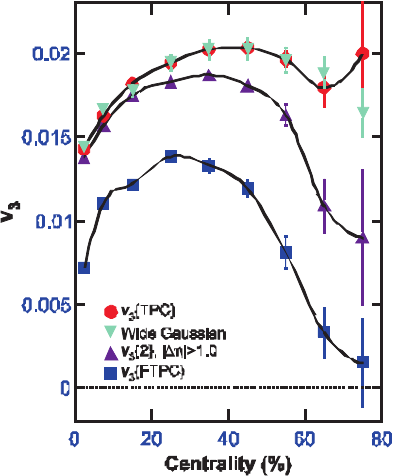} 
\caption{\label{starv3}
Left: Triangular flow measured by $v_3^2\{2\}$ inferred from Fourier fits to 1D projections of 2D angular correlations onto 1D azimuth from narrow $\Delta \eta$ bins~\cite{startriangle}. 
Right: Comparison of $v_3$ inferred from various methods. Data from the left panel (wide Gaussian) correspond to $v_3\{2\}$ inferred within the STAR TPC acceptance (upper-most points).
 }  
 \end{figure}

Reference~\cite{startriangle} reports extraction of $v_3^2\{2\}(\Delta \eta)$ by Fourier decomposition on azimuth of slices bin-by-bin across the eta acceptance. Figure~\ref{starv3} (left) shows $v_3^2\{2\}$ extracted for three \auau\ centrality bins and for CI, LS and US charge combinations. Narrow and broad peak structures are apparent. Those results can be compared directly with 2D angular correlations presented in Fig.~\ref{axialci2} (right) from Ref.~\cite{anomalous}.

Figure~\ref{starv3} (right) shows centrality trends for several $v_3$ methods. For some methods there is substantial triangular flow inferred even for very peripheral \auau\ collisions. It is notable that the $v_3$ values inferred from data in the left panel coincide with those labeled $v_3\{TPC\}$ (event-plane determined within the TPC $\eta$ acceptance, right topmost data) that describe all angular correlations including jets. A substantial role for nonflow is rejected because LS and US results are similar. Based on model comparisons it is concluded that the $v_3$ structure ``...is mainly due to $\Delta \eta$ dependent fluctuations''~\cite{startriangle}.

\section{Jet Angular correlations} \label{jets}

Several methods are employed to infer jet angular structure in \aa\ collisions, including triggered 1D azimuth correlations with background (ZYAM) subtraction, triggered 2D angular correlations on $(\eta,\phi)$, event-wise jet reconstruction and untriggered (minimum-bias) combinatoric 1D and 2D angular correlations.

\subsection{1D azimuth correlations and ZYAM subtraction}

Triggered dihadron correlations on azimuth are formed by selecting the highest-$p_t$ {\em trigger} particle in each event and forming the pair distribution on azimuth difference $\Delta \phi = \phi - \phi_{trig}$ relative to that particle. A subset of the remaining particles in the event is referred to as {\em associated}. Conditions are placed on accepted trigger and associated $p_t$ ranges.  The trigger particle is assumed to act as a proxy for the parent parton of a (triggered) jet. The result is jet angular correlations plus a combinatoric background including a sinusoid contribution interpreted as elliptic flow. The background is estimated using measured $v_2$ data relevant to collision and trigger conditions. The overall background amplitude is determined by the zero-yield-at-minimum (ZYAM) condition, which assumes that jet-related peaks do not overlap~\cite{tzyam}. 

 \begin{figure}[h]
  \hfill \includegraphics[width=3.3in]{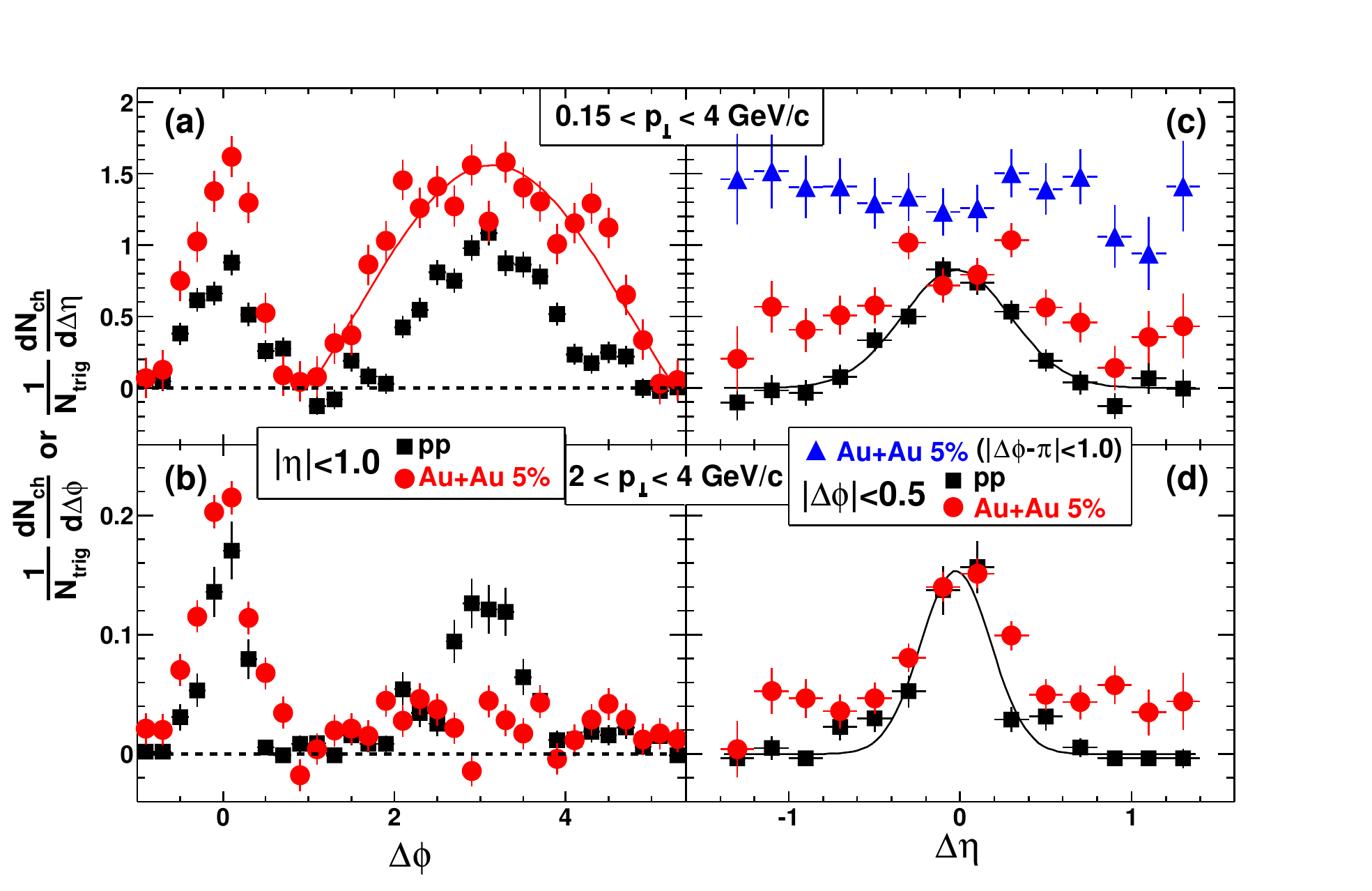} \hfill
\caption{\label{zyamx}
Jet-related dihadron correlations on $\Delta \eta$ and $\Delta \phi$ inferred from 0-5\% central 200 GeV \auau\ and \pp\ collisions by ZYAM subtraction and trigger-associated $p_t$ cuts~\cite{zyam1}.
 }  
 \end{figure}

Figure~\ref{zyamx} compares ZYAM-subtracted trigger-associated angular correlations for trigger particles in 4-6 GeV/c from 200 GeV central \auau\ and \pp\ collisions~\cite{zyam1}. The upper panels include associated particles within 0.15-4 GeV/c, the lower panels include only 2-4 GeV/c. In  the former all jet-related structure increases from \pp\ to central \auau\ and the AS peak is undistorted. In central \auau\ collisions it is observed that the away-side peak includes a softer particle complement than for \pp\ collisions. It is noted that the AS peak in the upper-left panel has a dipole $\cos(\Delta \phi)$ shape, suggesting global momentum conservation as the main source. However, the minimum-bias AS peak for \pp\ collisions has a similar shape~\cite{porter2,porter3,pptheory}. It is concluded that jets are modified in more-central \auau\ collisions.

Further exploration of dihadron correlation systematics with various trigger-associated cut combinations appeared to reveal strong distortion of the AS peak shape for some combinations. A possible source might be production of Mach cones as a medium response to interaction of energetic partons with the QCD medium~\cite{mach}.

 \begin{figure}[h]
  \hfill \includegraphics[width=3.3in]{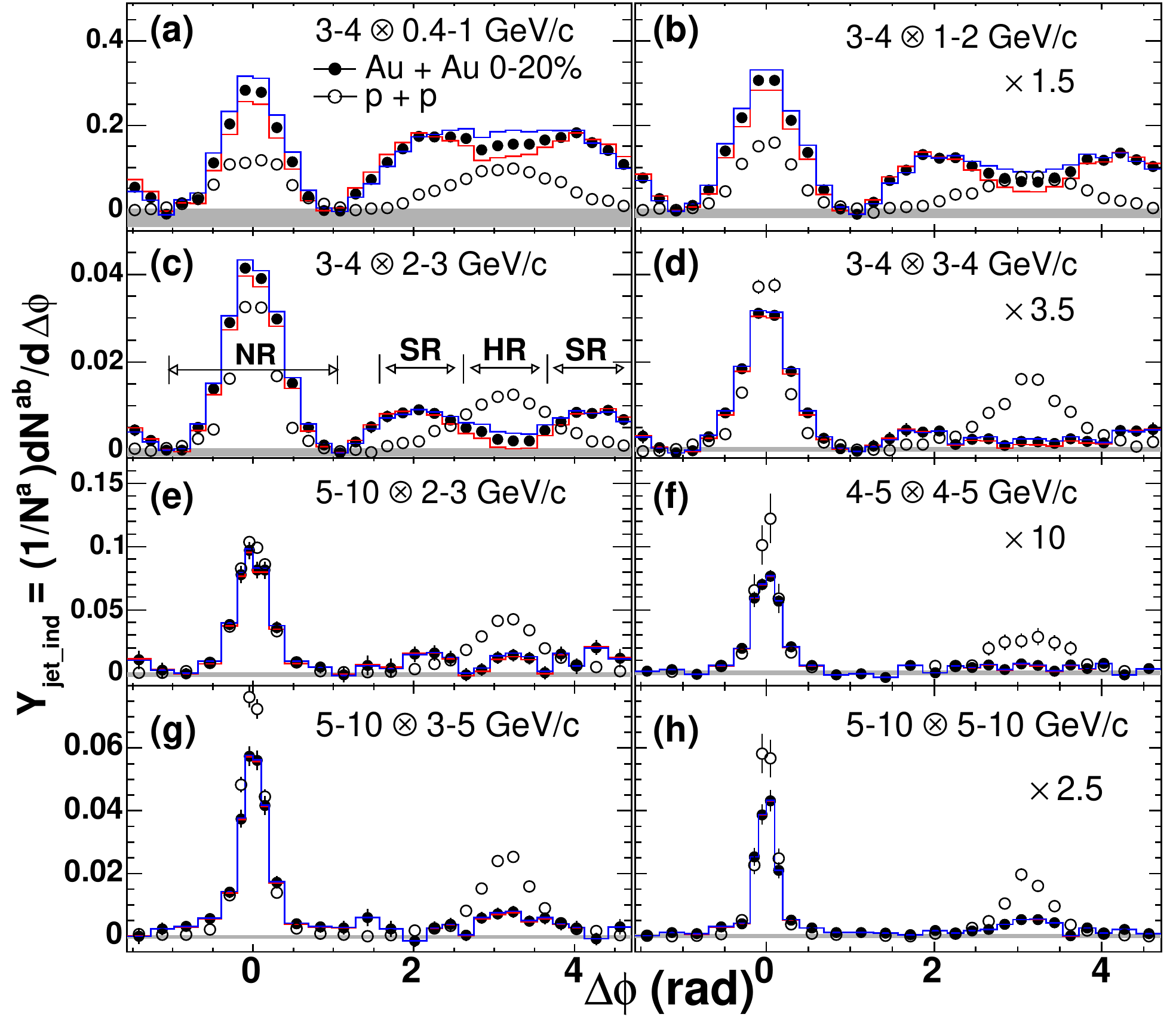} \hfill
\caption{\label{zyamy}
 Jet-related dihadron azimuth correlations from 200 GeV \auau\ collisions for various combinations of trigger-associated $p_t$ cuts~\cite{phenbigzyam}.
 }  
 \end{figure}

Figure~\ref{zyamy}  compares dihadron correlations from 200 GeV \auau\ and \pp\ collisions for several trigger-associated \pt\ cut combinations~\cite{phenbigzyam}. For the lower trigger cut 3-4 GeV/c AS double peaks are observed near $\Delta \phi = \pi \pm 1.1$. The double peaks are said to appear in a ``shoulder'' region (SR). Jet structure in more-central \auau\ is interpreted to have four components:  
(i) a jet fragmentation component near $\Delta \phi = 0$, 
(ii) a punch-through jet fragmentation component near  $\Delta \phi = \pi$, 
(iii) a medium-induced component near  $\Delta \phi = 0$ and 
(iv) a medium-induced components near  $\Delta \phi = \pi \pm 1.1$. The AS double-peak separation is said to be independent of $p_t$ as expected for conjectured Mach shocks (Mach cones). Similar structures were reported by Ref.~\cite{starbigzyamprc}  where AS peak evolution was described as follows: ``The transition from a broad away-side structure at low \pt\ to a narrow structure at higher \pt\ would then signal the change from away-side structures dominated by bulk particle production from the medium to a situation where jet-fragments dominate.''

\subsection{2D angular correlations and the ``ridge''}

Trigger particles can also be used to generate 2D angular correlations on $(\eta,\phi)$ providing more information about jet structure. The systematic variation of $\eta$ elongation first observed for minimum-bias correlations reported in Ref.~\cite{axialci}  is then revealed with trigger-associated cuts, and a same-side ``ridge'' structure is identified.

 \begin{figure}[h]
   \includegraphics[width=1.65in,height=1.65in]{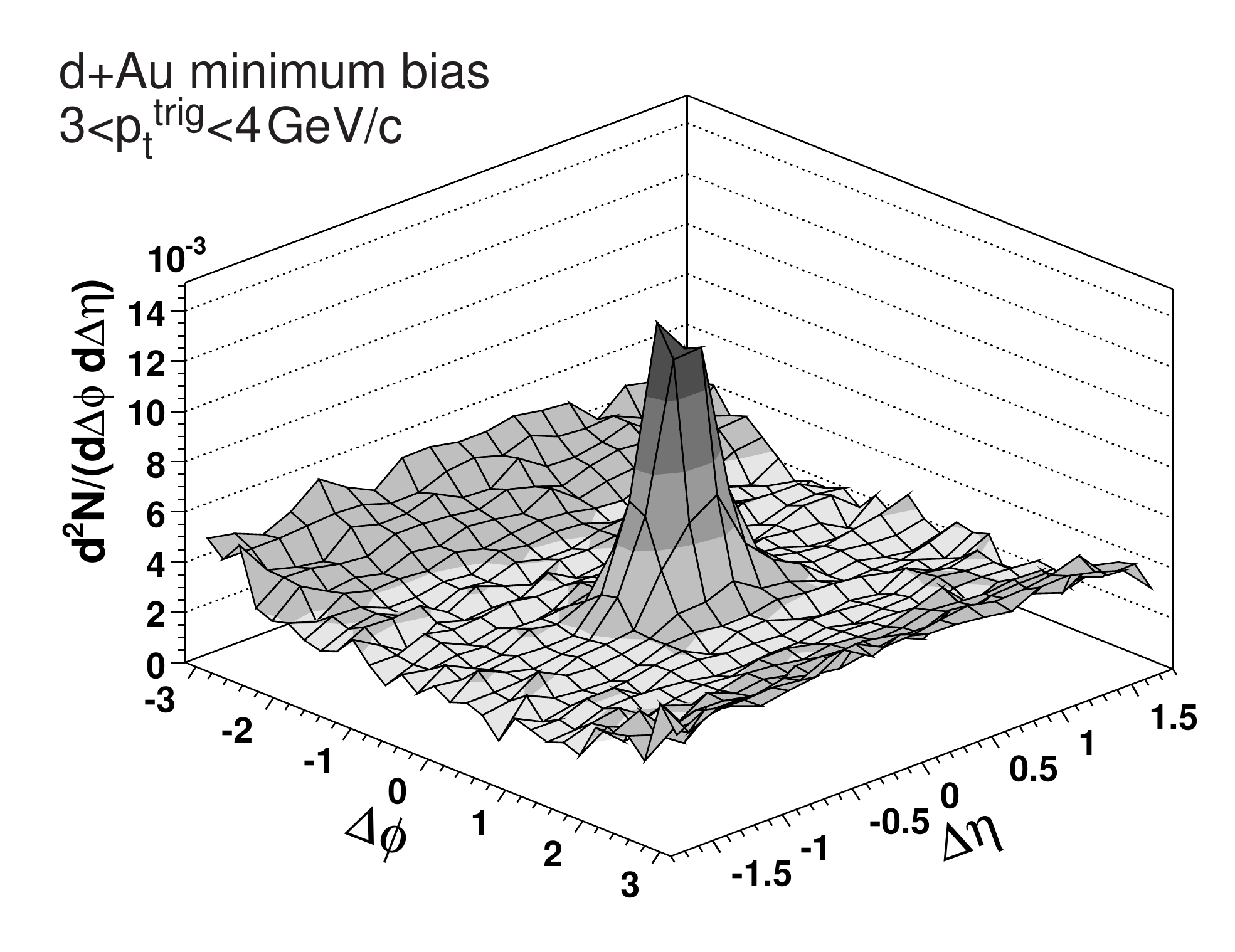}
 \includegraphics[width=1.65in,height=1.65in]{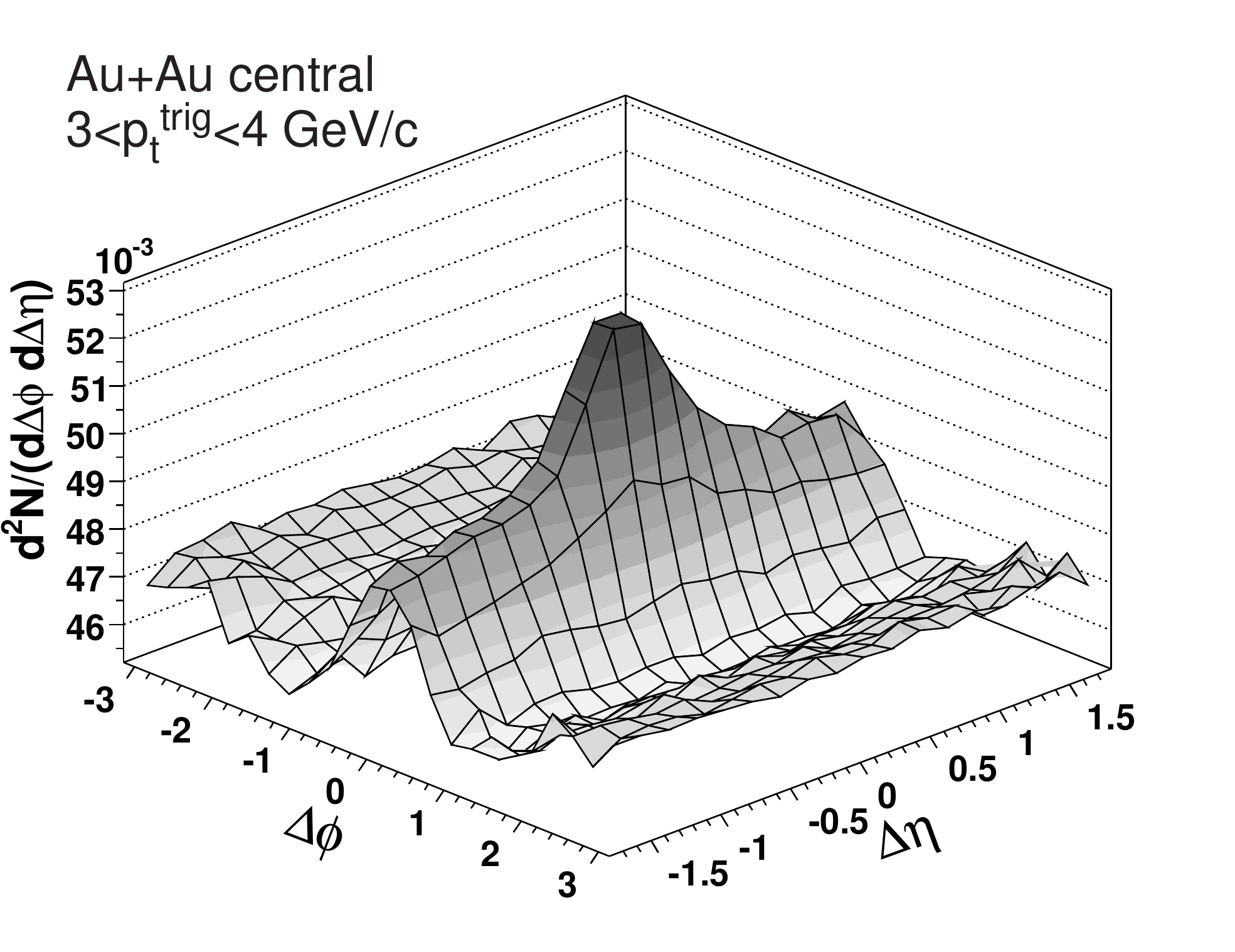}
\caption{\label{starridg}
2D angular correlations relative to a 3-4 GeV/c trigger-particle (angle relative to trigger) for 200 GeV d-Au collisions (left) and central \auau\ collisions (right)~\cite{starridgeprc}.
 }  
 \end{figure}

Figure~\ref{starridg} shows triggered 2D angular correlations from d-Au (left) and central \auau\ (right) collisions~\cite{starridgeprc}. The trigger \pt\ interval is 3-4 GeV/c and associated interval is 2-3 GeV/c. In the d-Au case the expected nearly-symmetric SS 2D peak attributed to jets is observed. In the \auau\ case ``long-range'' (on $\eta$) structure emerges that extends beyond the STAR TPC acceptance. The elongated structure is referred to as a ``ridge.'' Systematic studies are interpreted to conclude that ``jet-like'' and ``ridge-like'' structures come from distinct mechanisms. The latter (presumed nonjet) mechanism has received much theoretical attention~\cite{glasma}. In Ref.~\cite{dihadronzyamphob} it is shown that the same-side ridge appears to extend to large $|\Delta \eta| \approx 4$.

Triggered dihadron correlations may be used to probe the QCD medium (jet tomography). Strong jet quenching and apparent large reduction of AS jet yields suggests formation of an opaque colored medium in more-central collisions. The parton pathlength through the medium may then strongly influence jet modification.

In Ref.~\cite{stariaa} direct photons are used to establish a trigger direction relative to which away-side jet structure is studied. The photon energy then estimates the recoil energy of the parent parton of the away-side (partner) jet. Reduction of the away-side associated-hadron yield shows no significant dependence on photon (parton)  energy. It is concluded that dependence on parton species (light quark or gluon) and parton pathlength in the medium is also small.

In Ref.~\cite{starjetsurface} a novel {\em double}-trigger plus associated (2+1) technique is introduced: two high-\pt\ triggers are separated by $\pi$ on azimuth. The double trigger is said to bias toward tangential emission of jet partners outside an opaque core. Such dijets then do not interact with the core. Double-triggered same-side and away-side jet-related peaks in central \auau\ have similar characteristics to those in d-Au collisions.

 \begin{figure}[h]
   \includegraphics[width=3.3in,height=2.5in]{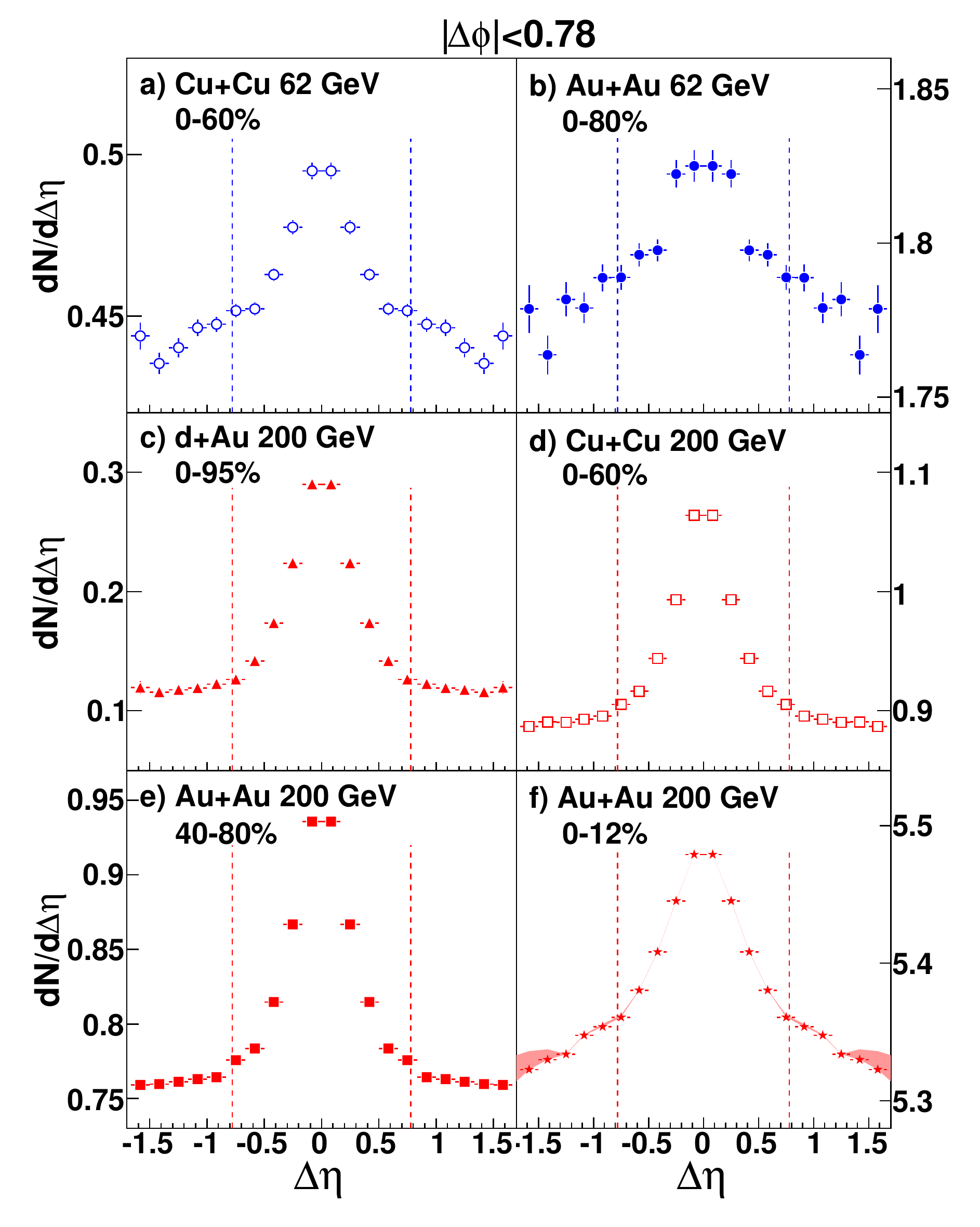}
 \includegraphics[width=3.3in,height=2.5in]{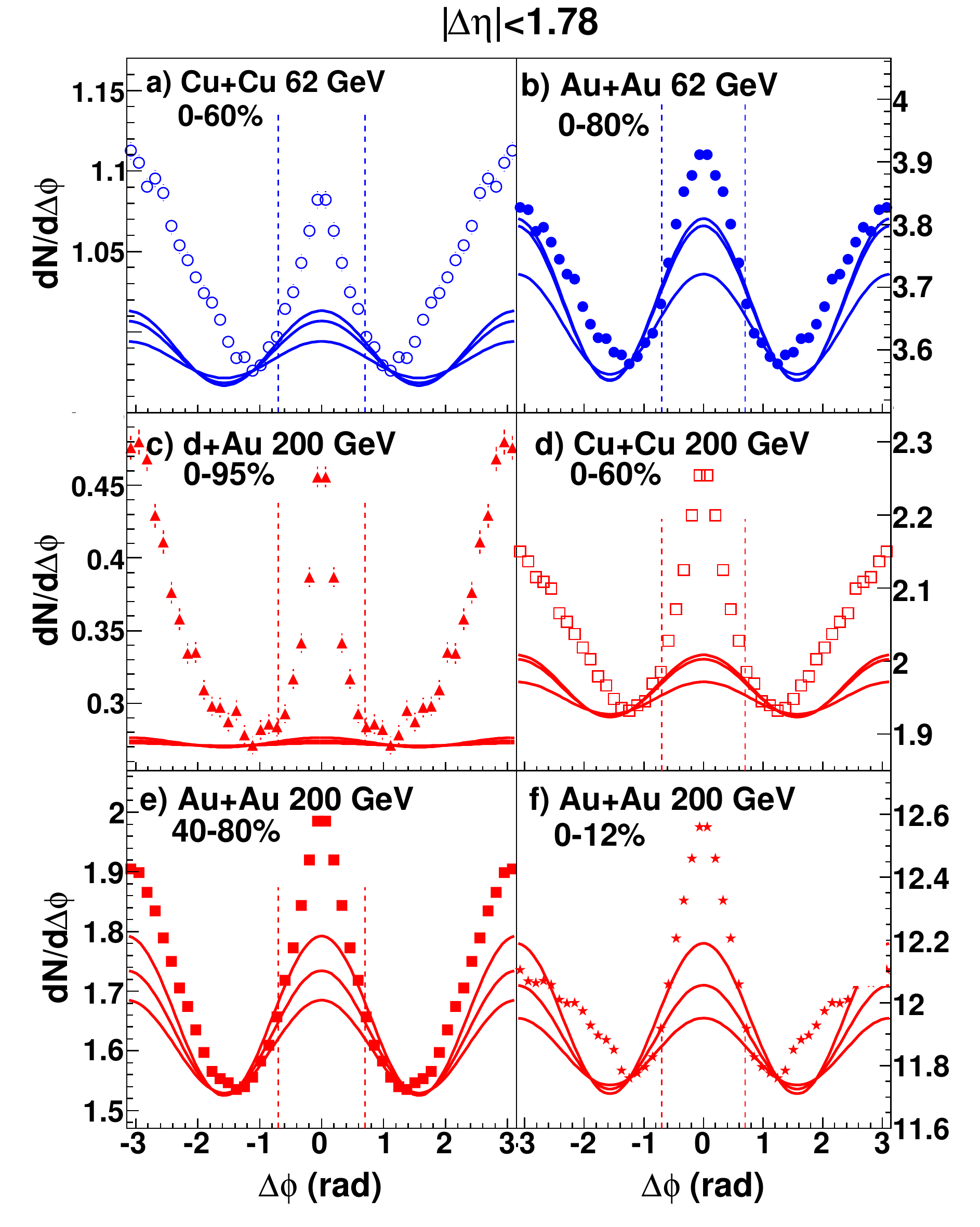}
\caption{\label{starnear}
Systematic study of same-side 2D peak shapes for different collision systems~\cite{starnearjets}.
Upper: Peak shapes on $\Delta \eta$ for specific $\Delta \phi$ acceptance cuts and \cucu, d-Au and \auau\ collisions at 62 and 200 GeV. 
Lower: Peak shapes on $\Delta \phi$ for the same systems showing ZYAM subtraction.
 }  
 \end{figure}

Figure~\ref{starnear} shows a study of the structure of the SS 2D peak vs several collision parameters (system size, centrality and energy)~\cite{starnearjets}. The upper panel shows SS peak shapes on $\Delta \eta$ for specific cuts on $\Delta \phi$ illustrating evolution of $\eta$ elongation with collision conditions. The lower panel shows SS peak shapes on $\Delta \phi$ illustrating the $\Delta \phi$ cuts and ZYAM subtraction (curves). The SS peak is described as having jet-like (narrow on $\Delta \eta$) and ridge-like (broad on $\Delta \eta$) components. The ridge structure is said to persist to $p_{t,trig} \approx 6$ GeV/c and $p_{t,assoc} \approx 3$ GeV/c. A 2D background is subtracted according to the ZYAM prescription. It is interesting that for 0-12\% central \auau\ (upper panel, lower right) the elongated ``ridge'' structure is consistent with monotonic decrease toward zero on $\Delta \eta$, possibly part of a {\em single} monolithic peak. In general, both jet-like peak and ridge share the same small azimuth width. The ridge persists in \cucu\ collisions and at 62 GeV. And the energy dependence of jet-like and ridge-like structures is equivalent.

\section{Fluctuations} \label{flucts1}

Fluctuation measurements were expected to reveal evidence for traversal of the QCD phase boundary in the case of a first-order transition. However, fluctuation signals are thought to be less important for a smooth crossover transition. More recently, the search for a conjectured QCD critical point (CP) in association with the RHIC beam energy scan (BES) program has been advocated. Fluctuations of event-wise mean \pt, multiplicity, net charge, net baryon number and hadron species ratios such as the $K/\pi$ ratio are reported. Several fluctuation analyses are summarized below.

Reference~\cite{starptfluctsergei} describes $\langle p_t \rangle$ fluctuations measured by $\langle \Delta p_{t,i} \Delta p_{t,j} \rangle$, where $\Delta p_t = p_t - \bar p_t$ is the particle $p_t$ deviation from the ensemble mean. The definition leads to a {\em per-pair} fluctuation measure. The general trend with \auau\ centrality is decrease approximately as $1/N_{part}$ or $1/n_{ch}$ as expected, and the increase with energy is quite slow. If that measure is multiplied by $dn_{ch}/d\eta$ (converting to a per-particle measure) the centrality dependence is similar to Fig.~\ref{mpt} (left) from Ref.~\cite{ebyept1} and there is a strong increase with collision energy. The paper concludes that ``...there are clear nonzero $p_t$ correlations. [...] The centrality dependence...may show signs of...thermalization, the onset of jet suppression, the saturation of transverse expansion in central collisions, or other processes.'' 

Reference~\cite{phenmultflucscale} addresses charged-particle density correlations on $\eta$ by measuring multiplicity fluctuations. An attempt is made to locate the QCD phase boundary in the context of the Ginzburg-Landau description of critical phenomena. Multiplicity distributions in $\eta$ windows of increasing size $\delta \eta$ and various \auau\ centralities are modeled with a negative binomial distribution (NBD). The NBD model parameter $k$ measures deviations from Poisson fluctuations (no correlations) and is parametrized by $1/k = 2\alpha \xi / \delta \eta + \beta$. $1/k$ trends on $\delta \eta$ and $N_{part}$ represented by parameters $\alpha \xi$ and $\beta$ tend to decrease as $1/N_{part}$ since $1/k$ is also a per-pair measure. The study concludes ``...The behavior [of product $\alpha \xi$] may be explained by the onset of a mixture of different types of particle production mechanisms which are not necessarily related to temperature or density correlations. However, interpreted within the Ginzburg-Landau framework the local maximum of the $\alpha \xi$ product [on $N_{part}$, at most a two-sigma effect] could be an indication of a critical phase boundary.''

Reference~\cite{phenbigmultflucprc} measures fluctuations with so-called {\em normalized variance} $\sigma^2_{n_{ch}} / \bar n_{ch}$ denoted by $w_{ch}$, a per-particle measure. The relation to NBD model parameter $k$ is $w_{ch} = 1 + \bar n_{ch}/k$. An estimate $w_{ch,dyn}$ of the excess fluctuations relative to a statistical reference is made. The trend of $w_{ch,dyn}$ is similar to that in Fig.~\ref{mpt} (left) from Ref.~\cite{ebyept1} describing $\langle p_t \rangle$ fluctuations. The study concludes ``...there is no evidence of critical behavior related to the compressibility observable in this dataset. There is also no significant evidence of dynamical fluctuations that are dependent on the transverse momentum or the charge of the particles measured. [...] Although
this analysis does not observe evidence of critical behavior, it does not rule out the existence of a QCD critical point.''


Reference~\cite{starkpifluct} reports an analysis of $K/\pi$ fluctuations motivated by the possibility to detect critical fluctuations in strangeness enhancement near the QCD phase boundary. Figure~3 of that reference shows $K/\pi$ fluctuations as measured by per-pair measure $\nu_{dyn}$. The trend is dominated by the expected $1/n_{ch}$ dependence. In Fig.~4 of the reference is plotted $(dn_{ch}/d\eta) \nu_{dyn}$ (for charge combinations LS - green, US - blue and CI - red), a per-particle measure which increases slowly with \auau\ centrality. The study concludes ``These results may indicate that, due to later stage hadronic rescattering, the decay products of resonances are less likely to survive in central collisions than in peripheral
collisions.''

\section{Alternative Methods and Results} \label{alternate}

Section~\ref{alt} summarizes alternative analysis methods and measures that may be applied without a priori assumptions about \aa\ collision mechanisms. In this section we review a selection of those results and consider some implications for interpretation of RHIC data.  

\subsection{Mid-rapidity hadron yields}

Mid-rapidity hadron yields are said to play a key role in constraining hydro calculations and indicating hadron production methods. Two alternative scenarios have been proposed: (a) The TCM including longitudinal projectile dissociation (soft) and transverse parton scattering and fragmentation (hard) and (b) a CGC glasma \aa\ initial state that may transition directly to a QGP. Arguments have been proposed based on HIJING {\em simulations} that the TCM and minijets are falsified by hadron yield systematics, that the TCM predicts much more yield increase with \aa\ centrality than is observed~\cite{liwang}. In fact the TCM with the correct in-vacuum dijet cross section (2.5 mb) {\em underpredicts} hadron yields for more-central \auau\ collisions. The \auau\ data from more-central collisions requires modified parton fragmentation---increase of mean fragment multiplicities by up to a factor 3.

\begin{figure}[h]
  \includegraphics[width=1.65in,height=1.65in]{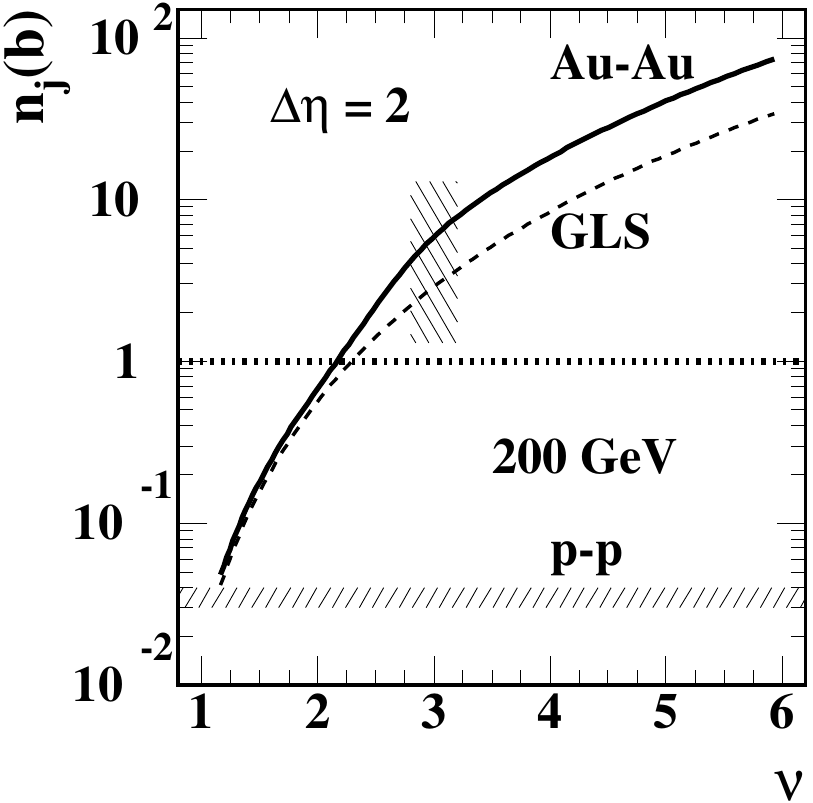}
  \includegraphics[width=1.65in,height=1.68in]{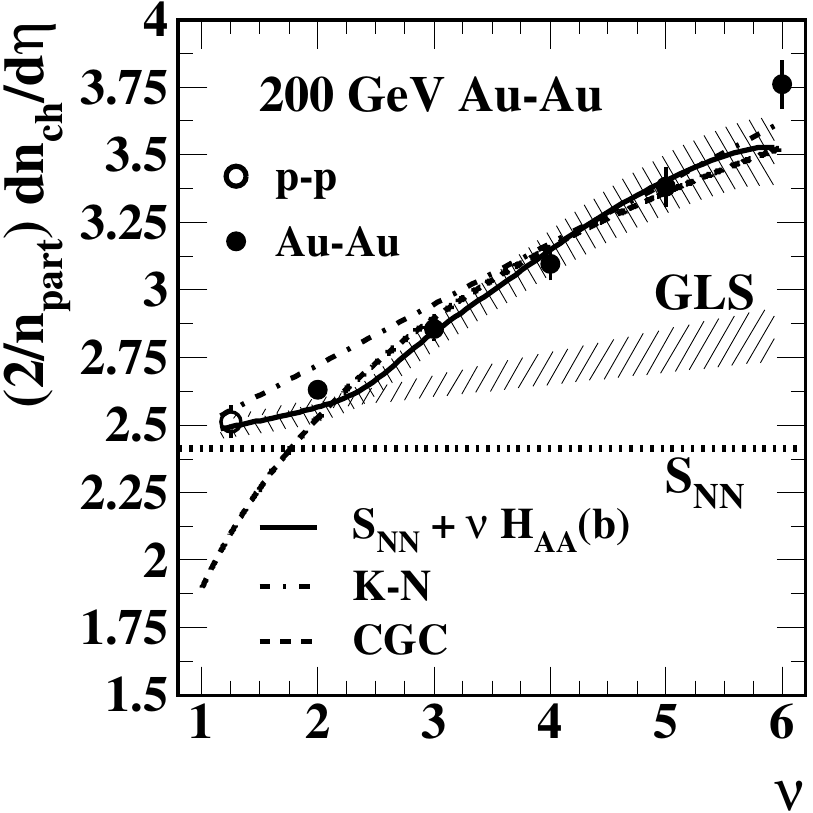}
\caption{\label{midrap}
Left: Dijet number $n_j$ per \auau\ collision within acceptance $\Delta \eta$ vs centrality measure $\nu$ for 200 GeV \auau\ collisions~\cite{jetspec}. Binary-collision scaling of \pp\ dijet cross section 2.5 mb is shown by the dashed curve. Possible increase of the dijet cross section for more-central collisions inferred from spectrum analysis~\cite{hardspec,fragevo} is describe by the solid curve. The upper hatched band indicates the location of the sharp transition in minijet characteristics, also where the dijet number per unit $\eta$ in \auau\ significantly exceeds unity.
Right: The TCM prediction for per-participant hadron production vs centrality based on minijet angular correlations (solid curve) compared to spectrum data (points)~\cite{jetspec}. The dashed curve shows a (scaled) CGC prediction~\cite{glasma1}. The GLS extrapolation (lower hatched band) is based on measured \pp\ trends~\cite{ppprd}.
 }  
 \end{figure}

Figure~\ref{midrap} (left) shows the centrality dependence of dijet production within $\Delta \eta = 2$ for 200 GeV \auau\ collisions~\cite{jetspec}. The GLS curve is based on a dijet total cross section of 2.5 mb consistent with \pp\ spectrum data~\cite{fragevo}. The solid curve corresponds to a 50\% increase above $\nu = 3$ inferred from \auau\ spectrum data~\cite{hardspec}.

Figure~\ref{midrap} (right) shows the total hadron yield inferred from $p_t$-integral 2D angular correlations (solid curve)~\cite{jetspec}. The volume of the SS 2D peak (number of jet-correlated pairs) is combined with the dijet number in the left panel to infer the mean fragment number per jet $\bar n_{ch,j}$.  The hard-component yield $\nu H_{AA} = n_j\, \bar n_{ch,j}$ is combined with fixed soft component $S_{NN}$ to predict the per-participant-pair hadron yield (solid curve)~\cite{jetspec}. The points are integrated from spectrum data~\cite{hardspec}. The dash-dotted line is a fitted two-component model with fixed parameter $x \approx 0.1$~\cite{kn}. The lower hatched region represents a GLS extrapolation of \pp\ systematics, with $x \approx 0.03$ corresponding to a fixed 2.5 mb pQCD dijet cross section. The yield increase above that value for more-central collisions results from the combination of a 50\% increase in the dijet cross section and a three-fold increase in the mean jet fragment multiplicity as inferred from \auau\ spectrum data. It is notable that the mean dijet number for central \auau\ collisions is 30-60 corresponding to a large jet fragment yield -- 30\% of all hadrons are contained within resolved jets~\cite{jetspec}.

The accurate TCM description of hadron production data for all \auau\ centralities is also notable. A CGC prediction (dashed curve) has the wrong functional form $\propto \log(8\nu)$ and {\em does not predict absolute yields}, has been scaled to pass through the more-central data points~\cite{glasma1}. Any successful description of hadron production must accommodate the interval of \aa\ {\em transparency} below $\nu = 3$ corresponding to 50\% of the \aa\ total cross section~\cite{anomalous}.

\subsection{$\bf y_t$ spectra from p-p and \auau\ collisions}

An accurate mathematical model of \pp\ spectrum structure provides an essential reference for \aa\ collisions and strongly suggests the underlying hadron production processes. The multiplicity dependence of \pt\ or \yt\ spectrum shapes leads directly to a two-component model or TCM. No a priori physical model is imposed.

\begin{figure}[h]
  \includegraphics[width=1.65in,height=1.65in]{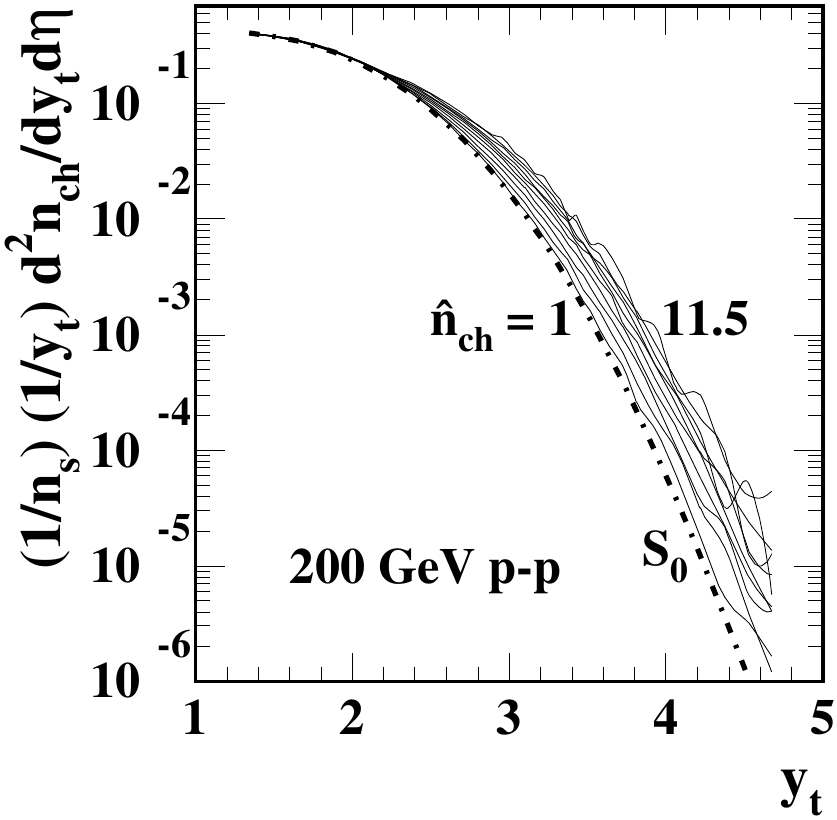}
  \includegraphics[width=1.65in,height=1.65in]{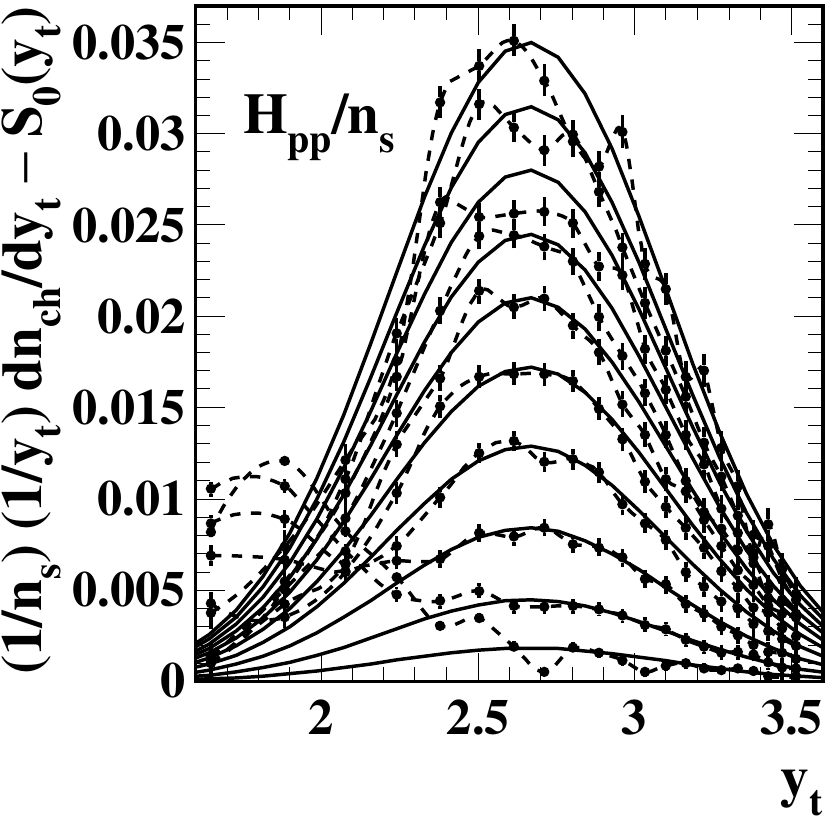}
\caption{\label{ppspec}
Left: Spectra for ten $n_{ch}$ classes from 200 GeV \pp\ collisions~\cite{ppprd}.
Right: Spectrum hard components extracted from the spectra in the left panel.
 }  
 \end{figure}

Figure~\ref{ppspec} (left) shows hadron spectra on \yt\ for ten multiplicity classes within acceptance $|\eta| < 0.5$~\cite{ppprd}. The spectra have been normalized by soft multiplicity $n_s$ defined iteratively in terms of inferred hard-component multiplicity $n_h$ and the relation $n_{ch} = n_s + n_h$. $\hat n_{ch} \approx n_{ch}/2$ is the uncorrected multiplicity expressed in terms of the corrected multiplicity (corrected for $p_t$ acceptance and tracking efficiency). The soft-component reference $S_0(y_t)$ is the limiting spectrum shape for $n_{ch} \rightarrow 0$.

Figure~\ref{ppspec} (right) is the hard component, obtained by subtracting $S_0$ from the spectrum data, described by a peaked (Gaussian) distribution $H_0(y_t)$ with amplitude approximately $\propto n_{ch}$ (solid curves). The spectrum systematics are summarized (within angular acceptance $2\pi$ and $\Delta \eta = 1$) by
\bea
\frac{dn_{ch}}{y_t dy_t} &=& n_s S_0(y_t) + n_h H_0(y_t),
\eea
where $n_h / n_s \propto n_{ch}$ and $H_{pp} = n_h H_0(y_t)$~\cite{ppprd}.

\begin{figure}[h]
  \includegraphics[width=1.65in,height=1.65in]{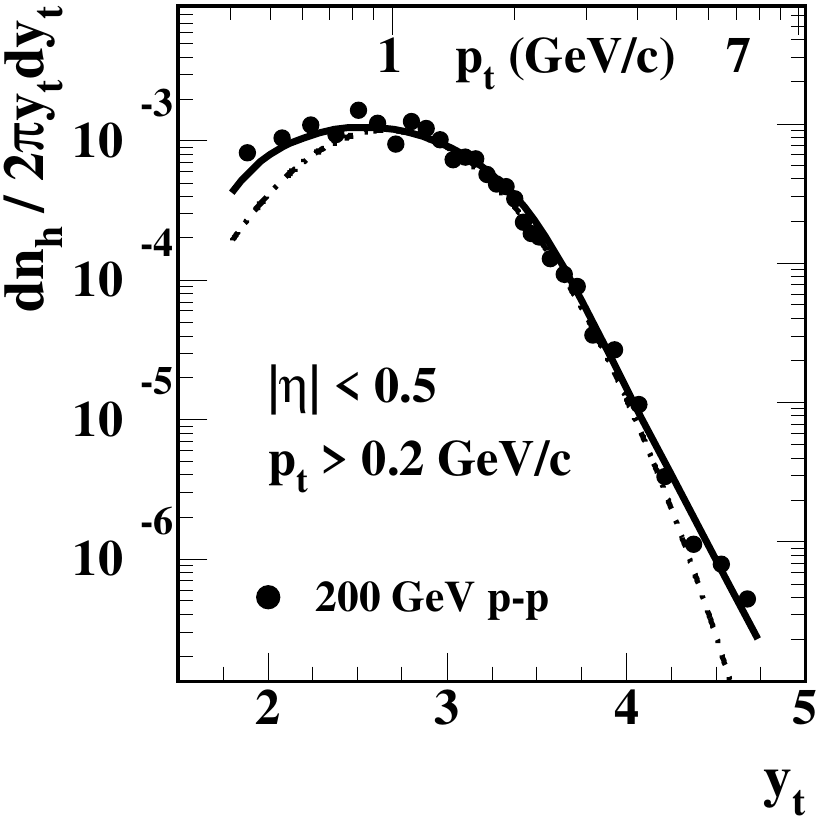}
  \includegraphics[width=1.65in,height=1.65in]{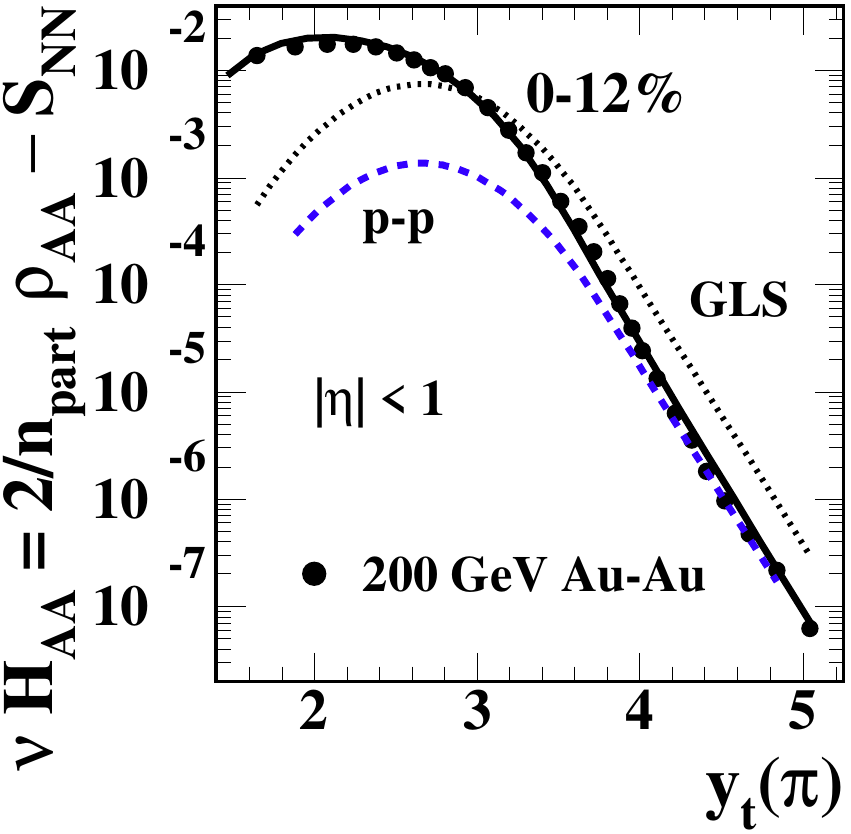}
\caption{\label{specfrag}
Left:  Spectrum hard component representing 200 GeV NSD \pp\ collisions (points)~\cite{fragevo}. The solid curve is a pQCD prediction for the corresponding fragment distribution derived from measured fragmentation functions and a dijet total cross section of 2.5 mb.
Right: Spectrum hard component for 0-12\% central 200 GeV \auau\ collisions (points)~\cite{hardspec}. The solid curve is a pQCD prediction based on a simple modification of fragmentation functions~\cite{fragevo}. The dotted curve is the prediction from the TCM extrapolated from \pp\ collisions.
 }  
 \end{figure}

Figure~\ref{specfrag} (left) shows the non-single-diffractive (NSD) average of the \pp\ spectrum hard components (points) from Fig.~\ref{ppspec}. The solid curve is a pQCD calculation of the \pp\ spectrum hard component based on a pQCD parton spectrum bounded below at 3 GeV and integrating to 2.5 mb and measured fragmentation functions from CDF, LEP and HERA~\cite{fragevo,eeprd}. The dash-dotted curve is the Gaussian approximation from Ref.~\cite{ppprd}. 

Figure~\ref{specfrag} (right) shows the spectrum hard component from 0-12\% central 200 GeV \auau\ collisions (points)~\cite{hardspec}. The dashed curve is the \pp\ Gaussian (with added power-law tail) from the left panel. The dotted curve is a GLS prediction for central \auau\ (\aa\ transparency). The solid curve is a pQCD description of the central \auau\ data based on a single modification of measured fragmentation functions (single-parameter change in a gluon splitting function)~\cite{fragevo}. Fragment reduction at larger $y_t$ is balanced by much larger fragment increase at smaller $y_t$ that conserves the parton energy {\em within resolved jets}.

\begin{figure}[h]
  \includegraphics[width=1.65in,height=1.65in]{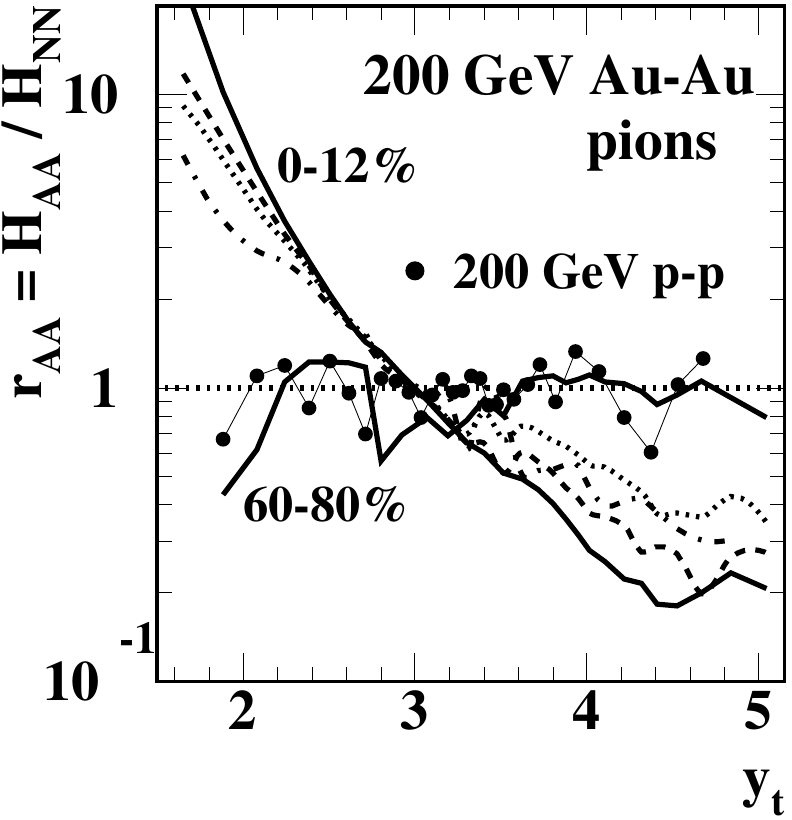}
  \includegraphics[width=1.65in,height=1.65in]{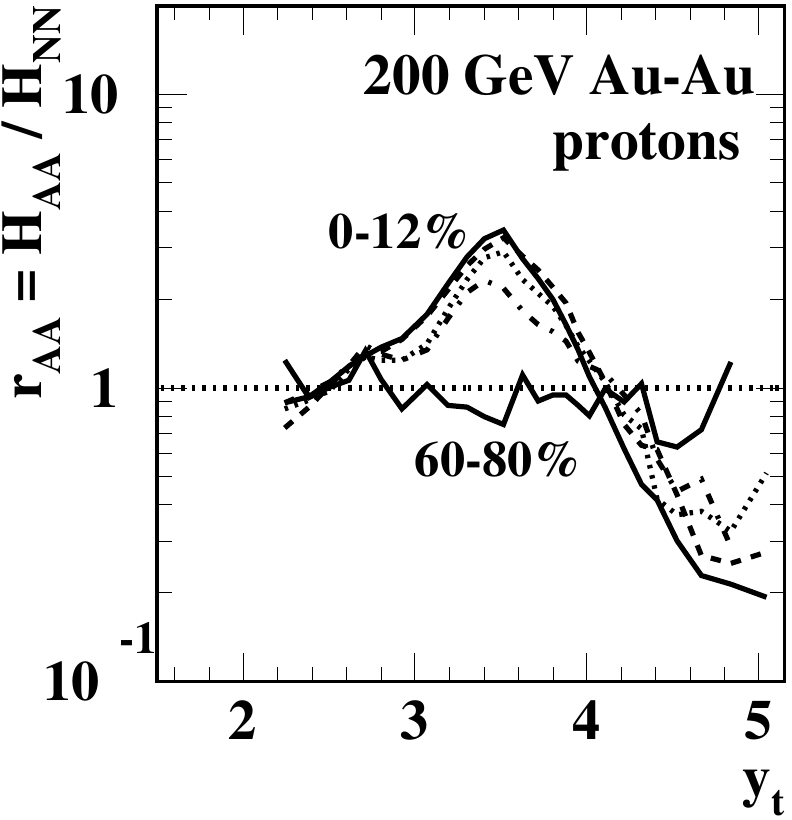}
\caption{\label{pidjet}
Hard component ratios $r_{AA}(y_t)$ for five centralities of 200 GeV \auau\ collisions (curves) for pions (left) and protons (right)~\cite{hardspec}. Also shown are \pp\ data (points, unidentified hadrons) compared to hard-component model $H_{NN}$.
 }
 \end{figure}

Figure~\ref{pidjet} shows hard-component ratios $r_{AA} = H_{AA} / H_{pp}$ for identified pions and protons from five centrality classes of 200 GeV \auau\ collisions~\cite{fragevo}. Whereas conventional spectrum ratio $R_{AA}$ includes the spectrum soft component that strongly biases the ratio below $y_t = 4$ ($p_t \approx 4$ GeV/c) hard-component ratio $r_{AA}$ accurately describes evolution of jet structure down to small momenta ($y_t \approx 2$ or $p_t \approx 0.5$ GeV/c) below which systematic uncertainties in soft-component subtraction become relatively large.

The results for low-mass pions (left panel) are consistent with the hard component in Fig.~\ref{specfrag} (right): suppression at larger $y_t$ and compensating enhancement at smaller $y_t$ for more-central \auau\ collisions~\cite{hardspec,fragevo}. The solid curve in this panel is derived from the central \auau\ points in the previous figure. The points in this panel are from the \pp\ data in Fig.~\ref{specfrag} (left) to provide a reference.  The result for protons (right panel) is surprising. The same factor-five suppression is observed at larger $y_t$ (near 10 GeV/c), but whereas the energy compensation for low-mass pions emerges near 0.5 GeV/c the compensation for massive protons peaks near $y_t = 3.5$ ($p_t \approx 2.5$ GeV/c), substantially above the proton mass ($y_t \approx 2.7$).

The $r_{AA}$ centrality trend is consistent with that for minijet correlations in Ref.~\cite{anomalous}: transparency (GLS, no ``high-$p_t$'' suppression) below $\nu = 3$ corresponding to a 50\% fractional cross section. The suppression/enhancement trend for spectrum hard components corresponds quantitatively to the increase of mean fragment multiplicity inferred from minijet correlations~\cite{jetspec}.

The relation between pion and proton spectrum hard components also relates to the ``baryon-meson'' puzzle~\cite{hardspec,nohydro}. The PID hard-component systematics shown in Fig.~\ref{pidjet} correspond quantitatively to the proton-pion puzzle as presented in Ref.~\cite{barmeson} and Fig.~\ref{baryonmeson}~\cite{hardspec}.

\begin{figure}[h]
  \includegraphics[width=1.65in,height=1.65in]{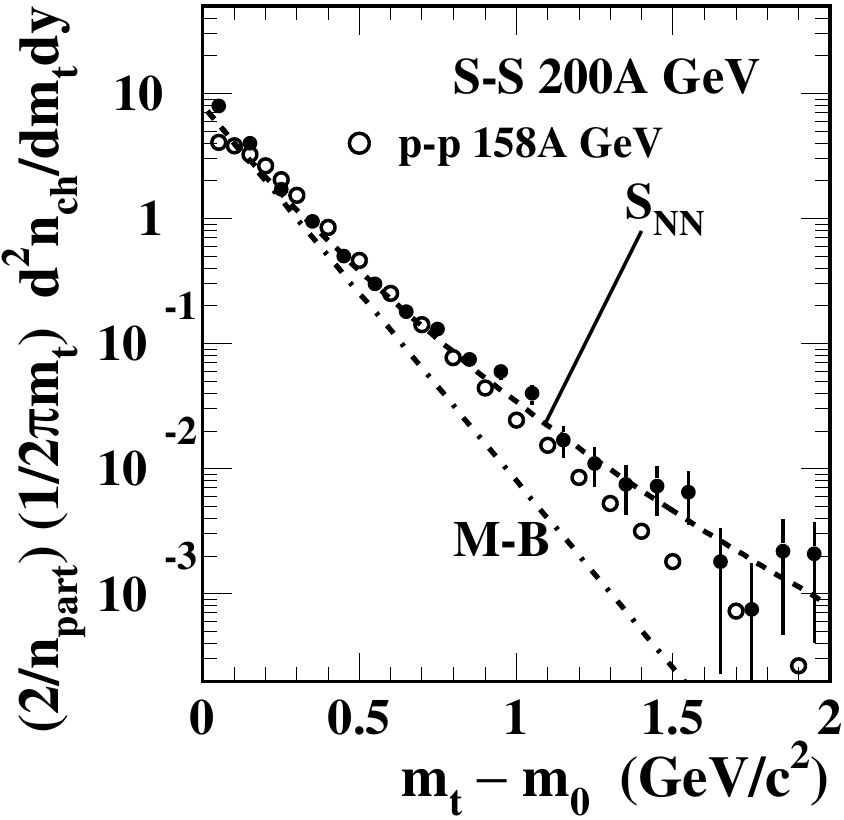}
  \includegraphics[width=1.65in,height=1.65in]{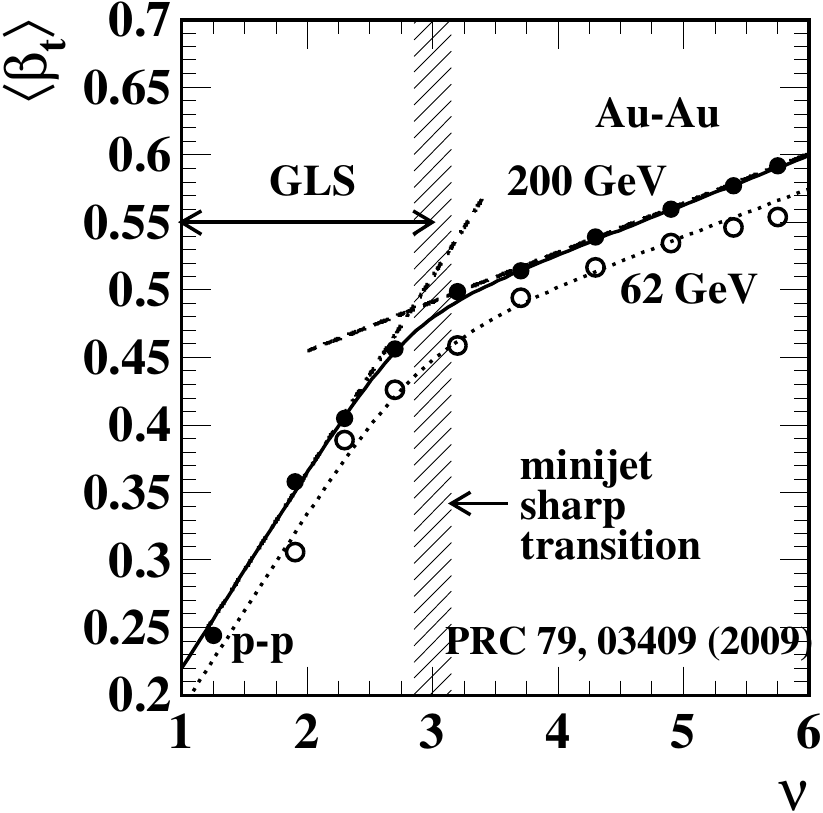}
\caption{\label{specstuff}
Left: 17 GeV \pp\ (open points) and 19 GeV \ss\ (solid points) \mt\ spectra compared to the soft component $S_{NN}$ of the TCM for 200 GeV \auau\ collisions (dashed curve)~\cite{nohydro}.
Right: Radial speed $\langle \beta_t \rangle$ inferred from blast-wave fits to 62 and 200 GeV \auau\ collisions~\cite{starbw}. When plotted on participant path length $\nu$ the relation of the $\langle \beta_t \rangle$ trends to the minijet sharp transition (hatched band) is interesting.
 }   
 \end{figure}

Figure~\ref{specstuff} (left) shows a comparison between soft component  $S_0$ for 200 GeV \pp\ and \auau\ spectra (dashed curve), a SPS S-S spectrum at 19 GeV (solid points), a \pp\ spectrum at the same energy (open points) and a Maxwell-Boltzmann distribution (M-B) with the same slope parameter $T = 145$ MeV~\cite{nohydro}. The S-S data have been interpreted to indicate radial flow based on deviation from the MB distribution~\cite{uli-sshydro}.

Figure~\ref{specstuff} (right) shows radial-flow mean speeds $\langle \beta_t \rangle$ inferred from blast-wave model fits to \pp\ and \auau\ spectra~\cite{starbw}. Substantial radial flow is inferred for \pp\ collisions. It is also notable that radial flow increases most rapidly for \auau\ collisions that exhibit {\em transparency to low-energy partons} (GLS, $\nu \leq 3$) and then less rapidly in a more-central interval where jet modification is strong.

The analysis in Ref.~\cite{hardspec} extracted spectrum hard components for identified hadrons (pions and protons) up to 11 GeV/c from 200 GeV \auau\ collisions. The hard components were later described quantitatively with a pQCD calculation including modified fragmentation functions~\cite{fragevo}. The same structure that is interpreted as radial flow with a blast-wave spectrum model is described by pQCD as a minimum-bias jet contribution to spectra when analyzed with the TCM. The correspondence is seen in Fig.~\ref{specstuff} (right panel). The break in the inferred $\langle \beta_t \rangle$ trend occurs at the sharp transition in minijet properties (hatched band) reported in Ref.~\cite{anomalous}.

\subsection{Jet-related angular correlations}

The minimum-bias angular correlations from 130 GeV \auau\ collisions reported in~\cite{axialci,axialcd} and shown in Figs.~\ref{axialci1} and~\ref{axialcd1} were unexpected and stimulated a followup analysis including detailed centrality systematics with high-statistics data~\cite{anomalous}. 

 \begin{figure}[h]
 \includegraphics[width=1.65in,height=1.65in]{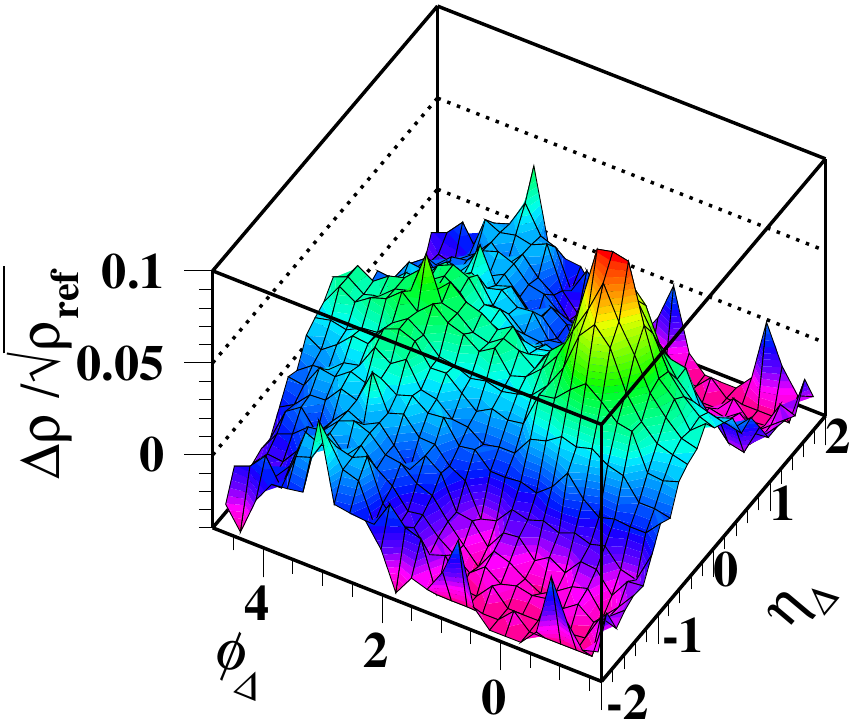}
 \includegraphics[width=1.65in,height=1.65in]{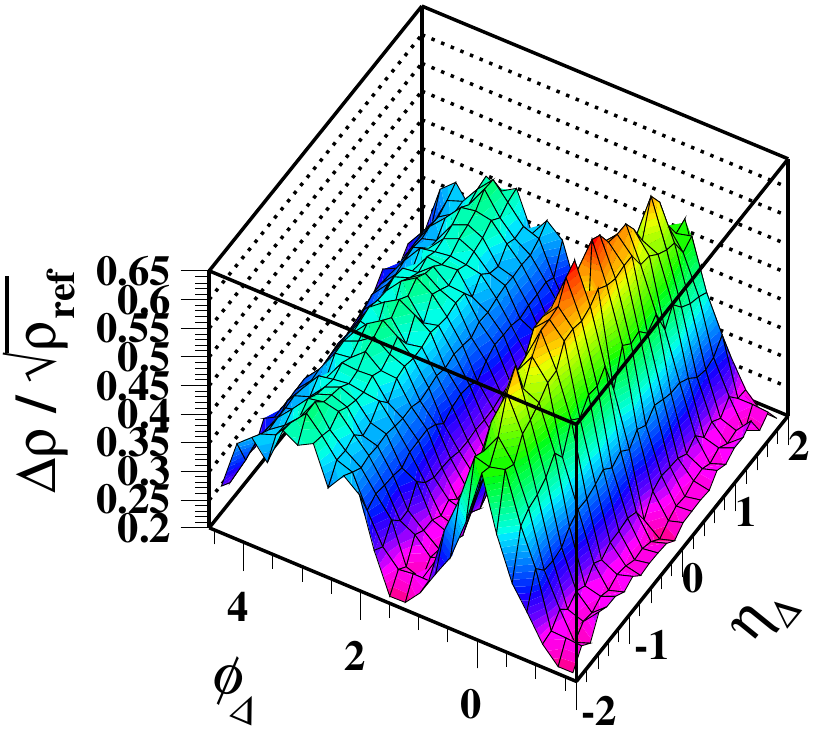}
\caption{\label{axialci2} 2D angular correlations for peripheral (left, $\approx$ \nn\ collisions) and central (right) 200 GeV \auau\ collisions~\cite{anomalous}.
 }   
 \end{figure}

Figure~\ref{axialci2} shows 2D angular correlations from 85-95\% and 0-5\% central 200 GeV \auau\ collisions~\cite{anomalous}. The former is approximately equivalent to \nn\ $\approx$ \pp\ collisions and the structure agrees with such measurements~\cite{porter2,porter3}. The data histograms are accurately described by a simple model function representing three principal data features: (a) a same-side (SS) 2D peak, (b) an away-side (AS) 1D azimuth dipole $\cos(\phi_\Delta -\pi)$ and (c) a nonjet azimuth quadrupole $\cos(2\phi_\Delta)$~\cite{anomalous}. Other model elements represent Bose-Einstein correlations, $\gamma$-conversion electron pairs and projectile-nucleon dissociation.

 \begin{figure}[h]
 \includegraphics[width=1.65in,height=1.65in]{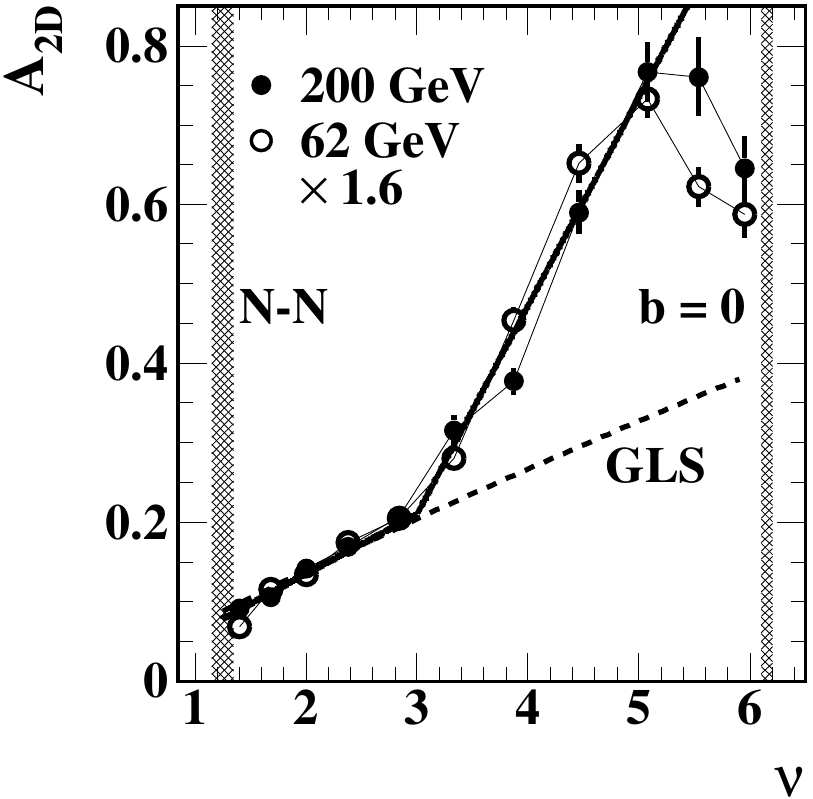}
 \includegraphics[width=1.65in,height=1.65in]{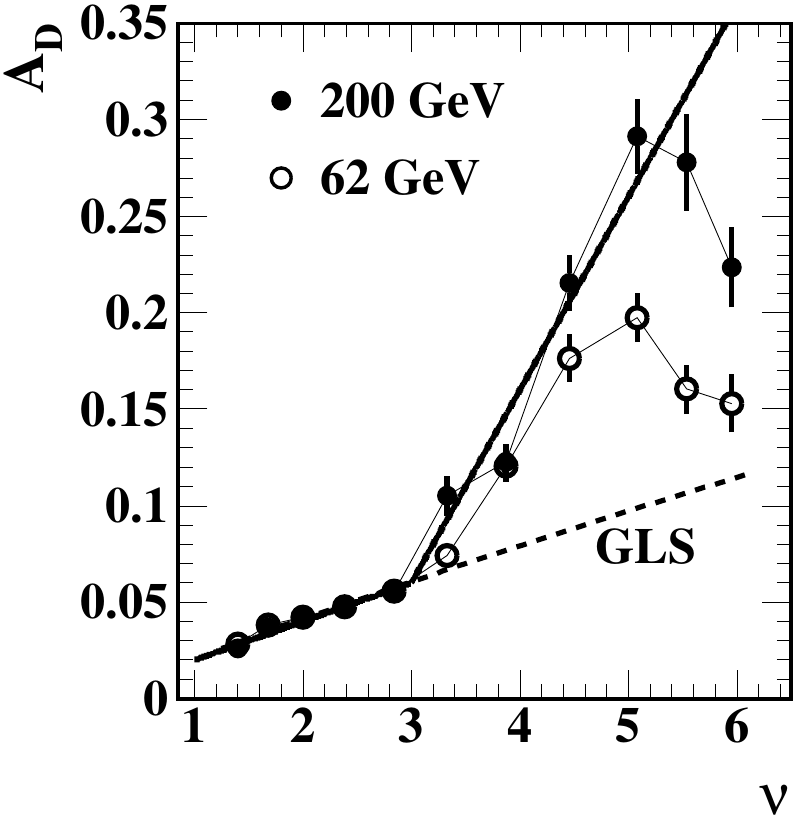}
\caption{\label{miniparams}
2D model fit parameters for the SS 2D peak amplitude (left) and AS 1D peak amplitude (right) from 200 GeV \auau\ collisions~\cite{anomalous}.
 }   
 \end{figure}

Figure~\ref{miniparams} shows the amplitudes for same-side 2D (left) and away-side 1D (right) peak amplitudes~\cite{anomalous}. Below $\nu = 3$ (the sharp transition) the data follow the GLS trend expected for transparent \auau\ collisions. The scaling with energy in the left panel (factor 1.6) is expected for jets  based on previous analysis~\cite{ptedep}. Note that the same {\em jet} energy scaling applies to all centralities, not just the GLS interval below $\nu = 3$. A direct mathematical link has been established between these jet angular correlations and $\langle p_t \rangle$ fluctuations as shown in Fig.~\ref{mpt} and Sec.~\ref{newflucts}.

\subsection{Nonjet azimuth quadrupole} \label{njquad}

Jet-related angular correlations are identified as (a) the SS 2D peak and (b) the AS 1D dipole. What remains is (c) the {\em nonjet quadrupole} plus some small structures (soft component from projectile dissociation, Bose-Einstein correlations and conversion-electron pairs). The nonjet (NJ) quadrupole can be identified with ``elliptic flow'' measurements~\cite{flowmeth}. Jet-related contributions to conventional $v_2$ measurements can be identified with ``nonflow''~\cite{2004}.

 \begin{figure}[h]
  \includegraphics[width=1.65in,height=1.65in]{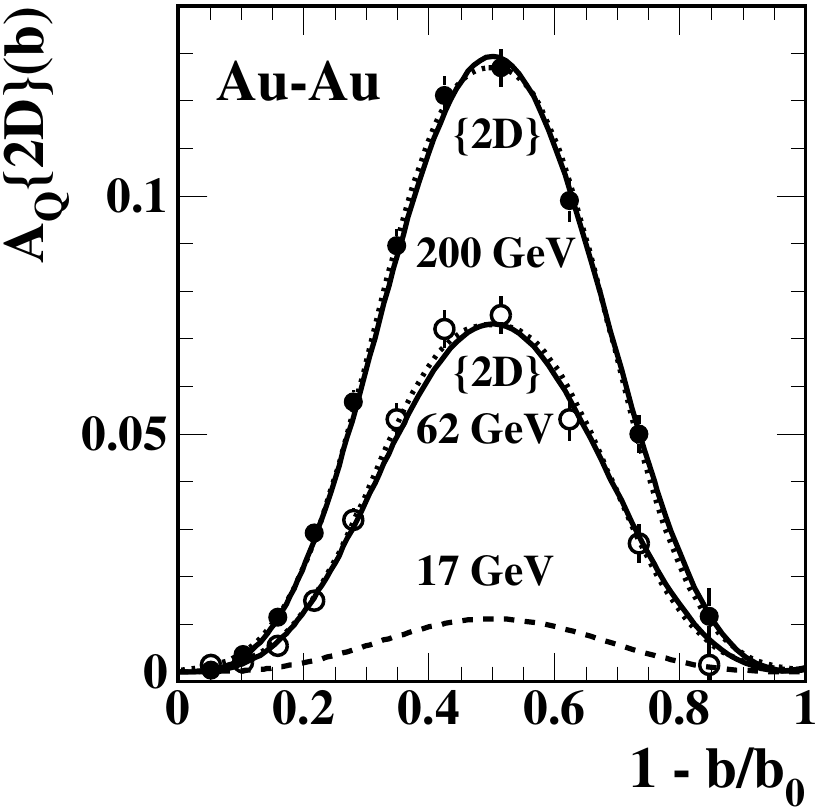}
  \includegraphics[width=1.65in,height=1.65in]{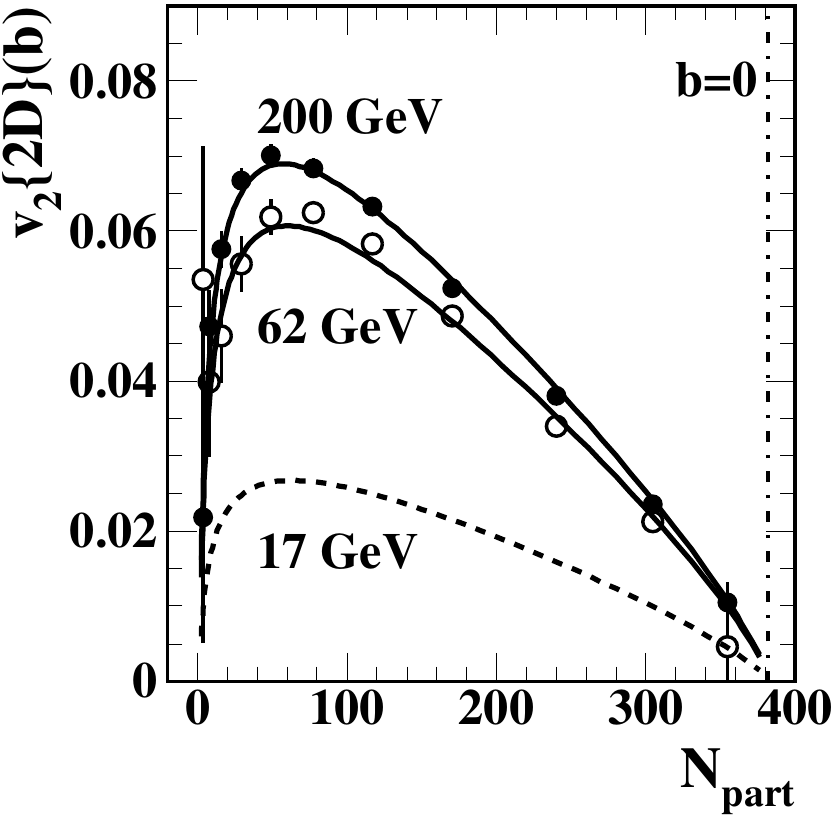}
\caption{\label{quad1}
Left: Nonjet azimuth quadrupole $A_Q\{2D\}(b)$ obtained with model fits to 2D angular correlations from 62 and 200 GeV \auau\ collisions~\cite{davidhq}.
Right: The same data transformed to conventional measure $v_2$. The curves are generated by Eq.~(\ref{quadeqq}).
 } 
 \end{figure}

Figure~\ref{quad1} (left) shows per-particle nonjet quadrupole amplitude $A_Q\{2D\}$ inferred from 2D model fits to angular correlations~\cite{davidhq}. It is notable that the centrality trend is approximately a Gaussian on relative impact parameter $b/b_0$, where $b_0 \approx 14.7$ fm for \auau\ collisions. 

The right panel shows the same data plotted vs centrality parameter $N_{part}$. The $A_Q$ data are converted to conventional measure $v_2$ via $A_Q = \rho_0 v_2^2$ where $\rho_0$ is the mean single-particle 2D angular density. As noted, half of the \auau\ cross section is obscured in the interval below $N_{part} \approx 50$.
The solid and dashed curves are defined by the relation
\bea \label{quadeqq}
A_Q &=& 0.0045 R(\sqrt{s_{NN}}) N_{bin} \epsilon_{opt}^2,
\eea
where $R(\sqrt{s_{NN}}) = \log(\sqrt{s_{NN}} / \text{13.5 GeV})/\log(200/13.5)$ \cite{davidhq}, and $\epsilon_{opt}$ is the eccentricity inferred from the {\em optical} Glauber model. A pQCD prediction based on a color-dipole model gives $v_2 = 0.02$ for \pp\ collisions, consistent with Eq.~(\ref{quadeqq}) extrapolated to \nn\ collisions~\cite{boris}.

 \begin{figure}[h]
  \includegraphics[width=1.65in,height=1.65in]{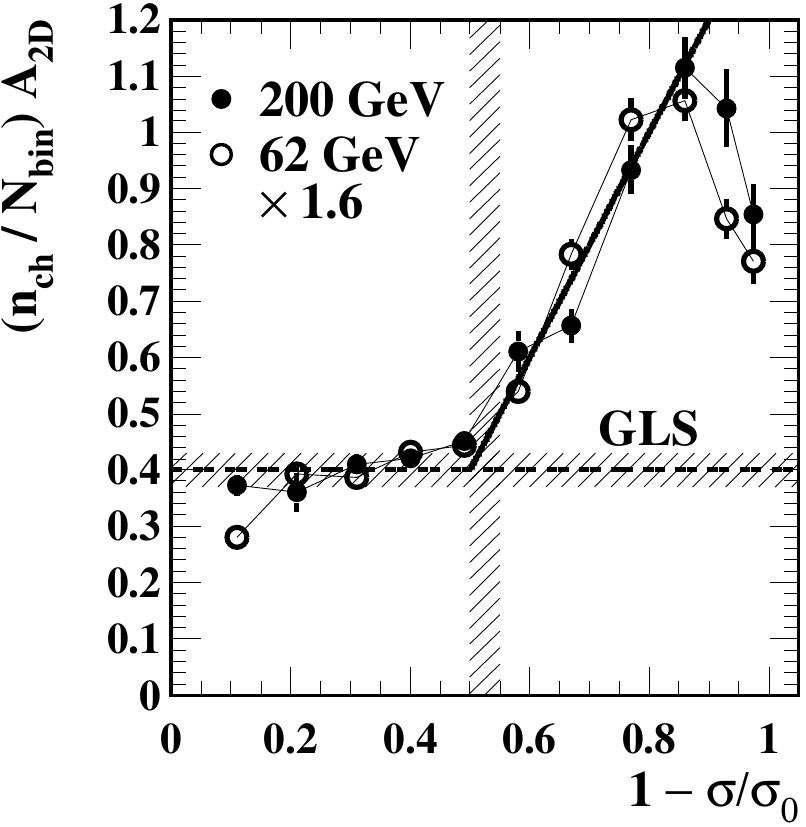}
  \includegraphics[width=1.65in,height=1.65in]{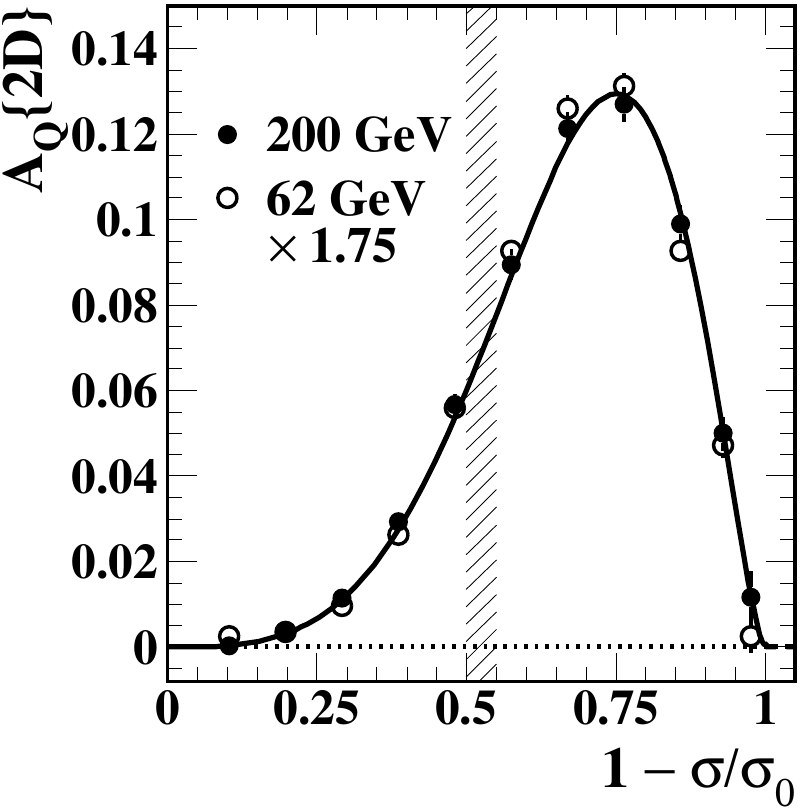}
\caption{\label{quad2}
Left: SS 2D peak amplitudes from Fig.~\ref{miniparams} (left) scaled by the number of \nn\ binary collisions showing strict agreement with the expectation for transparent \auau\ collisions (GLS) over 50\% of the total cross section~\cite{anomalous,nov2}.
Right: NJ quadrupole data from Fig.~\ref{quad1} (left) replotted on fractional cross section. The quadrupole amplitude increases to 60\% of its maximum over an interval where \auau\ collisions are transparent~\cite{davidhq,nov2}.
 } 
 \end{figure}

Figure~\ref{quad2} compares minimum-bias jet data (left panel) with the same nonjet quadrupole data plotted on fractional cross section~\cite{anomalous,davidhq}. It is notable that over the lower 50\% of the fractional cross section where \auau\ collisions appear to be transparent to 3 GeV partons (minijets follow binary collision scaling or GLS) the nonjet quadrupole (right panel) increases to 60\% of its maximum value. One can then ask how hydrodynamic flow is generated without secondary scattering of partons or hadrons~\cite{nohydro,nov2}? It is also notable that the observed energy scaling for jets (1.6, left panel) is similar to that for the nonjet quadrupole (1.75, right panel).

\begin{figure}[h]
  \includegraphics[width=1.65in,height=1.68in]{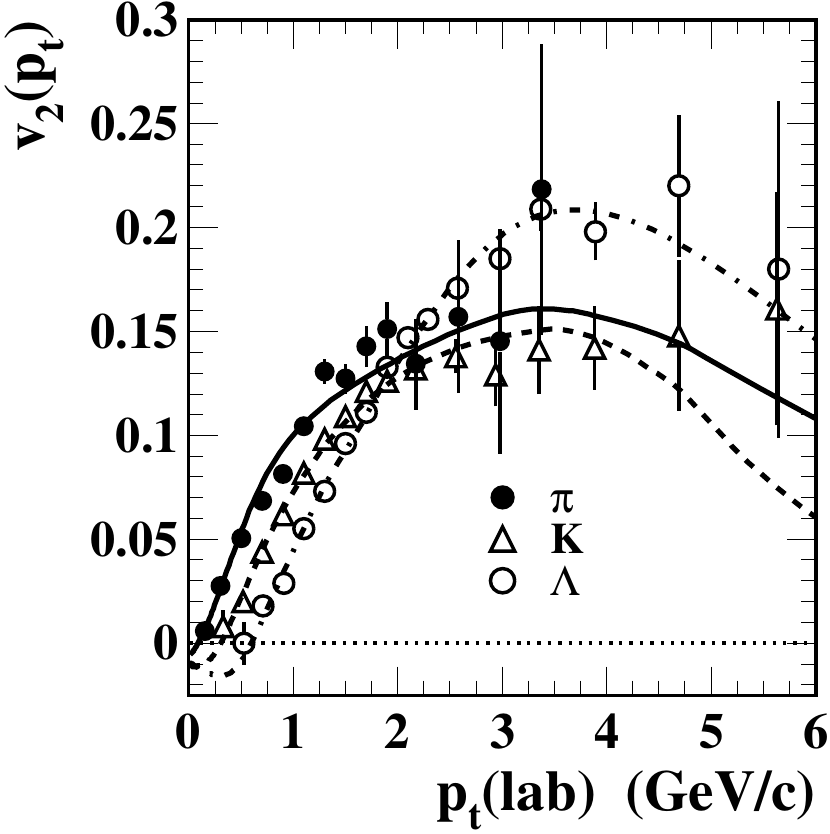}
   \includegraphics[width=1.65in,height=1.65in]{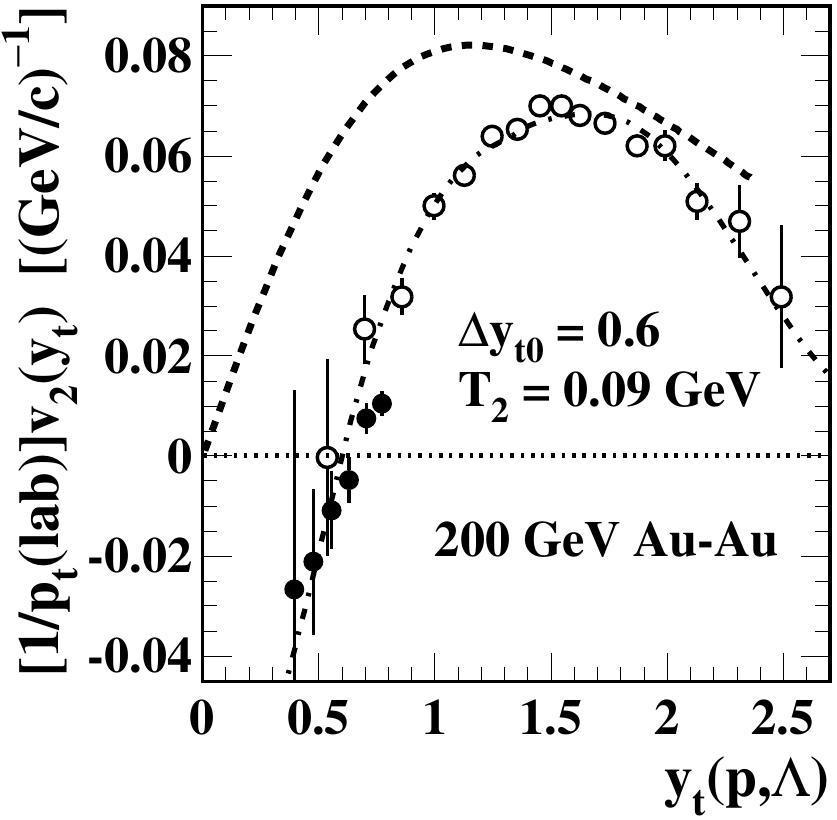}
\caption{\label{quad3}
Left: Published PID $v_2(p_t)$ data for three hadron species from Refs.~\cite{v2pions,v2strange} plotted in the conventional format. The curves are transformed from a common quadrupole spectrum described in the text.
Right:  $\Lambda$ data from the left panel (open points) are replotted in the form $v_2(p_t)/p_t$ vs transverse rapidity $y_t$ calculated with the proper hadron mass. The solid points are taken from the inset in Fig.~\ref{starpid2}~\cite{starpidv2}. The dashed curve is a viscous hydro prediction~\cite{rom}.
 }  
 \end{figure}

\begin{figure}[h]
  \includegraphics[width=1.65in,height=1.65in]{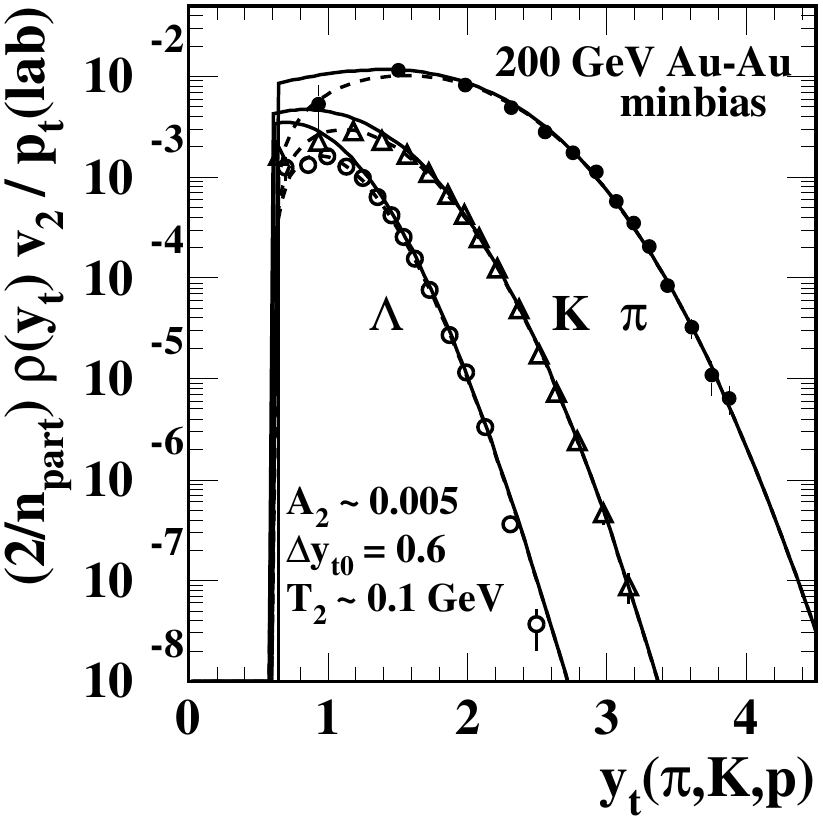}
 \includegraphics[width=1.65in,height=1.65in]{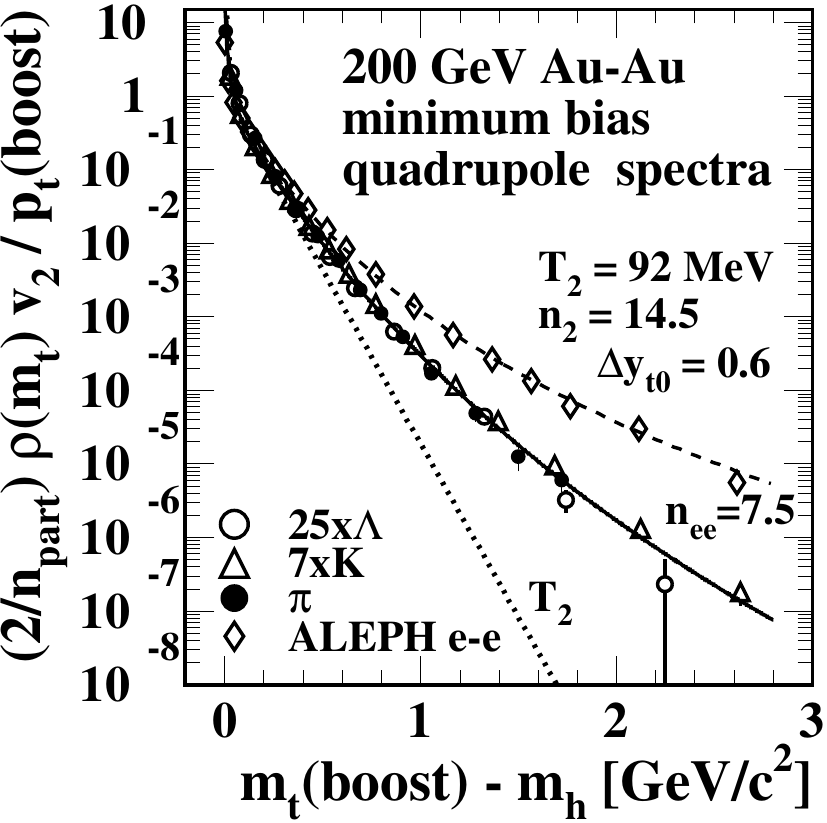}
\caption{\label{quad4}
Left: The data from Fig.~\ref{quad3} multiplied by the single-particle spectrum on $y_t$ for each hadron species revealing a common source boost (horizontal offset).
Right: The data from the left panel transformed to the boost frame (left shifted by $\Delta y_{t0}$), transformed to $m_t - n_h$ and scaled by expected hadron abundances (statistical model) to reveal a common quadrupole spectrum in the boost frame (solid curve).
 }  
 \end{figure}

Figure~\ref{quad3} (left) shows published $v_2(p_t)$ data for minimum-bias 200 GeV \auau\ collisions and identified $\pi$, K and $\Lambda$ (points) plotted in a conventional format~\cite{v2pions,v2strange,quadspec}. The three curves through data are obtained by transformation from a single universal {\em quadrupole spectrum model} described below and shown in Fig.~\ref{quad4}. The mass trend for different hadron species below 2 GeV/c (so-called mass scaling) is said to confirm a hydrodynamic interpretation of $v_2$ data as representing elliptic flow.

Figure~\ref{quad3} (right) shows the $\Lambda$ data from the left panel divided by $p_t$ measured in the lab frame and plotted vs transverse rapidity $y_t = \ln[(m_t + p_t)/m_h]$ (open points), where $m_h$ is the proper hadron mass. The data for all three hadrons species in the left panel then have a common zero intercept at $y_t = 0.6\pm 0.1$. That offset can be interpreted as a quadrupole source boost $\Delta y_{t0}$ (a kind of radial flow). The intercept is best defined by more-massive hadrons, why the $\Lambda$ data are featured. The data are consistent with a {\em narrow} source-boost distribution (an expanding cylindrical shell), whereas (dissipative) hydro theory (dashed curve) assumes Hubble expansion of a dense bulk medium~\cite{rom}. The published PID $v_2(p_t)$ data appear to falsify that model.

The solid points in the right panel are more-recent $\Lambda$ $v_2$ data for 0-10\% central \auau\ collisions shown in Fig.~\ref{starpid2} (inset)~\cite{starpidv2}. Although the maximum $v_2$ value of the more-central data should be smaller, the same zero intercept near 0.6 is indicated. The correspondence is consistent with the observation in Ref.~\cite{davidhq2} that quadrupole source boost $\Delta y_{t0}$ shows {\em no significant centrality dependence} over the most-central 70\% of the total cross section for 62 and 200 GeV \auau\ collisions. The more-recent negative-going $v_2$ data follow the trend (dash-dotted curve) inferred from minimum-bias 2002 and 2004 RHIC PID data in the quadrupole spectrum analysis of Ref.~\cite{quadspec}.

Figure~\ref{quad4} (left) shows the result of a further data transformation. $v_2(p_t)$ is a ratio -- the denominator is the single-particle spectrum represented by $\rho(y_t)$ that includes a strong jet contribution as demonstrated in~\cite{hardspec,fragevo}. The {\em numerator} of $v_2(p_t)$ represents only those hadrons that ``carry'' the nonjet quadrupole. The left panel shows the result of multiplying data in the $v_2/p_t(lab)$ format of Fig.~\ref{quad3} (right) by the corresponding single-particle hadron spectra in the per-participant form $(2/N_{part})\rho(y_t)$. The common source boost $\Delta y_{t0} = 0.6$ is again apparent. The solid curves correspond to a common source $m_t$ spectrum described below. The ratio of solid to dashed curves is a simple relativistic kinematic factor relating $p_t(boost)$ to $p_t(lab)$~\cite{quadspec}.

Figure~\ref{quad4} (right) shows the result of four further operations: (a) transform from lab to boost frame by shifting all spectra in the left panel to the left by $\Delta y_{t0} = 0.6$,  (b) transform spectra in the boost frame from $y_t$ to $m_t$, (c) multiply spectra by ratio  $p_t(lab)/p_t(boost)$, (d) rescale the resulting spectra by the constant factors (1,7,25) indicated in the panel, consistent with a statistical-model prediction for respective hadron species abundances at $T_{chem} \approx 150$ MeV.  Quadrupole data for three hadron species are then quantitatively described by a single L\'evy distribution (solid curve) to the uncertainty limits of the data. The quadrupole spectrum is cold ($T \approx 90$ MeV) and does not correspond to the spectrum representing most hadrons ($T \approx 145$ MeV). The \ee\ data (diamonds) form a \pt\ spectrum from 91 GeV LEP dijets (perpendicular to thrust axis) with a similar low temperature but smaller exponent $n_{ee} \approx 7.5$.

The data in Fig.~\ref{quad4} represent a minimum-bias average over \auau\ centrality. Subsequent analysis has determined that quadrupole source boost $\Delta y_{t0}$ is {\em independent of centrality} to the uncertainty limits of the data~\cite{davidhq2}. We then conclude that from all 200 GeV \auau\ PID $v_2(p_t,b)$ data we obtain two numbers: (a) a nonjet quadrupole amplitude $\Delta y_{t2}$ depending on energy and centrality as Eq.~(\ref{quadeqq}) and (b) a quadrupole source boost $\Delta y_{t0}$ common to all collision systems. The quadrupole spectrum is cold and appears to be universal for all \auau\ collision conditions. The solid curve in Fig.~\ref{quad4} (right) plus Eq.~(\ref{quadeqq}) may then predict all $v_2(y_t,b)$ data for all hadron species.

\subsection{Fluctuations} \label{newflucts}

Fluctuations of event-wise mean $p_t$ (as a proxy for local temperature), the $K/\pi$ ratio (strangeness enhancement) and other event-wise statistical quantities were expected to reveal critical fluctuations associated with a QCD phase boundary separating a hadronic phase from a partonic (or quark-gluon) phase. A number of fluctuation measures were proposed~\cite{tfluctmeth}. Considerable confusion emerged as to measure definitions, interpretations and the effects of limited detector acceptances.

Figure~\ref{mpt} (left) shows measure $\Delta \sigma_{p_t:n}$, a linearized alternative to r.m.s.\ fluctuation measure $\Phi_{p_t}$~\cite{phipt}. Further study~\cite{clt} led to the definition of per-particle {\em variance difference} $\Delta \sigma^2_{p_t:n}$~\cite{ptscale} directly related to 2D number angular correlations by an integral equation~\cite{inverse}.

\begin{figure}[h]
  \includegraphics[width=1.65in,height=1.5in]{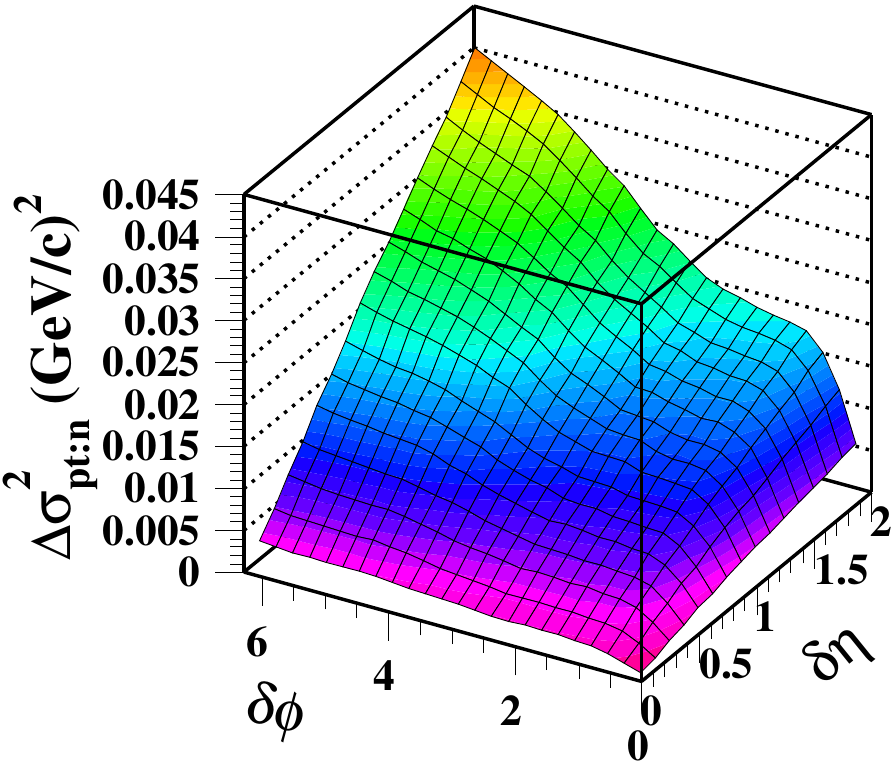}
  \includegraphics[width=1.65in,height=1.5in]{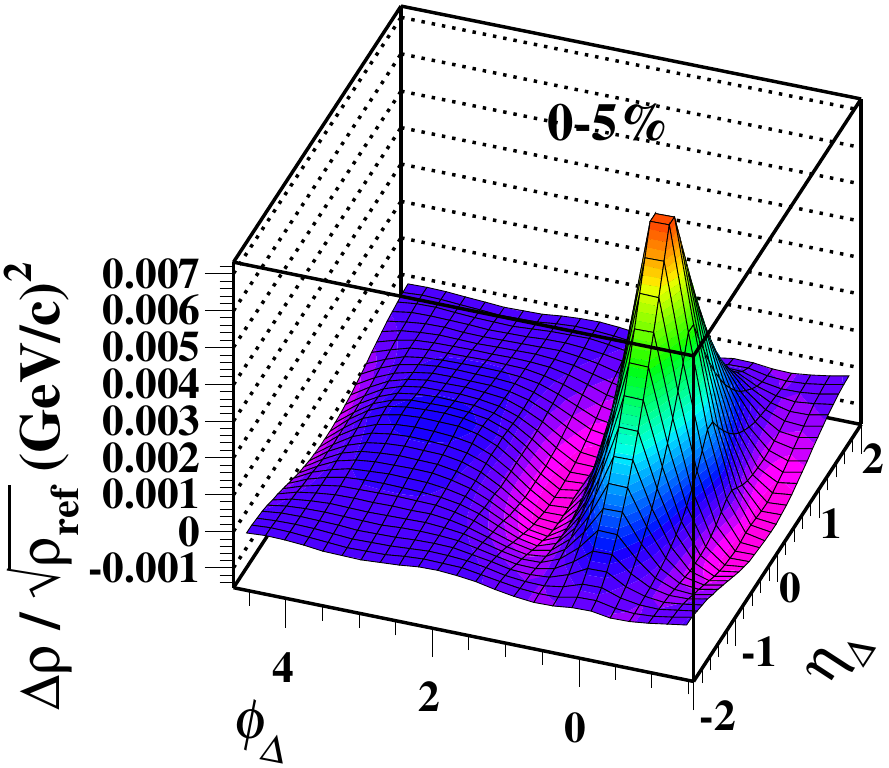}
\caption{\label{flucts}
Left: $p_t$ fluctuation measure $\Delta \sigma_{p_t:n}^2$ (variance difference) plotted vs angle bin sizes $(\delta \eta,\delta \phi)$ for 0-5\% central 200 GeV \auau\ collisions~\cite{ptscale}.
Right: Inversion of the scale (bin size) dependence in the left panel to reveal the underlying angular correlations. Fitted AS 1D dipole and nonjet quadrupole components have been subtracted to emphasize the jet-related SS 2D peak with its catenary shape on $\eta_\Delta$ and neighboring negative undershoots.
 }  
\end{figure}

Figure~\ref{flucts} (left) shows the scale dependence of fluctuation measure $\Delta \sigma^2_{p_t:n}(\delta \eta,\delta \phi)$ with the following properties: (a) The bin-wise $p_t$ variance at each angular scale $(\delta \eta,\delta \phi)$ is compared to a statistical reference, (b) the measure is sensitive to $p_t$ fluctuations {\em given} a bin-wise multiplicity $n$ condition ($p_t:n$)~\cite{ptscale}. The general trend is increase with increasing bin size or scale ($\delta \eta,\delta \phi$). The maximum value in this case corresponds to the STAR TPC acceptance $\Delta \eta,\Delta \phi = 2,2\pi$. Single fluctuation measurements from detectors having different angular acceptances correspond to different points on that surface.

Figure~\ref{flucts} (right) shows the result of {\em inversion} of the scale dependence of the fluctuation data in the left panel. The relation can be described by an integral equation, and the integral can be inverted by standard methods to reconstruct the underlying angular correlations that generated the fluctuations~\cite{inverse}. We find that $\langle p_t \rangle$ fluctuations supposedly related to a QCD phase boundary are actually driven by jet correlations. The angular correlations in the right panel inferred by fluctuation inversion were confirmed by direct pair-counting methods. The negative undershoots near the SS 2D peak on azimuth were also confirmed by direct correlation analysis. The result for peripheral \auau\ collisions corresponds to the expected jet structure obtained from \pp\ collisions. Fitted AS dipole and nonjet quadrupole components have been subtracted to emphasize the SS 2D peak as in Fig.~\ref{axialci1}.

Reference~\cite{ptedep} reports the energy dependence of \pt\ fluctuations and angular correlations from SPS to RHIC. A simple $\log(\sqrt{s_{NN}}/\text{10 GeV})$ trend is observed above 10 GeV, further supporting a jet interpretation. Again the similarity with the energy dependence of the nonjet quadrupole $\log(\sqrt{s_{NN}}/\text{13.5 GeV})$ is notable~\cite{davidhq} (Fig.~\ref{quad2}). The two intercept points on energy are consistent within systematic uncertainties and are kinematically consistent with the first significant penetration of projectile nucleons significantly below valence quarks on momentum fraction $x$.

\section{Discussion} \label{disc}

Two distinct approaches have been applied to RHIC heavy ion data. (a) Evidence for a conjectured quark gluon plasma has been sought in the form of ``signals'' recovered by model-dependent analysis methods designed  to detect them. Priority has been given to results that seem to confirm QGP formation. (b) Significant structures in yields, spectra and correlations are revealed by differential methods not depending on a priori {\em physical} models. Significant structures are represented by simple functional forms, and the systematic variation of model-function parameters with collision conditions is obtained. Model-parameter systematics are then compared with physical models, giving first priority to QCD descriptions of elementary collisions. The two approaches have led to quite different interpretations of RHIC data. In this section we consider the differences and their possible implications.

\subsection{Initial-state energy density}

Estimating the initial conditions (mid-rapidity gluon and energy densities) for \aa\ collisions at RHIC is said to be essential to support claims of densities and temperatures sufficiently high to produce deconfined quarks and gluons according to LQCD and to support a hydrodynamic description in terms of large initial pressure gradients~\cite{perfliq2}. The initial-state energy density is conventionally estimated by the Bjorken formula
\bea
\langle \epsilon \rangle &\approx& \frac{1}{\tau A} \frac{dE_t}{d\eta} \\ \nonumber
&\approx& \frac{1}{\tau A} \langle m_t \rangle \frac{dn_{ch}}{d\eta} \\ \nonumber
&\propto& N_{part}^{1/3} [1 + 0.1 (\nu - 1)],
\eea
where the last line assumes overlap area $A \propto N_{part}^{2/3}$ and we invoke the two-component expression for $dn_{ch}/d\eta$ from Ref.~\cite{kn}. Reference~\cite{whitephen} estimates $\langle \epsilon \rangle \approx$ 15 GeV/fm$^3$ for central 200 GeV \auau\ collisions based on $\langle m_t \rangle \approx 0.85$ GeV/$c^2$, seemingly well above the threshold for producing a QGP. But what is the energy density at the sharp transition reported in Ref.~\cite{anomalous}? In that case $N_{part} \approx 50$ rather than 380 and $\nu \approx 3$ rather than 6. Based on the centrality scaling above the energy density is reduced by only a factor 0.4, to 6 GeV/fm$^3$, nominally well above the critical density required for QGP formation. But spectrum and angular correlation data indicate that \auau\ collisions are still transparent at that centrality, with no evidence for parton or hadron secondary scattering~\cite{anomalous}. Applying the Bjorken formula assuming that all produced $E_t$ is subject to thermalization may greatly overestimate the relevant energy density. Much of the final-state $E_t$ may be carried by hadrons (and parent partons) that experienced {\em no secondary interactions}.

\subsection{Initial conditions, the TCM and minijets} \label{nomini}

The HEP two-component model based on projectile dissociation and minijet (minimum-bias jet) production has been discounted as a description of initial conditions at RHIC based on comparisons between HIJING simulations and measured hadron production systematics at mid-rapidity~\cite{perfliq2,global3,2compphob}. Reference~\cite{liwang} states that ``...the available RHIC experimental data...can already rule out the simple two-component model without nuclear modification of the parton distributions in nuclei....'' The basis for that conclusion is a TCM for \pp\ collisions given by
\bea \label{kneq}
\frac{dn_{ch}}{d\eta} &=& n_s + n_h \frac{\sigma_{jet}}{\sigma_{inel}},
\eea
where $n_s = 1.6$ and $n_h = 2.2$ (dijet fragment multiplicity) are assumed~\cite{liwang}. The measured charge density for 200 GeV NSD \pp\ collisions is about 2.5, implying that the cross-section ratio must be about 0.4. Given that $\sigma_{inel} \approx 60$ mb or more in that reference the calculated $\sigma_{jet}$ must be at least 25 mb, more than half the actual \pp\ inelastic cross section 42 mb. It is not clear how a pQCD calculation can determine such a nonperturbative cross section.

The generalization of Eq.~(\ref{kneq}) to \aa\ collisions is given by~\cite{kn}
\bea \label{knaa}
\frac{2}{N_{part}} \frac{dn_{ch}}{d\eta} &=& n_{pp} [1 + x(\nu - 1)],
\eea
where $n_{pp} \approx 2.5$ represents the 200 GeV \pp\ NSD $\eta$ density and $x = n_h \sigma_{jet} / n_{pp} \sigma_{inel}$ is the same parameter appearing in the TCM from Ref.~\cite{kn}. The value  $x = 0.35$ inferred from Ref.~\cite{liwang} can be compared with $x \approx 0.1$ inferred empirically from {\em more-central} 200 GeV \auau\ data. Reference~\cite{liwang} concludes that minijet production must be suppressed by ``shadowing'' in \auau\ collisions.

But the HIJING (PYTHIA) Monte Carlo characterization of \pp\ collisions is quantitatively inconsistent with \pp\ data. Reference~\cite{ppprd} reported a TCM for \pp\ spectra from which a dijet production probability could be inferred corresponding to a 2.5 mb dijet total cross section~\cite{fragevo} (Fig.~\ref{ppspec} of this paper). That cross section is consistent with a parton spectrum lower bound near 3 GeV (dijet energy $Q = 6$ GeV) and also consistent with the original minijet observation by UA1~\cite{ua1} and theoretical descriptions thereof~\cite{sarc,kll}. In Eq.~(\ref{knaa}) the corresponding value is $x \approx 0.03$ (GLS extrapolation), more than ten times smaller than assumed in Ref.~\cite{liwang}. The jet cross section is overestimated in part because the steep power-law parton spectrum is assumed to extend down to $p_0 = 2$ GeV, well below 3 GeV inferred from spectra and correlations~\cite{ppprd,fragevo,jetspec}. The fragment production in more-central \auau\ collisions is actually {\em underestimated} by extrapolation from measured \pp\ {\em data} (GLS). Modified fragmentation leads to the larger $x \approx 0.1$ in more-central collisions (Figs.~\ref{midrap} and~\ref{specfrag} of this paper). The parton spectrum cutoff, and therefore jet  cross section, is nonperturbative and must be inferred from data.

Figure~\ref{phob1} (left) compares HIJING predictions (dotted lines) with data (points) at 20 and 200 GeV. The HIJING slope at 200 GeV (hard component) is inconsistent with the data, one basis for discounting the minijet TCM. But setting aside the Monte Carlo simulations the hard component inferred from \pp\ {\em data}~\cite{ppprd} is ten times smaller and underpredicts the \auau\ data. Within the lower 50\% of the cross section (below $N_{part} \approx 50$, not shown in those plots) the \pp\ extrapolation accurately describes more-peripheral \auau\ data~\cite{jetspec}. Above that point (sharp transition~\cite{anomalous}) fragment production increases by a factor 3-4 as jet production is modified by the medium~\cite{hardspec,fragevo}.
The two-component minijet model accurately describes hadron
production (yields, spectra and correlations) in terms of pQCD parton
scattering and nonperturbative (measured) fragmentation to jets. The relation between HIJING simulated yields and correlations and \auau\ measured correlation data is discussed in detail in Sec.~VIII-I of Ref.~\cite{anomalous} .

\subsection{The CGC vs HIJING and data}

Reference~\cite{perfliq2} concludes ``If the initial conditions were not well constrained at RHIC, then the conclusion that a QGP was formed could not be sustained. [...] ...the surprising very weak centrality and beam energy dependence...is most satisfactorily explained and predicted by the CGC.... This is one of the strongest lines of empirical evidence...that the CGC initial state...is formed....'' The reference is to HIJING simulations compared to \auau\ data as in Fig.~\ref{phob1} of this paper. As noted, HIJING with parton spectrum lower bound $p_0 = 2$ GeV greatly overpredicts the gluon density or minijet production compared to what is inferred directly from \pp\ collisions.  A CGC estimate based on a saturation scale $Q_s \approx$ 1-1.5 GeV is {\em even larger}, leading to inference of an opaque (gluonic) medium. And the centrality variation of the CGC estimate is seen to disagree strongly with the measured centrality trend when observed over the entire centrality interval as in Fig.~\ref{midrap}, not just more-central collisions~\cite{glasma1}. 

\subsection{Jet contributions to spectra and correlations}

Blast-wave (BW) fits to \pt\ spectra are employed to infer bulk-matter properties such as temperatures and radial flow speeds (Figs.~\ref{bulk} and~\ref{lowpt}). The assumption that spectrum structure actually represents such quantities can be questioned~\cite{nohydro}. In Fig.~\ref{specfrag} (right) we observe that a pQCD description of the pion spectrum hard component provides an accurate and quantitative description for all \auau\ centralities. In Fig.~\ref{specstuff} (right panel) we find that the radial speed inferred from BW fits is rising most rapidly over a centrality interval (GLS) where \auau\ collisions appear to be transparent according to jet angular correlations~\cite{anomalous}. The nonzero $\langle \beta_t \rangle$ inferred for 200 GeV \pp\ collisions characterizes a \pt\ spectrum shape similar to that for 19 GeV \ss\ collisions. Does the departure from a M-B distribution signal radial flow or a universal aspect of longitudinal nucleon or parton fragmentation?

Baryon-meson spectrum ratios (Fig.~\ref{baryonmeson}) are interpreted to favor hadronization from a thermalized partonic medium by recombination of constituent quarks. However, the spectrum hard components in Fig.~~\ref{pidjet} produce the same B/M anomaly but can be associated with parton fragmentation to jets~\cite{hardspec}. The most-central pion data from Fig.~~\ref{pidjet} are the same data (points) described by pQCD (solid curve) in Fig.~\ref{specfrag} (right).

Jet quenching is inferred from spectrum ratios (Fig.~\ref{highptraa}) that reveal jet structure only above 4 GeV/c. In contrast, spectrum hard-component ratios (Fig.~\ref{pidjet}) reveal jet structure down to 0.5 GeV/c ($y_t \approx 2$). Hadron {\em suppression} at 10 GeV/c is accompanied by much greater {\em enhancement} at 0.5 GeV/c (for pions) consistent with a simple mechanism for modified fragmentation~\cite{fragevo}. The low-$p_t$ enhancement remains within the resolved jet as in Fig.~\ref{axialci2}~\cite{anomalous,jetspec,nohydro}, with jet parameters as in Fig.~\ref{miniparams}.

Jet correlation structure on 1D azimuth is inferred by ZYAM subtraction of a combinatorial background with certain trigger-associated \pt\ cuts imposed (Figs.~\ref{zyamx} and~\ref{zyamy}). The ZYAM subtraction method assumes that SS and AS jet peaks do not overlap and that published $v_2$ data are unbiased. However, we observe that minimum-bias SS and AS jet peaks overlap strongly on azimuth and that much of the published $v_2$ data include strong biases from jet contributions (nonflow)~\cite{davidhq,davidhq2} leading to (a) oversubtraction of the constant offset and sinusoid and (b) artificial suppression and distortion of ZYAM-inferred jet structure, including an AS double peak and inference of ``Mach cones''~\cite{tzyam}.

Jet correlation structure on 2D $(\eta,\phi)$ with trigger-associated \pt\ cuts imposed reveals substantial changes in the SS 2D jet peak shape with increasing \auau\ centrality (Figs.~\ref{starridg} and~\ref{starnear}). The $\eta$-elongated SS peak in more-central collisions is separated into a ``jet-like'' part narrow on $\Delta \eta$ and a ``ridge-like'' part broad on $\Delta \eta$. It is notable that both parts share the same azimuth width and the same energy systematics. But the ``ridge'' is attributed to nonjet mechanisms. Minimum-bias 2D angular correlations (no trigger-associated \pt\ cuts) show a single monolithic SS peak well-described by a 2D Gaussian for all centralities~\cite{anomalous}. The peak volume corresponds to pQCD predictions for jet systematics~\cite{jetspec}. \pt\ angular correlations from fluctuations (Fig.~\ref{flucts}) indicate jet-like elevated $p_t$ values in the elongated SS peak region on $\eta_\Delta$ smoothly connected to the central region with no discontinuity. The SS 2D peak structure resulting from trigger-associated \pt\ cuts represents a small fraction of the minimum-bias jet peak.

\subsection{The nonjet quadrupole and elliptic flow}

``Elliptic flow'' $v_2$ is conventionally measured by several nongraphical numerical methods (NGNM) as in Fig.~\ref{v2nf}. The range of results for different analysis methods is the basis for estimating systematic uncertainties. The source of such variation may be dominated by jets~\cite{davidhq2}. An alternative method based on model fits to 2D angular correlations as in Fig.~\ref{quad1} returns reproducible data that distinguish accurately between jet structure and a nonjet azimuth quadrupole~\cite{davidhq}.

So-called mass scaling of $v_2(p_t)$ below 2 GeV/c is said to confirm a hydrodynamic phenomenon. However, in Fig.~\ref{quad2} the nonjet quadrupole amplitude is observed to rise to 60\% of its maximum value over the GLS interval (lower 50\% of total cross section) where \auau\ collisions are {\em transparent to 3 GeV partons}: the SS peak amplitude scales with the number of \nn\ binary collisions while the peak shape is unchanged from \pp\ collisions~\cite{nov2}.

In Figs.~\ref{phencq} through \ref{starpid2} attempts are made to demonstrate  so-called constituent-quark scaling above 2 GeV/c interpreted to demonstrate flow of a quark-gluon medium in which constituent quarks play a prominent role. But Figs.~\ref{quad3} and \ref{quad4} demonstrate that $v_2(p_t)$ for all hadron species can be predicted from a single universal spectrum shape with these characteristics: (a) fixed source boost $\Delta y_{t0} \approx 0.6$, (b) temperature $T_2 \approx 90$ MeV, (c) L\'evy shape with exponent $n \approx 14.5$. Similar analysis of centrality dependence reveals that $\Delta y_{t0}$ does not vary significantly with \auau\ centrality over the interval 70-0\%~\cite{davidhq2}. In Fig.~\ref{quad3} (right) we find that the boost distribution inferred from $v_2(p_t)$ data is very different from that assumed in hydro predictions for a Hubble-expanding medium.

So-called $v_2$ ``scaling'' methods tend to obscure the sought-after phenomenon. Hydro theory can only predict the velocity (rapidity boost) profile of a flowing medium. Hadron details must be supplied by a separate model. If $v_2$ data can be manipulated to reveal the hadron source boost then {\em direct comparisons} can be made between data and hydro. In Figs.~\ref{quad3} (right) and \ref{quad4} the common source boost for all hadrons can be read directly as intercept $\Delta y_{t0}$ with good sensitivity. On \pt\ the intercept points for different hadron masses become $p_{t0} = m_h \sinh(\Delta y_{t0}) \approx m_h \Delta y_{t0} \approx 0.6 m_h$ (so-called mass scaling). If the data are transformed to ``kinetic energy'' $KE_t$ (or equivalently transverse mass) the intercept points become $m_{t0}-m_h = m_h [\cosh(\Delta y_{t0})-1] \approx 0.5 m_h (\Delta y_{t0})^2 \approx 0.18 m_h$. The sensitivity to boost is further reduced by a factor 3. Plotting on $y_t$ maximizes sensitivity to hydro boost predictions while plotting on $m_t - m_h$ {\em minimizes} any sensitivity to the nominal theory.

Certain differences between $v_2$ methods interpreted to indicate nonflow and/or flow fluctuations as in Fig.~\ref{v2flucts} (right) are quantitatively predicted by the properties of the SS 2D peak from jet correlations~\cite{anomalous,multipoles}. The $v_3$ (triangular flow) data in Fig.~\ref{starv3} correspond exactly to the SS 2D jet peak in 2D angular correlations as in Fig.~\ref{axialci2}. A similar correspondence is shown for other ``higher harmonics''~\cite{multipoles,sextupole}.

$v_2(p_t)$ is a ratio---the denominator is the single-particle spectrum $\rho(p_t)$ including a strong jet contribution over a broad \pt\ interval; the numerator includes as a factor the quadrupole spectrum $\rho_2(p_t)$ with no significant jet contribution~\cite{quadspec}.  From Fig.~\ref{quad4} we learn that $\rho_2(p_t)$ is a cold spectrum very different from the single-particle spectrum describing most hadrons. It is possible that the nonjet quadrupole is actually ``carried'' by a small fraction of all hadrons. Ideal hydro calculations implicitly assume a common spectrum for almost all hadrons which should cancel in the $v_2$ ratio. The falloff from ideal hydro in Fig.~\ref{starpid1} and similar results reflects the differences in the two spectra combined in ratio. Only the numerator of $v_2(p_t)$ has significance for ``elliptic flow.'' Hydro theory cannot in principle describe the $v_2$ ratio over significant \pt\ intervals.

\subsection{Gluons in dijet and  quadrupole production}  \label{energy}

The possible relations among soft gluons released during projectile nucleon dissociation, midrapidity dijets and the nonjet quadrupole are intriguing. If one sets aside a hydrodynamic mechanism the sharply-peaked $\eta$ structure in Fig.~\ref{highway} (right) suggests a low-$x$ glue origin for the nonjet quadrupole~\cite{gluequad}. In Fig.~\ref{starbes} (lower right) the $\eta$ dependence is shown to scale closely with beam rapidity.

The energy dependence of dijet production is determined largely by the depth on $x$ of the nucleon PDF probed by a given collision energy. Once the longitudinal gluon flux is so determined dijet production by scattering is determined by the dijet energy scale $Q$ with fixed lower bound near 6  GeV. Soft and hard hadron production should then vary similarly with energy. 
We observe that the nonjet quadrupole scales with energy in essentially the same way, suggesting low-$x$ glue as a common origin for soft, hard and quadrupole. pQCD calculations based on a color-dipole model predict $p_t$-integral $v_2 = 0.02$ for \pp\ \cite{boris} consistent with peripheral \auau\ collisions~\cite{davidhq}. Thus, dijets and nonjet quadrupole may have a common gluonic origin but represent two limiting cases of QCD: short-wavelength and long-wavelength gluonic radiation~\cite{gluequad}.

\section{Summary} \label{summ}

The RHIC accelerator complex has been a major technical success. The development of computing, software and analysis methods to deal with petabytes of particle data have required an enormous effort by well over 1000 persons. The product has been a massive inventory of analysis results reported in hundreds of published papers relating to a very complex question: what physical mechanisms dominate ultrarelativistic heavy ion collisions and what state of matter (if any) results?

Strong theoretical arguments have been advanced to support the claim that a {\em strongly-coupled} quark gluon plasma (sQGP) is formed in more-central \auau\ collisions with a remarkably small shear viscosity, warranting the designation ``perfect liquid.'' The principal experimental observations invoked are large-amplitude elliptic flow $v_2$ data described to good approximation by ideal hydrodynamics (indicating small dissipation) and strong jet quenching, apparent five-fold reduction of jet fragment yields by a dense QCD medium nearly opaque to colored quarks and gluons. Other evidence includes strong radial flow inferred from $p_t$ spectra and mid-rapidity hadron yields indicating large initial gluon (color glass condensate) and transverse-energy $E_t$ densities that may drive the hydrodynamic flows.

However, a two-component model (TCM) of \aa\ collisions based on {\em measured} properties of \pp\ collisions is found to describe accurately the centrality trends from \auau\ collisions across the more-peripheral half of the total cross section. And the systematics of jet-related structure in the more-central half is still described in part by perturbative QCD (pQCD). Within the same more-peripheral interval a ``quadrupole'' measure, alternative to $v_2$ but describing the same correlation structure identified as elliptic flow, increases to 60\% of its maximum value within peripheral \auau\ collisions where no secondary scattering of partons or hadrons is observed. And in more-central collisions the same simple quadrupole trend remains unchanged, although jet structure (parton fragmentation) is strongly modified there.

Differential analysis of $v_2(p_t)$ data reveals a quadrupole {\em source boost} distribution and unique quadrupole $m_t$ spectrum that may be compared directly with hydro predictions. The comparison contradicts hydro models based on Hubble expansion of a bulk medium as assumed for sQGP. The quadrupole spectrum is quite different from the spectrum describing most final-state hadrons, suggesting that the quadrupole identified with elliptic flow is actually carried by a small minority of all final-state hadrons. Other results for hadron production, $p_t$ spectrum structure and jet correlations also conflict with the sQGP paradigm.

The correct model for high-energy heavy ion collisions may lie on a continuum between production of projectile and scattered-parton fragments (hadrons) described by the two-component model and late hadronization of a thermalized weakly-coupled QGP as limiting cases. The sQGP conjecture already retreats along that continuum from the asymptotic weakly-coupled QGP. However, further paradigm change may be necessary.  Three major questions emerge from a review of RHIC experimental results: (a) What is the mechanism for the nonjet azimuth quadrupole? (b) What mechanism modifies parton fragmentation to jets in more-central \auau\ collisions? (c) Is there any {\em necessary} role for hydrodynamics in RHIC collisions? Some of the exceptions to the standard heavy-ion paradigm emerging in recent years suggest that QCD is a richer field theory than was previously imagined.

I thank Edward Sarkisyan-Grinbaum for proposing this review project. I appreciate extensive contributions over years from Duncan Prindle, David Kettler, Jeff Reid and Dhammika Weerasundara from the University of Washingon and from Lanny Ray, Aya Ishihara, Michael Daugherity and Elizabeth Oldag from the University of Texas at Austin.
This work was supported in part by the Office of Science of the U.S.\ DOE under grant number DE-FG03-97ER41020. 

\end{document}